\documentclass[twocolumn,amssymb,superscriptaddress,floatfix]{revtex4}
\usepackage[maxfloats=45]{morefloats}
\usepackage{subfigure}
\usepackage{graphicx}
\usepackage{rotating} 
\usepackage{color}
\usepackage{lscape}
\usepackage{rotating}
\usepackage{xfrac}
\usepackage{amsmath,amsthm} 
\usepackage{epstopdf}
\newcommand\abs[1]{\left|#1\right|}

\usepackage{xr}
\usepackage{afterpage}
\usepackage{bm}
\usepackage{epstopdf}
\vspace*{0.2in}

\begin{document}
\title{Decoding Single Molecule Time Traces with Dynamic Disorder} 
\author{Wonseok Hwang}
\affiliation{Korea Institute for Advanced Study, Seoul 02455, Republic of Korea}
\author{Il-Buem Lee}
\affiliation{Department of Physics, Korea University, Seoul 02841, Republic of Korea}
\author{Seok-Cheol Hong}
\affiliation{Department of Physics, Korea University, Seoul 02841, Republic of Korea}
\author{Changbong Hyeon}
\thanks{hyeoncb@kias.re.kr}
\affiliation{Korea Institute for Advanced Study, Seoul 02455, Republic of Korea}

\date{\today}
\begin{abstract}
Single molecule time trajectories of biomolecules provide glimpses into complex folding landscapes that are difficult to visualize using conventional ensemble measurements. 
Recent experiments and theoretical analyses have highlighted dynamic disorder in certain classes of biomolecules, whose dynamic pattern of conformational transitions is affected by slower transition dynamics of internal state hidden in a low dimensional projection. 
A systematic means to analyze such data is, however, currently not well developed. 		
Here we report a new algorithm -- Variational Bayes-double chain Markov model (VB-DCMM) -- to analyze single molecule time trajectories that display dynamic disorder. 
The proposed analysis employing VB-DCMM allows us to detect the presence of dynamic disorder, if any, in each trajectory, identify the number of internal states, and estimate transition rates between the internal states as well as the rates of conformational transition within each internal state.
Applying VB-DCMM algorithm to single molecule FRET data of H-DNA in 100 mM-Na$^+$ solution, followed by data clustering, we show that at least 6 kinetic paths linking 4 distinct internal states are required to correctly interpret the duplex-triplex transitions of H-DNA.
\end{abstract}

	\maketitle
\section*{Author Summary}
We have developed a new algorithm to better decode single molecule data with dynamic disorder. 
Our new algorithm, which represents a substantial improvement over other  methodologies, can detect the presence of dynamic disorder in each trajectory and quantify the kinetic characteristics of underlying energy landscape. 
As a model system, we applied our algorithm to the single molecule FRET time traces of H-DNA. 
While duplex-triplex transitions of H-DNA are conventionally interpreted in terms of two-state kinetics, slowly varying dynamic patterns corresponding to hidden internal states can also be identified from the individual time traces. 
Our algorithm reveals that at least 4 distinct internal states are required to correctly interpret the data. 

\section*{Introduction}
	Recent technological advances in single molecule experiments on biomolecules have provided an unprecedented chance to investigate dynamics of proteins and nucleic acids at single molecule (SM) level, which has previously been elusive in conventional experiments \cite{Liphardt2001_Science,XieSci03,Greenleaf2008_Science,Zhuang2002_Science,Rhoades2003,Zhuang2003_CurrOpinSturctBiol,Stigler2011_Science}.
	Folding/unfolding pathways gleaned from individual SM trajectories indicate rugged folding landscapes inherent to biomolecules \cite{Thirumalai2005_biochem,Mickler2007_PNAS,Zhuang2002_Science}.
	Long time trajectories from SM measurements, which now can be extended more than hundreds seconds, allow us to address how a rugged conformational landscape is sampled over time \cite{Stigler2011_Science,Rognoni2014PNAS,Qu2008_PNAS}. 
	One of the striking findings from these measurements is that even under the same folding condition, conformational dynamics of individual molecules differ substantially from one another while still maintaining their biological functions.
	Cofactor-induced conformational transitions of \emph{T}. ribozymes \cite{Solomatin2010}, Holliday junctions \cite{Hyeon:2012aa}, TPP-riboswitch \cite{Haller12032013}, and preQ$_1$-riboswitch \cite{Rinaldi:2016iw} are the recent seminal examples that exhibit molecular heterogeneity at equilibrium.
	The variation in the velocities of individual RecBCD helicase motors along the dsDNA \cite{Liu2013} is a good example of the molecular heterogeneity out of equilibrium, driven by ATP hydrolysis.
	Together with other reports   \cite{Lu_1998_Science,English_2005_NCB,Zhuang_2002_Science,van_Oijen_2003_Science,ANIE_2004_Angew,Flomenbom_2005_PNAS,Yang_2003_Science,Min_2005_PRL,Cremer_2007_JACScomm,Rissin_2008_JACS,piwonski2012PNAS,Wu_2012_BPJ}, these could be merely a subset of more widespread, yet unrecognized cases that exhibit dynamical heterogeneity in SM time traces. 
	
	The chance of conformational frustration increases with the system size ($N_{\text{sys}}$). 
	For a given $N_{\text{sys}}$, the time for conformational sampling ($\tau_{sample}$) is expected to scale as $\tau_{sample} \sim e^{N_{sys}}$ \cite{Palmer82AP}. 
	Suppose that $T_{obs}$, which is in practice limited by several factors \cite{Hubner_2001_JCP,Elenko_2010_RSI,Hwang:Autofocusing}, is long enough to observe many (more than hundreds) transitions along a trace generated from SM measurement.
	Two distinct scenarios arise depending on the length of $\tau_{sample}$ relative to $T_{obs}$,
	(i) If the sampling time is shorter than $T_{obs}$ ($\tau_{sample} \ll T_{obs}$), then the conformational space of biomolecule is fully sampled. 
	In this case, the ergodicity of the system is ensured such that for any molecule $\alpha$ (or time trace $\alpha$) the time average of an observable $O_{\alpha}$, $\langle O\rangle_T = \frac{1}{T_{obs}} \int_0^{T_{obs}}  O_{\alpha}(\tau) ~ d\tau$, is equivalent to the ensemble average of $O_{\alpha}(t)$ over all $\alpha$'s ($1\leq \alpha \leq N_{ens}$) at any moment $t$, $\frac{1}{N_{ens}} \sum_{\alpha=1}^{N_{ens}}O_{\alpha}(t) = \langle O \rangle_{ens}$, i.e., $\langle O \rangle_{T} = \langle O \rangle_{ens}$; thus thermodynamic properties of the system can be read out by analyzing a single time trace. 
	(ii) In contrast, if $\tau_{sample} \gg T_{obs}$ is satisfied due to ruggedness of conformational space characterised with a number of deep local basins of attraction, then each time trace can sample only a local region of the conformational space.
	In this case, dynamic pattern from each time trace would look different, 
	and a change in the dynamic pattern from one time interval to another would be observed only occasionally.  
	
	
	To be more precise about the second scenario ($\tau_{sample} \gg T_{obs}$), suppose that 
	the average time scale for each local basin of attraction to be ``sampled" by the conformational dynamics of molecule is $\tau_{conf}$ and that the time for the molecule to make transitions between different superbasins of attraction is $\tau_{int}$ (Fig. \ref{fig_r_landscape}). 
	    In principle the relaxation rates and energy barrier heights of biomolecules span continuous spectra. 
	    So, the clear time scale separation may not always be waranteed. 
	However, to be able to grasp the presence of dynamic disorder, if any, in SM time traces straightforwardly, 
	a separation between two distinct time scales is required such that $\tau_{conf} \ll \tau_{int}$ (or $\Delta G_{conf}^{\ddagger}\ll\Delta G_{int}^{\ddagger}$). 
	 If $\tau_{conf}$ and $\tau_{int}$ were comparable (or the spectra of relaxation rates were uniform and continuous), an algorithm we will propose here as well as others could hardly be of any help to conceive a concrete landscape model as the one illustrated in Fig.\ref{fig_r_landscape}.
	Therefore, here we consider $\tau_{conf}$ and $\tau_{int}$ as two disparate time scales as illustrated in Fig. \ref{fig_r_landscape}.
	$\tau_{conf}$ is the time at which the time average of an observable $\langle O \rangle_\tau = \frac{1}{\tau} \int_{0}^{\tau} O(t) dt$ reaches its steady state value when $\tau > \tau_{conf}$, corresponding to a time scale in which to fully sample the local basin of attraction. 
	Alternatively, $\tau_{conf}$ is limited by a kinetic barrier with the greatest $\Delta G^{\ddagger}_{conf}$ within the local basin of attraction, so that $\tau_{conf}\gtrsim e^{\Delta G^{\ddagger}_{conf}/k_BT}$. 
	On the other hand, $\tau_{int}$ is the transition time that is expected to scale with the height of kinetic barriers ($\Delta G^{\ddag}_{int}$) between the two superbasins as $\tau_{int} \sim e^{\Delta G^{\ddag}_{int} / k_B T}$.
	When measurements are conducted with a finite duration of observation time ($T_{obs}$), 
	we can conceive two entirely different dynamic patterns depending on the relationship between $\tau_{conf}, \tau_{int}$, and $T_{obs}$:   
			\begin{figure}[h!]
				\centering
				\includegraphics[width=0.6\columnwidth]{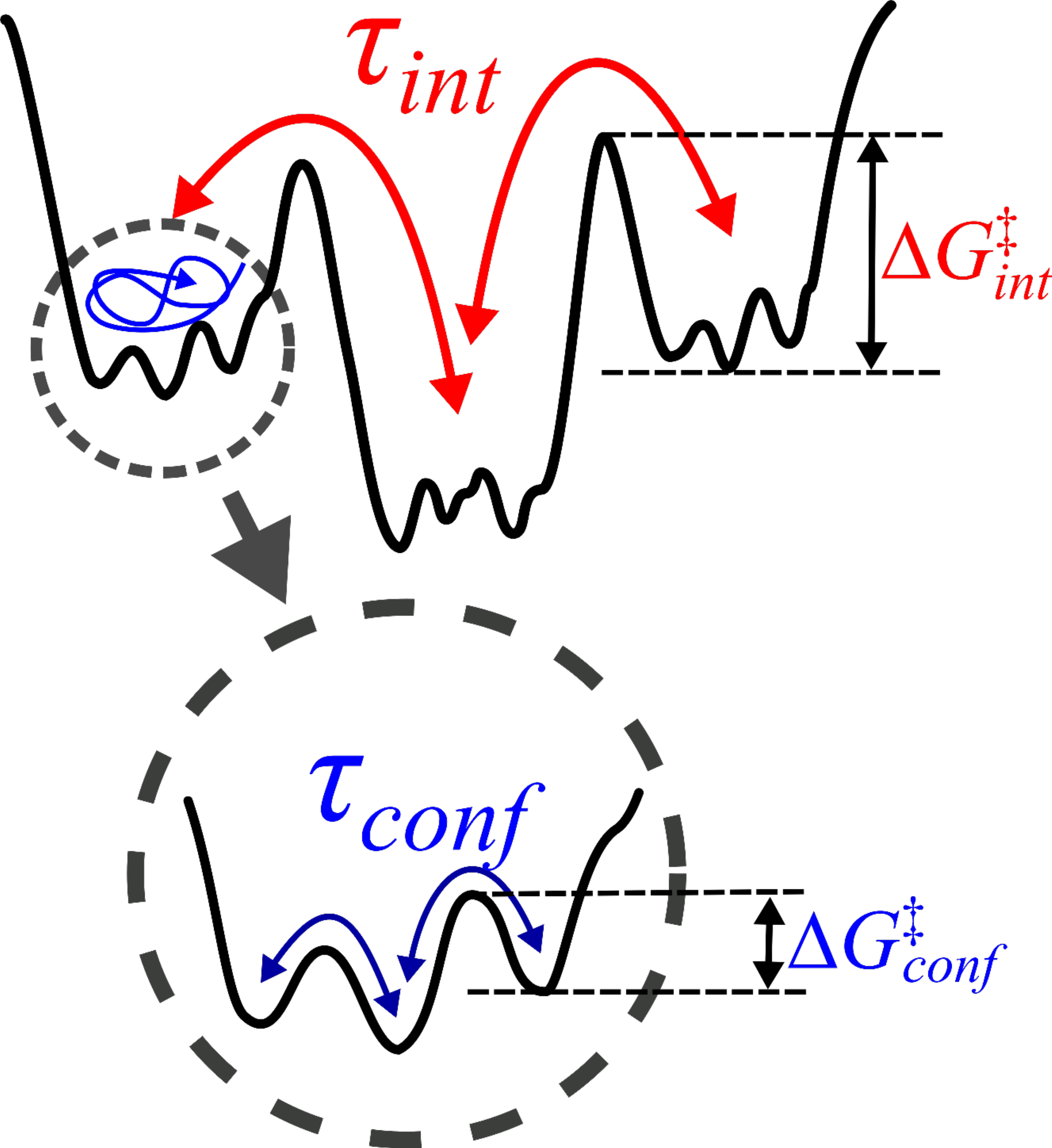}
				\caption{{\bf A rugged energy-landscape with hierarchical structure and an emergence of multiple time scales of transitions.}
					$\tau_{int}$ is the transition time between different superbasins of attraction whereas $\tau_{conf}$ is the time scale of conformational dynamics of molecule \emph{within} each basin.
					Due to large difference in kinetic barriers ($\Delta G^{\ddag}_{int} \gg \Delta G^{\ddag}_{conf}$), $\tau_{int} \gg \tau_{conf}$.
				}
				\label{fig_r_landscape}
			\end{figure}

	\begin{itemize}
	\item $\tau_{conf} \ll T_{obs} \ll \tau_{int}$: 
	The interconversion time between distinct basins of attractions is far longer than the observation time. 
	The dynamic patterns from individual trajectories that sample distinct basin of attraction are expected to differ from each other. 
	Since $T_{obs} \ll \tau_{int}$, there is few chance to observe  an exchange of dynamic pattern in a single time trace, which corresponds to a case with quenched disorder that each SM time trace looks entirely different.
	Such cases are reported in Holliday junction \cite{Hyeon:2012aa}, \emph{T}. ribozyme \cite{Solomatin2010}, and RecBCD \cite{Liu2013}.
	\item $\tau_{conf} \ll \tau_{int} \lesssim T_{obs}$: The interconversion time between basins of attraction is shorter than or comparable to the observation time. 
	In this case, it is possible to observe a few rounds ($\sim T_{obs}/\tau_{int}$) of pattern exchanges in a single time trace. 
	Such SM time traces are called to have a dynamic disorder \cite{Hyeon2014_PhysRevLett.112.138101,Hinczewski2016PNAS,Zwanzig1992_JCP,Zwanzig1990,Rinaldi:2016iw,Wu_2012_BPJ}.
	\end{itemize}

	While the most interesting and physically relevant question to ask about the heterogeneity in single molecule time traces is its molecular origin, detection and quantification of such heterogeneity should precede such question for a further analysis.
	For SM time traces with quenched disorder, it is relatively straightforward to analyze as one can use the criterion of ergodicity and partition each time trace into its dynamic subensembles \cite{Hyeon:2012aa}. 
	It is, however, more challenging to analyze time traces with dynamic disorder.

	In the ion-channel community, ion currents across a single ion-channel measured with patch-clamp technique often demonstrate time series that switch between multiple dynamic patterns, and such a phenomenon is called `mode-switching' \cite{Wilson_1993gr} or `modal gating' \cite{Siekmann_2014dy}.
An algorithm (aggregated Markov model, AMM) developed by ion-channel community to analyze time series exhibiting dynamics disorder is in principle of use, but when applied to our synthetic data, we found that the algorithm tends to overpredict the transitions between hidden states (see Fig. \ref{fig_r_iAMM} and discussion related to it below).  
	Thus, here we have developed a more reliable and systematic algorithm -- Variational Bayes-Double Chain Markov Model (VB-DCMM) -- which combined variational Bayes method with Double Chain Markov Model (DCMM) \cite{DCMM,Poritz_1982,Kenny_1990,Wellekens_1987,Paliwal_1993}, to analyze SM time traces with dynamic disorder in which dynamic pattern of conformational transition changes at much longer time scale than apparent conformational fluctuations due to a slower transition of a \emph{hidden} variable. 
	
	We first explain the algorithm for VB-DCMM, and next apply our VB-DCMM method to synthetic data as a blind test to show that our method can accurately identify the hidden internal states and determine the kinetic rate constants associated with the data. 
	The results from our analysis using VB-DCMM are reliable as long as a clear separation in time scales exists between the apparent conformational transition ($\tau_{conf}$) and the interconversion times ($\tau_{int}$).

\begin{figure}[h!]
			\centering
			\includegraphics[width=0.73\columnwidth]{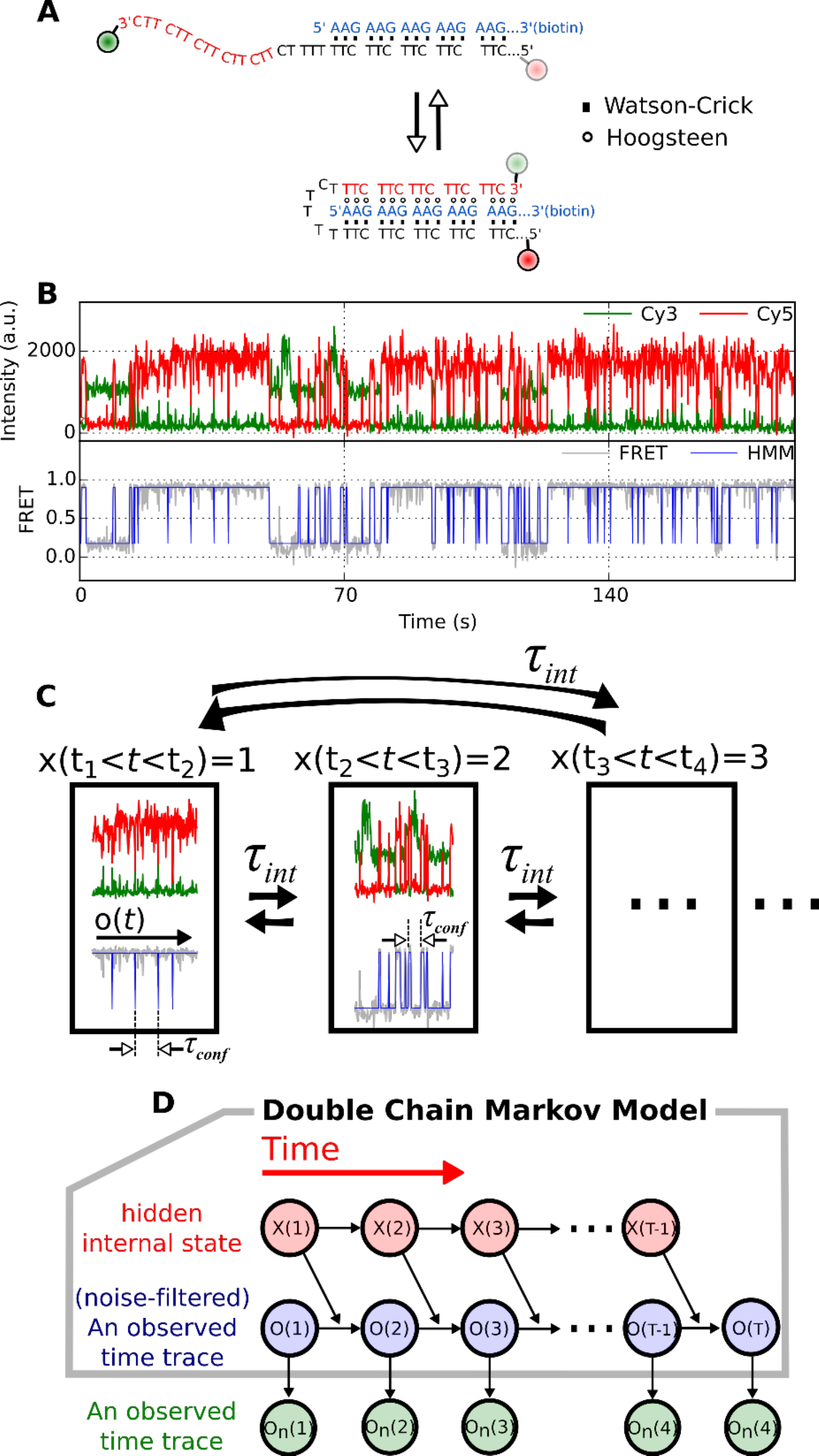}
			\caption{{\bf Duplex-triplex transitions of H-DNA with dynamic disorder.} 
				(A) Illustration of H-DNA dynamics. 
				The sequences in blue and black form duplex via Watson-Crick base pairing; the sequences in red extended from 3'-end region of the black sequence can pair with the sequences in blue via Hoogsteen base pairs to form the triplex helix. 
				(B) A time trace of H-DNA displaying dynamic disorder. 
				(Top) The fluorescence signals from Cy3 (green) and Cy5 (red) dyes. 
				(Bottom) FRET signal (gray) was calculated using the signals from Cy3 and Cy5. 
				Blue line is the noise-filtered FRET signal obtained using HMM. 
				The low-FRET ($\sim$0.1) and high-FRET state ($\sim$0.9) correspond to the duplex  and triplex states, respectively. 
				The dynamic pattern of the time trace changes occasionally from one time interval to another. 
				For example, the transitions from low to high FRET state around 70 s are much slower compared with those around 140 s. 
				(C) The model for H-DNA dynamics with dynamic disorder. Hierarchical transitions, (1) transitions within $x(t)=i$, and (2) interconversion between $x(t)=i$ and $x(t')=j $ ($i \neq j$), can be described using Double Chain Markov Model (DCMM). 
				(D) Graphical representation of DCMM. 
				$x(t)$, $o(t)$, and $o_n(t)$ represent internal state, noise-filtered observable (blue line in (B)), and the original observable at time $t$ (gray line in (B)), respectively. 
				The black arrows signify how each state is determined by others. For example, the state of observable at time $t$, $o(t)$ is determined by the previous observable state at time $t-1$, $o(t-1)$, and the state of the previous internal state, $x(t-1)$.
			}
			\label{fig_r_intro}
		\end{figure}

	As a prototypical example of single molecule time traces with dynamic disorder, data from H-DNA \cite{Lee2012,Lee2012bpj} that undergoes duplex-to-triplex conformational transitions (Fig. \ref{fig_r_intro}A) are analyzed.
	A kinetic pattern of two-state like conformational transitions between duplex (low FRET $\sim$ 0.1) and triplex form of H-DNA (high FRET $\sim$ 0.9) observed in one time interval changes to another  pattern in the next time interval (Fig. \ref{fig_r_intro}B). 
	DCMM models this peculiar dynamic pattern of H-DNA in Fig. \ref{fig_r_intro}B by assuming a slowly varying dynamics of a hidden internal state. 
	Fig. \ref{fig_r_intro}C illustrates how the dynamic pattern of the original time trace of \emph{observable state}, $o_n(t)$ (gray traces in Fig. \ref{fig_r_intro}C), 
	changes with the \emph{internal state} $x(t)$ at a given time $t$. 
	The dynamic pattern of $o_n(t)$, displaying multiple transitions, is slave to the slowly changing value of $x(t)$.
	DCMM implements this idea into an algorithm and allows us to extract the information of $x(t)$ from $o_n(t)$. 
	Finally, we apply VB-DCMM to an ensemble of H-DNA time traces obtained from smFRET experiments and show that the dynamics of H-DNA at [Na$^+$]=100 mM should be modeled using at least 4 large basins of attraction. 

\section*{Algorithm}      
Here, we provide a general overview of the VB-DCMM algorithm, defining terms and parameters. More technical details of derivation and implementation of the algorithm are given in the Supplementary Information. \\ 
    
\subsection*{Modeling time series with dynamic disorder.}
Markov chain approach is ubiquitously used in modeling biological systems. 
For example, reversible conformational transitions of biomolecules probed by single molecule fluorescence resonance energy transfer (smFRET) or force spectroscopy are often modeled as a homogeneous Markov process in which the transition rates between experimentally discernible conformational states are uniquely decided.
    To decipher time series with dynamic disorder that change their dynamic pattern from one time interval to another we assume that there are hidden ``internal states", each of which determines the rate of conformational transitions.
    A signature of the transition between internal states, which gives rise to dynamic disorder in time series, are difficult to detect using the value of FRET efficiency or end-to-end distance alone when the values observed along the time series are indiscernible even if the internal state is altered. 
    By assuming that the transition between internal states is described by a homogeneous Markov process, and that transition between observable (in this study, FRET) follows non-homogeneous Markov process, whose transition rates are slaved to the internal state at each time, we model time trajectories made of these two layers of Markov chains.
    This algorithm corresponds to the Double Chain Markov Model (DCMM)  \cite{DCMM,Poritz_1982,Kenny_1990,Wellekens_1987,Paliwal_1993} (Fig. \ref{fig_r_intro}C,D).

    DCMM is characterized by the following model parameters:
    (i) Transition matrix $\bm{A}$ for homogeneous Markov chain, which describes the transition probability between the $K$-distinct internal states along the time series ($\bm{x}=(x(1), x(2), ...,x(t), \cdots,  x(T-1))$).
    Here $K$ is a total number of internal states in the model, and $x(t)$, specifying internal state at time $t$, takes one of the values between 1 and $K$. 
    $T$ is the total  observation time.
    The internal state at time $t$+1 ($x(t+1)$) is determined by the previous internal state at time $t$ ($x(t)$), whose transition to $x(t+1)$ is determined by a $K \times K$ Markov transition matrix $\bm{A}$ as $P(x(t+1)=\mu) = \sum_{\nu=1}^{K} A_{\mu,\nu} P(x(t)=\nu)$ where $P(x(t)=\nu)$ denotes the probability of $x(t)$ being in the $\nu$-th internal state; 
    (ii) $K$ transition matrices $\bm{B}^{(\mu)}$ with $\mu \in \{1, 2, ..., K\}$ for non-homogeneous Markov chain describes the transition probability between the observable states along the time series ($\bm{o}=(o(1), o(2), ..., o(T))$).
$o(t)$ specifies the state of the observable among $N$ possible states $\{1,2,\ldots,N\}$ at time $t$.
    Transition from $o(t)$ to $o(t+1)$ is determined by an $N \times N$ transition matrix $\bm{B}^{x(t)}(t)$, the matrix elements of which are slave to the value of $x(t)(=\mu\in\{1,2,\ldots,K\})$.
    
    For example, if there are two ($K=2$) internal states, and each internal state has three ($N=3$) observables in a given time trace recorded with time resolution $\Delta t$, then two transition matrices for $\bm{o}$ with $\mu = 1, 2$ can be considered (i.e., $\bm{B}^{(1)}$ and $\bm{B}^{(2)}$):
    \begin{align}
    \bm{B}^{(\mu)} = 
        \begin{pmatrix}
        k^{ (\mu) }_{1 \rightarrow 1}  \Delta t & k^{ (\mu) }_{1 \rightarrow 2}  \Delta t & k^{ (\mu) }_{1 \rightarrow 3}  \Delta t\\
        k^{ (\mu) }_{2 \rightarrow 1}  \Delta t & k^{ (\mu) }_{2 \rightarrow 2}  \Delta t & k^{ (\mu) }_{2 \rightarrow 3}  \Delta t\\
        k^{ (\mu) }_{3 \rightarrow 1}  \Delta t & k^{ (\mu) }_{3 \rightarrow 2}  \Delta t & k^{ (\mu) }_{3 \rightarrow 3}  \Delta t \nonumber
        \end{pmatrix}.
    \end{align}
      Next, the transition matrix $\bm{A}$ for the interconversion between two internal states is: 
    \begin{align}
    \bm{A} = 
    \begin{pmatrix}
    \gamma^{ (1) \rightarrow (1) }  \Delta t & \gamma^{ (1) \rightarrow (2) }  \Delta t \\
    \gamma^{ (2) \rightarrow (1) }  \Delta t & \gamma^{ (2) \rightarrow (2) }  \Delta t \nonumber
    \end{pmatrix}. 
    \end{align}
  In the above matrices, the matrix elements must satisfy,
    $\sum_{j=1}^3k^{ (\mu) }_{i \rightarrow j}\Delta t= 1 $ for each $i=1,2,3$ in $\bm{B}^{(\mu)}$, and $ \gamma^{ (1) \rightarrow (1) }  \Delta t + \gamma^{ (1) \rightarrow (2) }  \Delta t =  \gamma^{ (2) \rightarrow (1) }  \Delta t + \gamma^{ (2) \rightarrow (2) }  \Delta t = 1 $ in $\bm{A}$.
    More detailed descriptions about DCMM are available in the original papers \cite{DCMM,Poritz_1982,Kenny_1990,Wellekens_1987,Paliwal_1993} particularly in ref. \cite{DCMM} (see also {\bf SI}).
A similar but more general version of DCMM, which can accommodate inputs variables as well as multiple number of internal state sequences, has been suggested by extending the factorial hidden Markov model \cite{Ghahramani_1997_MachLearn,Jordan_1997_NIPS}. 
    \\
    
    \subsection*{Determining the number of internal states.} 
    DCMM can estimate the transition matrices $\bm{A}$ and $\bm{B}^{(\mu)}$ quantitatively,  and hence determine the most probable sequence of internal state and associated kinetic rates, $\{k^{(\mu)}_{a\rightarrow b}\}$ and $\{\gamma^{(\mu)\rightarrow (\nu)}\}$. 
    However, the likelihood (the probability of observing data for given model parameters), maximized by DCMM, $P(\bm{o}|\bm{\pi},\bm{A},\bm{B})$ where $\bm{\pi}\equiv (\pi_1,\pi_2,\ldots,\pi_K)$ with $\pi_{\mu}=P(x(1)=\mu|\bm{o},\bm{A},\bm{B})$, is prone to increase when more number of parameters are used in the model. 
    DCMM can select the best set of parameters for a given model, but not suited to select the best model (i.e., cannot determine the optimal number of internal states $K$ for a given time trace).
    To overcome this limitation, often used is the maximum evidence method, where the \emph{evidence} ($P(\bm{o}|K)$, also called marginal likelihood) is defined as the conditional probability of observing data ($\bm{o}$) for a given model ($K$), so that 
    \begin{align}
    P(\bm{o}|K)=\int P(\bm{o}|\bm{\lambda})P(\bm{\lambda}|K)d\bm{\lambda}
    \label{eqn:DCMM}
    \end{align}
    where $\bm{\lambda}\equiv(\bm{\pi},\bm{A},\bm{B})$ represents the parameter space. 
    In this method, the  penalty against model complexity is naturally incorporated during the calculation, allowing to select the best model (see {\bf SI}).
    By calculating the evidence for each different model (different $K$, the number of internal states in data), 
    one can select the best model with an optimal number of internal states that maximizes the evidence. 
    The calculation of the evidence, however, involves a massive computational cost to explore the entire parameter space for a given model. 
    \\
     
    \subsection*{Variational Bayes Double Chain Markov Model.}
    To alleviate the computational cost in employing the maximum evidence method in Eq.\ref{eqn:DCMM}, we employ the Variational Bayes \cite{Bishop:2006}, a method that effectively uses a mean-field approximation.
    The method has previously been used to determine the number of observable states (FRET states) from smFRET data \cite{Bronson20093196,Bronson2010Graphical,Okamoto2012}, the number of diffusive states from single molecule tracking data \cite{Persson2013}, and the number of DNA-protein conformations from tethered particle motion data \cite{Johnson15092014}. It has also been used inside the empirical Bayes method which can analyze several smFRET time series simultaneously \cite{Meent2013,Meent20141327}.
    In our study, the variational Bayes method combined with DCMM (VB-DCMM) was used to analyze single molecule time traces with dynamic disorder. 
    The analytical expression of the lower bound of the \emph{evidence} ($F$), offered by VB-DCMM, makes clear where the model penalty comes from, thus providing guidelines to choose the prior parameters to incorporate a prior knowledge of data (see {\bf SI}).
    Once prior parameters are selected, VB-DCMM iteratively increases the lower bound of $\log{ (evidence) }(=\log{P(\bm{o}|K)})$ by identifying a better approximation to the true probability distribution. 
	\begin{align}
    \log{P(\bm{o}|K)}&=\int q(\bm{Z})\log{P(\bm{o}|K)}d\bm{Z}\nonumber\\
    &=F[q]+D_{KL}(q||p)\geq F[q^*].
    \label{eqn:evidence}
    \end{align}
    where $q(\bm{Z})$ is an arbitrary probability distribution of a set of variables, $\bm{Z}(\equiv (\bm{x},\bm{\lambda}))$ consisting of parameters and hidden variables of model, 
    \begin{align}
    F[q]\equiv \int q(\bm{Z})\log{\left(P(\bm{o},\bm{Z}|K)/q(\bm{Z})\right)}\nonumber
    \end{align}
    and 
    \begin{align}
    D_{KL}(q||p)\equiv \int q(\bm{Z})\log{\left(q(\bm{Z})/P(\bm{Z}|\bm{o},K)\right)}\geq 0, \nonumber
    \end{align}
    where $D_{KL}(q||p)$ is the Kullback-Leibler divergence of $q(\bm{Z})$ from $P(\bm{Z}|\bm{o},K)$, which we want to minimize. 
    Once the solution from the algorithm converges, the approximate value of $\log{P(\bm{o}|K^*)}(\simeq F[q^*])$ and the (locally) best model parameters (a set of the best kinetic rates), $\bm{\pi}^*$, $\bm{A}^*$ and $\bm{B}^*$, which determines all the rate constants to describe the given time traces ($\{k_{a\rightarrow b}^{(\mu)}\}$ and $\{\gamma^{(\mu)\rightarrow(\nu)}\}$), are acquired from an approximated probability distribution (See SI for the mathematical details). 
	The performance of VB-DCMM is quite robust over a wide variation of prior parameters (Fig. \ref{fig_r_simul_ub}, \ref{fig_r_simul_ua}).
	\\
	
	 \subsection*{Implementation of the algorithm.}
    The observable sequence $\bm{o}$ is obtained by filtering the noise in the experimental data ($\bm{o}_n$) using Hidden Markov Model (HMM) following a similar procedure as the previous studies \cite{McKinney2006,Bronson20093196} using a custom code written based on the code from Sagemath software \cite{sage}. 
    Next, the $\bm{o}$ is analyzed using VB-DCMM to select the best model and to estimate the best model parameters. 
    The optimal sequence of internal states $\bm{x}$ is determined by using Viterbi algorithm \cite{DCMM}.
    All the implementations and data analysis are done by using our custom code. VB-DCMM is freely available at ``https://github.com/TBiophysG/VBDCMM"
   	 \\

	\section*{Results and Discussion}
    \subsection*{Validation of VB-DCMM.}
\begin{figure}[h!]
    	\centering
    	\includegraphics[scale=0.48]{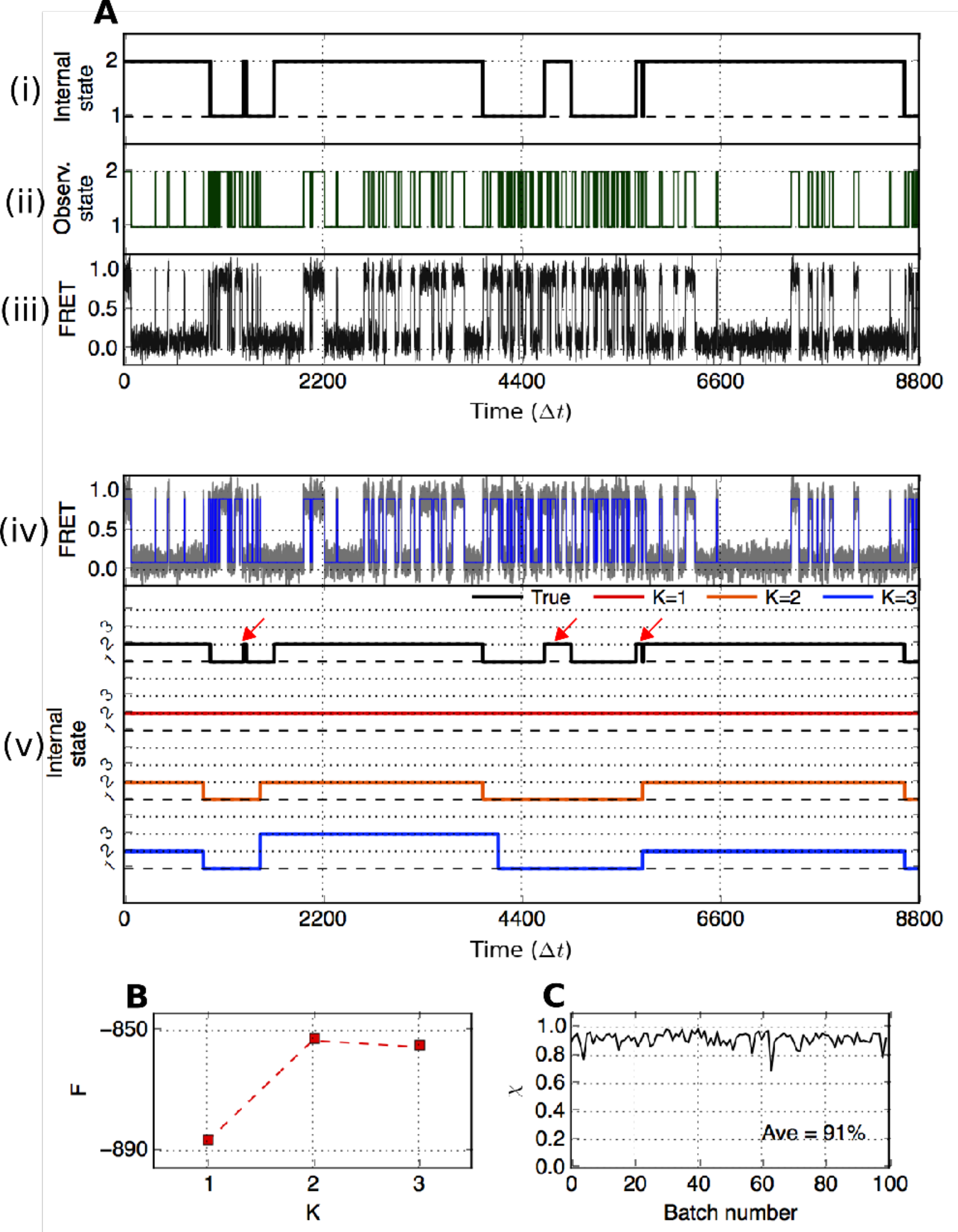}  	
    	\caption{{\bf Validation of VB-DCMM on synthetic data.}
	    	(A) 
	    	(i) A time trace of internal state generated with $\gamma^{(1) \rightarrow (2)}\Delta t=\gamma^{(2) \rightarrow (1)} \Delta t=0.001$. 
	    	(ii) An observable time trace generated based on the trace of internal state in (i) by using internal state-dependent parameters $k^{(1)}_{L \rightarrow H} \Delta t = k^{(1)}_{H \rightarrow L} \Delta t =0.05, k^{(2)}_{L \rightarrow H} \Delta t = 0.00625, k^{(2)}_{H \rightarrow L} \Delta t = 0.025$.
	    	(iii) An synthetic FRET data with Gaussian noise overlaid on the trace in (ii). 
	    	(iv) Noised filtered FRET state by HMM (blue line). 
	    	(v) Traces of internal state with different $K$, estimated using VB-DCMM on the noise-filtered FRET trace from (iv) (black line is the true internal state trace while red, orange, and blue are internal state estimated from the model with $K=1, 2,$ and $3$, respectively. The indices of internal state were determined by comparing $\bm{B}^{(\mu)}$ estimated for each internal state with $\bm{B}^{(\mu),\text{true}}$ which is used to generate the synthetic data). 
	    	(B) Estimated lower bound of the evidence function $F(K)$ of DCMM models with $K=1, 2$, and $3$. 
	    	(C) Accuracy of detecting internal states. The overlap function $\chi$ calculated for 100 synthetic FRET traces generated under the identical condition used for generating the trace of internal state shown in (A).
	    	}
    	\label{fig_r_demonstration}
    \end{figure}

    To first validate the efficacy of VB-DCMM in identifying internal states in a given SM time trace,
    we applied VB-DCMM algorithm on synthetic data that mimic a SM time trajectory with dynamic disorder (see {\bf Methods}). 
    To generate a synthetic SM time trajectory, we first produce a time trajectory specifying the value of internal state from $t=1$ to $t=T-1$. 
    The time trajectory of internal state is represented with a symbol $\bm{x} \equiv (x(1), x(2), \cdots, x(t), \cdots, x(T-1)$). 
    When the total number of distinct internal states in the model is $K$, one of the values in $\{1, 2, \cdots, K\}$ is assigned to $x(t)$.
    Thus, for $K=2$ a typical time trajectory of internal state $\bm{x}$ looks like
    (1, 1, 1, $\cdots$, 1, 1, 2, 2, 2, $\cdots$, 2, 2, 1, 1, $\cdots$, 1, 1), 
    (1, 1, 1, $\cdots$, 1, 2, 2, $\cdots$, 2, 2),
    (2, 2, $\cdots$, 2, 1, 1, $\cdots$, 1, 1, 2, 2, $\cdots$, 2, 2, 2), etc.
    The time trajectory given in Fig. \ref{fig_r_demonstration}A(i) is an example generated with $K=2$ and $T$=8801.
    Next, similar to the structure of $\bm{x}$, the time trajectory of noiseless observables is represented using $\bm{o}\equiv (o(1),o(2),\cdots,o(t),\cdots,o(T))$.
    In Fig. \ref{fig_r_demonstration}A-(ii), a trajectory of $\bm{o}$ is shown, also demonstrating the influence of $\bm{x}$ on $\bm{o}$. 
    Finally, Gaussian noise was added on $\bm{o}$ and the range of signal was adjusted to produce the final trajectory $\bm{o}_n \equiv (o_n(1), o_n(2), \cdots, o_n(T))$ which now resembles a time trajectory of SM FRET signal (Fig. \ref{fig_r_demonstration}A-(iii)).
    
    Deciphering the information of internal states from an observed time trace involves solving an inverse problem, i.e., decoding $\bm{o}_n$ to obtain $\bm{x}$. 
    To decode the trace of internal states from the synthetic data, we follow a 3-step procedure: 
    (1) Filter the noise from $\bm{o}_n$ to obtain $\bm{o}$ using Hidden Markov Model (HMM) \cite{McKinney2006} (Fig. \ref{fig_r_demonstration}A-(iv), blue line); 
    (2) Analyze $\bm{o}$ by applying VB-DCMM algorithm with different models $1, 2,\ldots, K$ (again, $K$ is the total number of internal states assumed in each model);
    (3) To select the best model we calculated the conditional probability of observing data for a given model parameter $K$, $P( \bm{o} | K )$, 
    which is often called {\it evidence} or {\it marginal likelihood} in 
    machine learning community (Eq.\ref{eqn:evidence}) \cite{Bishop:2006}.    
    Calculation of $P(\bm{o} | {K})$ is conducted using the Variational Bayes (VB) method, which gives the lower bound of $\log{ P(\bm{o}|{K})}$ denoted by $F(K)$.
    Details of the evidence function $F(K)$ and approximation procedure are provided in the Supplementary Information (SI).
    Finally, we select the best model $K^*$ which maximizes $F(K)$, i.e., $K^*=\arg\max F(K)$.  
    
    To be specific, in order to identify the best model parameter $K$ for the time trace $\bm{o}(t)$ given in Fig.\ref{fig_r_demonstration}A-(iv), we varied  $K$ from 1 to 3.
    The most probable trace of internal states, $\bm{x}^{\text{model}}_{(K)}$, was calculated for each model with $K=1$ (red), $K=2$ (orange), $K=3$ (blue) (see Fig. \ref{fig_r_demonstration}A-(v)).
    The evidence $F(K)$ calculated using VB method was maximized at $K=K^*=2$, and  
    the resulting time trace of the internal states, $\bm{x}^{\text{model}}_{(K^*=2)}$, most closely recovers the trajectory of $\bm{x}$ (black trace in Fig. \ref{fig_r_demonstration}A(v)) except at the time interval where the transitions of $x(t)$ between 1 and 2 occur only transiently or at the boundaries of transitions (red arrows on Fig.\ref{fig_r_demonstration}A-(v)). This result shows that VB-DCMM can avoid the over-fitting problem that other methods based on maximum likelihood are often fraught with \cite{Bishop:2006}.
    \\
    
    \subsection*{Conditions required for an accurate recovery of internal states.}
    VB-DCMM detects a signature of change in internal state ($\bm{x}$) from a given observable time trace ($\bm{o}$) by evaluating the statistical difference in transition rates. 
    Thus, in the absence of an enough number of transitions in the trace $\bm{o}$, the algorithm becomes less reliable. 
    For example, we obtained $F (K=2) \approx F (K=3)$ although $F(2) \gg F(3)$ is more desirable (Fig. \ref{fig_r_demonstration}B. See another example in Fig. \ref{fig_r_simul_nondistinguishable}). 
    This is due to the lack of statistics in transition events in this particular test trace given in Fig. \ref{fig_r_demonstration}A. 
    For example, when only a part of the time trace is selected and analyzed using HMM, 
    the estimated rates of transition from high ($H$) to low ($L$) FRET value are $k^{est}_{H \rightarrow L} ~ \Delta t= 0.016$ in $1500 \lesssim t \lesssim 4000$, 
    and $k^{est}_{H \rightarrow L} ~ \Delta t= 0.026$ in $5700 \lesssim t \lesssim 8700$.
    Thus, in ($K$=3)-model the two time intervals, originally generated by using the same kinetic parameter ($k^{(2)}_{H \rightarrow L} ~ \Delta t= 0.025$), are determined to be distinct from each other (blue trace in Fig. \ref{fig_r_demonstration}A-(v)). 
    By contrast, in ($K=2$)-model, $k^{est}_{H \rightarrow L}~ \Delta t= 0.020$ was estimated over these two time intervals.
    This type of statistical error is unavoidable for a small  $T_{obs}$. 
    A more systematic evaluation on the accuracy of the algorithm as a function of $T_{obs}$ and transition rate between distinct internal states will be discussed in the next section.
    
    To assess the accuracy of the best model $\bm{x}^{\text{model}}_{(K^*)}$ predicted by VB-DCMM against the solution $\bm{x}$, 
the following overlap function can be used.  
    \begin{equation} \label{eq: chi}
    \begin{aligned}
    \chi = \frac{1}{T-1} \sum_{t=1}^{T-1} \delta_{x(t), x_{(K^*)}^{\text{model}}(t)}
    \end{aligned}
    \end{equation}
    where $\delta_{i,j}$ is the Kronecker delta and 
    $T=T_{obs}/\Delta t$ is the total number of data in the traces ($\Delta t$ denotes the temporal resolution of the data). 
    For 100 synthetic time traces, generated under the identical parameters used for producing the time trace in Fig. \ref{fig_r_demonstration}A, we found that $\chi\approx 0.9$ on average (Fig. \ref{fig_r_demonstration}C). 
    Note, however, that $x(t)$, only available for the case of ``synthetic data". 
    Thus, to assess the accuracy of our method against a real time trace from SM experiments, we devised other metrics.  

    For a given time trace with dynamic disorder, our algorithm quantifies the kinetic features of the time trace in terms of the transition rate between the observable states $a$ and $b$ within the $\mu$-th internal state $k^{(\mu)}_{a \rightarrow b}$ and the transition rate from the $\mu$-th internal state to $\nu$-th internal state $\gamma^{(\mu) \rightarrow (\nu)}$ ($ 1 \leq \mu, \nu \leq K$, $1 \leq a, b \leq N$. Here, $\mu$ is the index for internal state whereas $a$ and $b$ are indices for observable (In FRET displaying low/high two state transitions, these states correspond  to the low and high FRET values). 
    $K$ is the total number of hidden internal states, and $N$ denotes the total number of observables).
    To be able to extract the information of multiple internal states reliably from a time trace using VB-DCMM, 
    two general conditions are required for the time trace being analyzed. 
    \begin{enumerate}
    \item A large time scale separation should be present in the kinetics within each internal state, i.e., $k^{ (\mu) }_{a \rightarrow b}$ and $k^{ (\nu) }_{a \rightarrow b}$ ($\mu\neq\nu$) should be disparate.  
    \item There should be a clear time scale separation between intra-basins and inter-basin transitions (i.e., $\tau_{conf}$ and $\tau_{int}$). 
    More precisely, the intra-basin transition probability $k^{(\mu)}_{a \rightarrow b} ~ \Delta t $ should be much \emph{greater} than the transition probability from the $\mu$-th to any other internal state $\sum_{\nu \neq \mu} \gamma^{ (\mu) \rightarrow (\nu) } ~ \Delta t$ ($=1 - \gamma^{(\mu) \rightarrow (\mu)} ~ \Delta t$). 
    \end{enumerate}
    To substantiate the above-mentioned conditions 1 and 2, we define  two metrics $D_{\text{conf}}$ and $D_{\text{int}}$, which compute the average Hamming-like distances between the distinct rate constants extracted from a given time trace using VB-DCMM analysis:
    \begin{equation} \label{eq: D_conf}
    \begin{aligned}
    D_{\text{conf}} =    \frac{2}{ K (K-1) }  \sum_{\substack{ \mu, \nu=1 \\ \mu > \nu } }^{K}\frac{1}{ N (N-1) } \sum_{\substack{a, b=1\\ a \neq b} }^{N} \abs{\log_2{ \frac{ k^{  (\mu)  }_{a \rightarrow b } }{  k^{  (\nu)  }_{a \rightarrow b} } } }
    \end{aligned}
    \end{equation}
    and    
    \begin{equation} \label{eq: D_int}
    \begin{aligned}
    D_{\text{int}} =    \frac{1}{ K }  \sum_{\mu=1 }^{K} \frac{1}{ N(N-1) } \sum_{\substack{a, b=1 \\ a \neq b}}^{N} \abs{ \log_2{ \frac{ k^{  (\mu)  }_{a \rightarrow b } }{ \sum_{\nu \neq \mu} \gamma^{ (\mu) \rightarrow (\nu) } }} }.
    \end{aligned}
    \end{equation}    
    $D_{\text{conf}}$ measures the dissimilarity between distinct internal states in terms of the intra-basin transition rates.
Two distinct internal states ($\mu$, $\nu$ ($\mu\neq\nu$)) can be better discerned if the intra-basin transition rate of one internal state (say, $k_{a\rightarrow b}^{(\mu)}$) differs greatly from that of other internal state ($k_{a\rightarrow b}^{(\nu)}$), so that $|\log_2{\left(k_{a\rightarrow b}^{(\mu)}/k_{a\rightarrow b}^{(\nu)}\right)}|$ is maximized. 
    $D_{\text{int}}$ measures the average number of intra-basin transitions in each internal state using the ratio between the transition probabilities, $k^{(\mu)}_{a \rightarrow b} ~ \Delta t$ and $\sum_{\nu \neq \mu} \gamma^{ (\mu) \rightarrow (\nu) }\Delta t\left(=1 - \gamma^{(\mu) \rightarrow (\mu)} ~ \Delta t\right)$. 
    A greater $D_{\text{int}}$ ensures a large time scale separation in dynamics between intra-basin and inter-basin transitions, which improves the reliability of our method to decode the internal state from a given time trace. 
    In general, $D_{\text{int}}$ or $D_{\text{conf}}$ shows a good correlation with $\langle\chi\rangle$ (see below); thus, one can use $(D_{\text{int}},D_{\text{conf}})$ to assess the accuracy of predicted internal states. 
    Note that the metrics $D_{\text{int}}$ and $D_{\text{conf}}$ can be estimated for real data, while $\langle \chi \rangle$ can be calculated only against the synthetic data. 
    Since there is a good correlation between ($D_{\text{int}}$,$D_{\text{conf}}$) and $\chi$, one can evaluate ($D_{\text{int}}$,$D_{\text{conf}}$), alternative to $\chi$, to assess the reliability of a predicted result of $x_{(K^*)}^{\text{model}}(t)$. 
      \\

    To be more concrete, we applied VB-DCMM algorithm to analyze synthetic data generated with $N=2$ (transitioning between high and low  FRET values) and $K=2$ (two internal states; $\mu=1$ and $2$) under various scenarios. 

\begin{itemize}
    \item We fixed the transition rates in the state $\mu=1$ as $k^{(1)}_{L \rightarrow H}~\Delta t=k^{(1)}_{H \rightarrow L} ~ \Delta t= 0.05 $, and varied the rates associated with the state $\mu=2$ 
    over the range of $ 0.125 \leq k^{(2)}_{L \rightarrow H} / k^{(1)}_{L \rightarrow H},~ k^{(2)}_{H \rightarrow L} / k^{(1)}_{H \rightarrow L} \leq 8$ (Fig. \ref{fig_r_test_k_gamma1}A, left). For the interconversion probability between the two internal states we set $\gamma^{(1) \rightarrow (2)} ~\Delta t  = \gamma^{(2) \rightarrow (1)} ~\Delta t = 0.001$. 
    The accuracy of the model prediction ($\langle \chi \rangle$, Eq.(\ref{eq: chi})) is on average greater than 0.9 as long as the transition rates $k_{L\leftrightarrow H}^{(\mu)}$ and $k_{L\leftrightarrow H}^{(\nu)}$ ($\mu\neq \nu$) differ more than the factor of 4. Note that in Fig. \ref{fig_r_test_k_gamma1}A (left), the value of $\langle \chi \rangle$ is greater for $k^{(2)}_{L \rightarrow H} / k^{(1)}_{L \rightarrow H}$, $k^{(2)}_{H \rightarrow L} / k^{(1)}_{H \rightarrow L} \gg 1$ than for $k^{(2)}_{L \rightarrow H} / k^{(1)}_{L \rightarrow H}$, $k^{(2)}_{H \rightarrow L} / k^{(1)}_{H \rightarrow L} \ll 1$; this is because
    a statistically sufficient number of transitions make the detection of internal states more reliable.
    In contrast, when $k^{(2)}_{L \rightarrow H} / k^{(1)}_{L \rightarrow H}$, $k^{(2)}_{H \rightarrow L} / k^{(1)}_{H \rightarrow L}  \simeq 1$, i.e. when the kinetics inside the two internal states are essentially identical, 
    it is difficult to discern the two internal states.
    In this case, $K$=1 instead of $K$=2 is effectively the correct number of internal states. 
    Indeed, when $K=1$ is assumed (i.e., assuming true internal state $x(t)=1$ for all $t$ in Eq.(\ref{eq: chi})), the re-calculated $\langle \chi \rangle$ is close to 1 (see Fig. \ref{fig_r_test_k_K1}).
        \begin{figure}[t]
        	\centering
           	\includegraphics[scale=0.76]{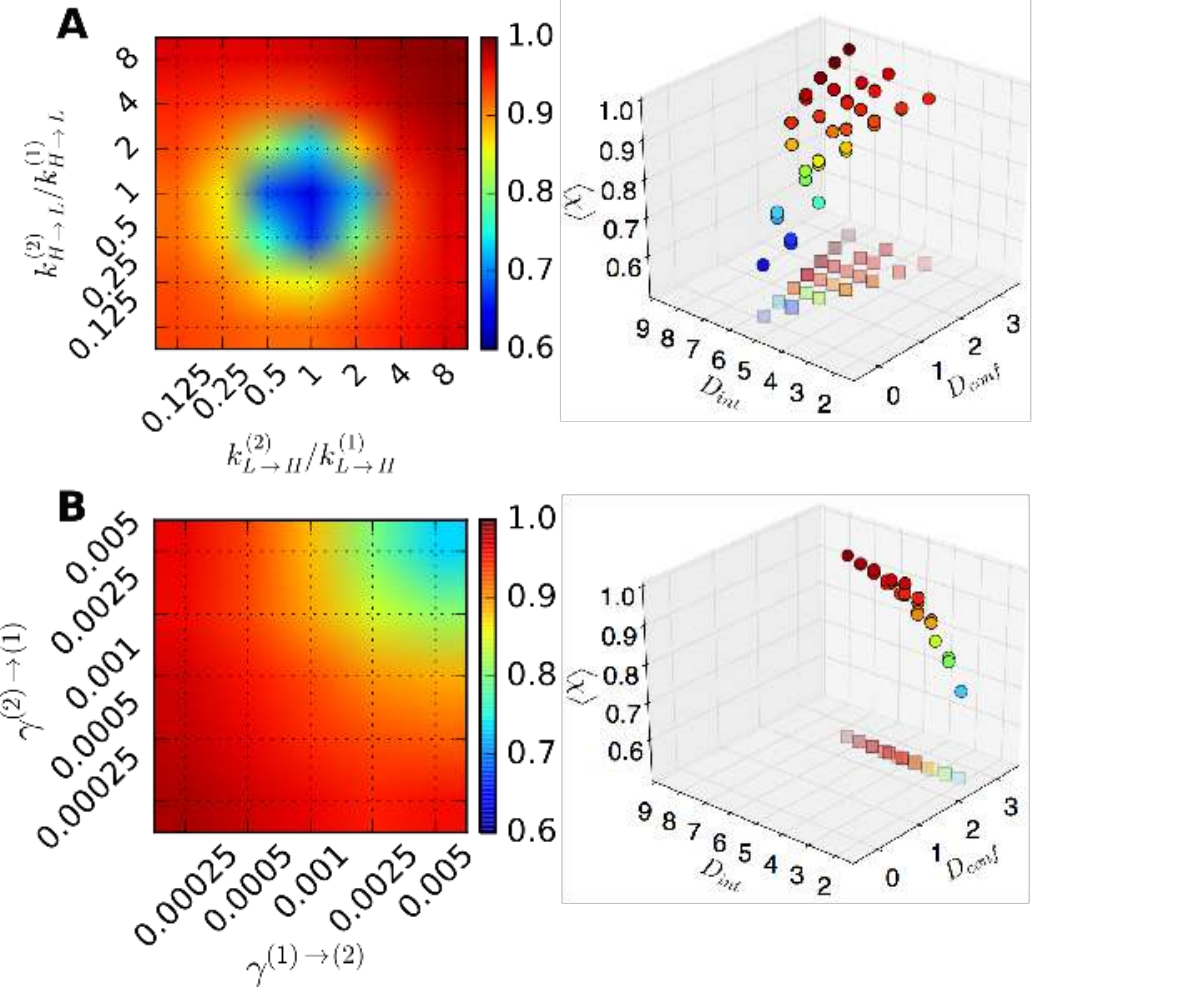}  
        	
        	\caption{{\bf Accuracy of VB-DCMM in detecting internal states under various conditions of $k_{L\leftrightarrow H}^{(\mu)}$ and $\gamma^{(1)\leftrightarrow(2)}$ with $T_{obs} / \Delta t = 8800$.}
        		(A) The color bar denotes the accuracy of analysis in terms of $\langle \chi \rangle$  under varying $k^{(2)}_{L \rightarrow H}, k^{(2)}_{H \rightarrow L}$ with $K=2$, $k^{(1)}_{L \rightarrow H} \Delta t = 0.05$, $k^{(1)}_{H \rightarrow L}\Delta t=0.05$, and $\gamma^{(1) \rightarrow (2)}\Delta t=\gamma^{(2) \rightarrow (1)}\Delta t=0.001$.
        		(B) $\langle \chi \rangle$ under varying $\gamma^{(1) \rightarrow (2)}$ and $\gamma^{(2) \rightarrow (1)}$ with $K=2$, $k^{(1)}_{L \rightarrow H} \Delta t = k^{(1)}_{H \rightarrow L} \Delta t =0.05, k^{(2)}_{L \rightarrow H} \Delta t = 0.00625, k^{(2)}_{H \rightarrow L} \Delta t = 0.0125$.
        		$\langle\chi\rangle$ was calculated by averaging over the results from analysis of 100 traces in each condition. 
        		The panels on the right show the relation between the value of $\langle\chi\rangle$ and pairs of $D_{\text{int}}$ and $D_{\text{conf}}$ values which are evaluated at varying kinetic parameters. 	
        		Results from the analysis over the data with the same parameters but different length of time trace (or different number of data points $T_{obs} / \Delta t = 2200, 4400$) are provided in Fig. \ref{fig_r_test_T}.
        	}
        	\label{fig_r_test_k_gamma1}
        \end{figure}

    \item To explore the effect of interconversion between distinct internal states on the performance of algorithm, we generated synthetic data with $k^{(1)}_{L \rightarrow H} ~ \Delta t=k^{(1)}_{H \rightarrow L}~ \Delta t = 0.05 $, $k^{(2)}_{L \rightarrow H} / k^{(1)}_{L \rightarrow H} = 0.125$, and $k^{(2)}_{H \rightarrow L} / k^{(1)}_{H \rightarrow L}= 0.25$ by, this time, varying $\gamma^{(1) \rightarrow (2)}~ \Delta t$ and $\gamma^{(2) \rightarrow (1)}~ \Delta t = 0.00025 \sim 0.005$ (Fig. \ref{fig_r_test_k_gamma1}B, left). 
    The results clearly show that the case with smaller $\gamma^{(\mu) \rightarrow (\nu)}$ results in a higher $\langle\chi\rangle$, which is expected because each internal state can have more number of transitions in the traces $\bm{o}$ when the interconversion is slower (Fig. \ref{fig_r_test_k_gamma1}B). 
    Re-plotting $\langle\chi\rangle$ as a function of $D_{\text{conf}}$ and $D_{\text{int}}$ reveals clear dependence of the accuracy on $D_{\text{int}}$ (Fig. \ref{fig_r_test_k_gamma1}B, right).
    Similar trends are observed for other conditions of $k^{(2)}_{L \rightarrow H} /  k^{(1)}_{L \rightarrow H}$ and $k^{(2)}_{H \rightarrow L}  /  k^{(1)}_{H \rightarrow L}$ (Fig. \ref{fig_r_test_gamma2_gamma3}).
    
    \item Analyses on synthetic data generated using the same input parameters with those in Fig. \ref{fig_r_test_k_gamma1}, but with a different number of data points in each trace, $T_{obs}/\Delta t=4400$, and $2200$ (Fig. \ref{fig_r_test_T}) show a similar trend as observed in Fig. \ref{fig_r_test_k_gamma1} with $T_{obs}/\Delta t=8800$ but with slightly smaller $\langle \chi \rangle$ values. 
    
    \item Extension of VB-DCMM algorithm to a more complicated case for $K>2$ (Fig. \ref{fig_r_simul_K3}, Fig. \ref{fig_r_simul_K4L4}) or $N>2$ (Fig. \ref{fig_r_simul_L3}, Fig. \ref{fig_r_simul_K4L4}) is straightforward.
    Application of VB-DCMM to a trajectory in which each internal state trajectory has different $N$ is also straightforward (Fig. \ref{fig_r_simul_L24}).
    In the latter case, the data is analyzed by assuming that all internal states have the same number of possible observables, $N$; but the analysis would indicate that transition associated with a small transition rate is essentially disallowed. 
    In all situations considered for various $K$ and $N$, VB-DCMM can be used for the reliable recovery of the sequence of true internal states.
           
    \item Analyses of synthetic traces show that the accuracy of the algorithm improves with both $D_{\text{conf}}$ and $D_{\text{int}}$ (Fig. \ref{fig_r_test_k_gamma1} right panels and Fig. \ref{fig_r_Dconf_Dint_simul}). 
    Thus, these two metrics allow one to 
    judge the reliability of the information on internal states extracted from a given time trace.
    Alternatively, a single parameter $D_{\text{tot}} (= D_{\text{conf}} + \alpha D_{\text{int}})$ with an empirically acquired coefficient $\alpha \approx 0.8$) can be used to judge the reliability of the extracted information. 
    Note that $\langle \chi\rangle$ remains similar as long as $D_{\text{tot}}$ remains constant (Fig. \ref{fig_r_Dconf_Dint_simul}B). 
    Hence, when $\langle \chi\rangle$ is plotted against $D_{\text{tot}}$, all synthetic data generated using different parameters approximately collapse onto a single universal curve (Fig. \ref{fig_r_simul_D}). 
     \begin{figure}[t]
     	    	\centering
     	    	\includegraphics[scale=0.48]{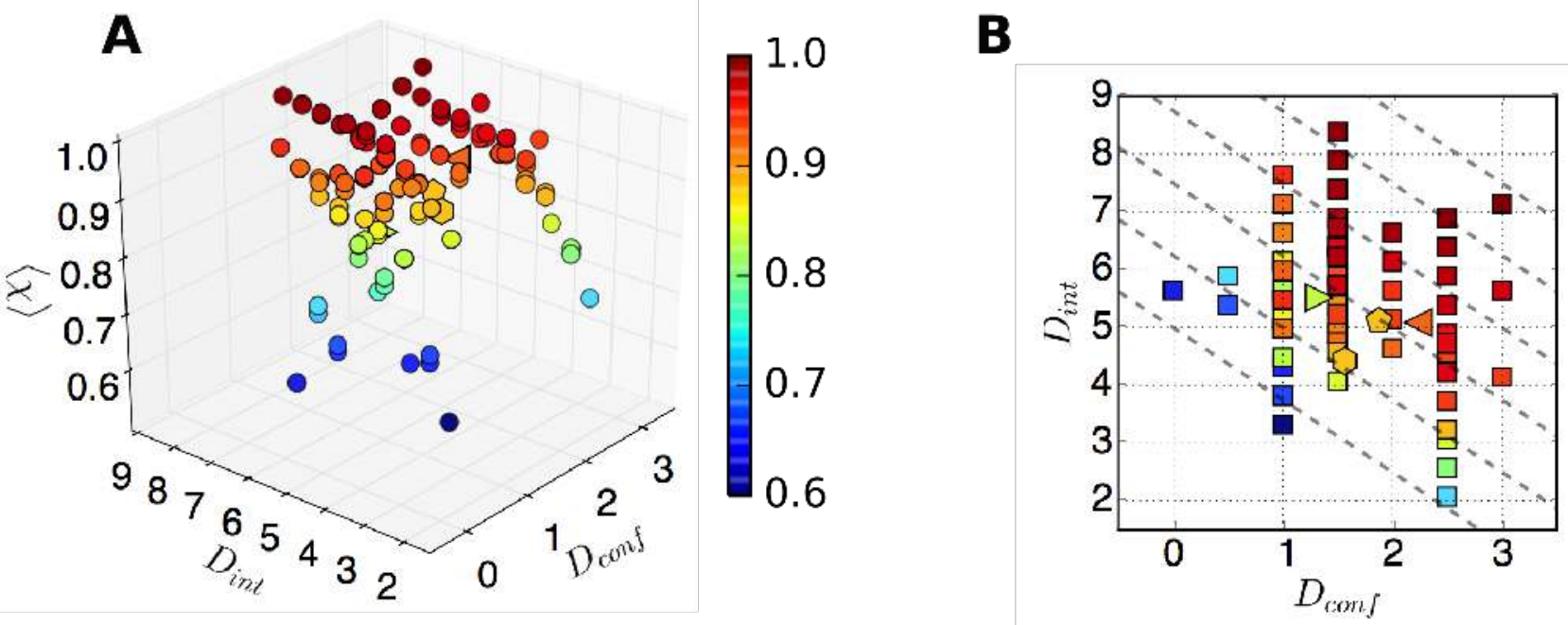}
     	
     	\caption{{\bf Average accuracy of internal state detection as a function of $D_{\text{conf}}$, and $D_{\text{int}}$.} To construct this diagram, we employed various synthetic data in Fig. \ref{fig_r_test_k_gamma1} (circle, two internal states ($K=2$), two FRET states ($N=2$)), Fig. \ref{fig_r_simul_K3}  (left triangle, $K=3$, $N=2$), and Fig. \ref{fig_r_simul_L3} (hexagon, $K=2$, $N=3$). 
	The right triangle symbol denotes the result from the similar analysis shown in Fig. \ref{fig_r_simul_K3} with $K=3$ but with smaller relative differences in the transition rates, $k$'s. Pentagon represents the result obtained with $K=4$ and $N=2$.
     		(A) Color code denotes the accuracy of internal states predictions in terms of $\langle \chi \rangle$, averaged over 100 traces for each condition.
     		(B) The dashed lines corresponding to $\Delta = D_{\text{conf}} + 0.8 D_{\text{int}} = 4, 5,\ldots 9$ are overlaid on the 2-D scatter plot of $\langle\chi\rangle(D_{\text{conf}},D_{\text{int}})$ calculated in Fig. (A).
     	}
     	\label{fig_r_Dconf_Dint_simul}
     \end{figure}    
     
    \item There are multiple ways of assessing the efficacy of VB-DCMM in decoding the internal states.
     In addition to $\langle \chi\rangle$, $D_{\text{conf}}$, $D_{\text{int}}$, and $D_{\text{tot}}$ as the possible measures for the assessment, one can also use the statistical property that the dwell times of homogeneous Markov process satisfies $\sqrt{\langle\tau^2\rangle - \langle \tau \rangle^2} / \langle \tau \rangle \sim 1$  (see {\bf SI} for details).   
     \end{itemize}

    \subsection*{Application of VB-DCMM on H-DNA data.}
\begin{figure*}[h!]
      	\centering
        \includegraphics[scale=0.63]{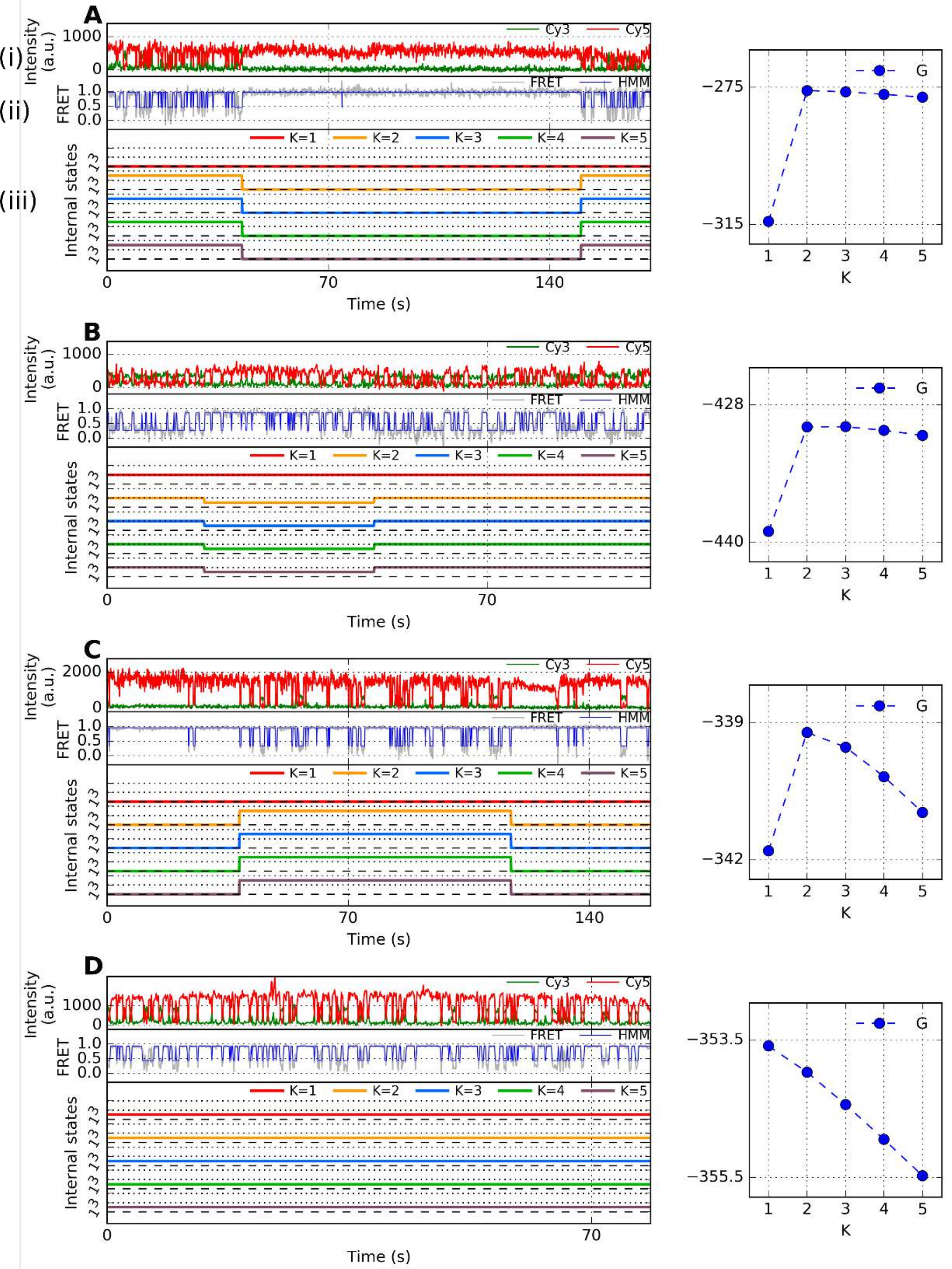}
      	
      	\caption{{\bf Representative time traces of H-DNA at [Na$^+$]= 100 mM and their analysis.}
      		(A) (i) Fluorescence signal and (ii) their FRET state. (iii) Internal states estimated for $K=1, 2, \ldots, 5$.
      		Right panel shows $G(K)$ (blue circle) where $K_{obs}$ specifies the number of detected internal states in individual traces (blue). 
      		(B, C, D) Other representative time traces and their $G(K)$ obtained under the same experimental condition.
      	}
      	\label{fig_r_traces_100mM}
      \end{figure*}

    Now, to analyze the duplex-triplex transitions of H-DNA (Fig. \ref{fig_r_traces_100mM}), we obtain $\bm{o}$ by filtering the noise from FRET signal (Fig. \ref{fig_r_traces_100mM}-(ii), blue line) and apply the VB-DCMM algorithm to decode the hidden internal state in the signals. 
    Fig. \ref{fig_r_traces_100mM}-(iii) shows time series of internal state, $\bm{x}_{(K)}^{\text{model}}$, calculated from the VB-DCMM by varying $K$ from 1 to 5.
	It is of note that
    the number of actually observed internal states in the $\bm{x}^{\text{model}}_{(K)}$ for a given input parameter $K$
    does not change after some $K_{obs}(\leq  K)$ 
    ($K_{obs}=$ 2 (Fig. \ref{fig_r_traces_100mM}A), 2 (Fig. \ref{fig_r_traces_100mM}B), 2 (Fig. \ref{fig_r_traces_100mM}C), and 1 (Fig. \ref{fig_r_traces_100mM}D)).
    (See also other time traces of synthetic data and H-DNA analyzed in SI: Fig. \ref{fig_r_simul_nondistinguishable}A ($K_{obs}=$ 1), Fig. \ref{fig_r_traces_50mM}A ($K_{obs}=$ 3), Fig. \ref{fig_r_traces_50mM}B ($K_{obs}=$ 2), Fig. \ref{fig_r_traces_50mM}C ($K_{obs}=$ 3), Fig. \ref{fig_r_traces_50mM}D ($K_{obs}=$ 3), Fig. \ref{fig_r_traces_26mM}A ($K_{obs}=$ 4),  Fig. \ref{fig_r_traces_26mM}B ($K_{obs}=$ 2), Fig. \ref{fig_r_traces_26mM}C ($K_{obs}=$ 2), Fig. \ref{fig_r_traces_26mM}D ($K_{obs}=$ 3), and Fig. \ref{fig_r_test_G} ($K_{obs}=$ 3)). 
    A similar behavior is also observed when analyzing data using the variational Bayes Gaussian mixture model \cite{Bishop:2006}.

    To account for the contribution due to degeneracy in labeling the internal states,   $\log{ K! }$ term is conventionally considered in formulating the evidence function $F(K)$ (See {\bf SI} for the details); however, in our problems, the actual number of degeneracy in labeling internal states should be $_KC_{K_{obs}}\times K_{obs}!$ instead of $K!$. 
    Therefore, we replace the $\log{ K! }$ term in $F(K)$ with $\log{\left[K!/(K-K_{obs})!\right]}$,
    and considered a modified evidence function, $G(K)$, to identify an optimal $K$ for a given time trace: 
    \begin{align}
    G(K) \equiv F(K) - \log{  (K-K_{obs})! } 
    \end{align} 
    $G(K)$ shows a clear peak, allowing us to identify the optimal $K(=K^*)$ with ease (blue circles on the right side of Fig. \ref{fig_r_traces_100mM}, \ref{fig_r_traces_50mM}, and \ref{fig_r_traces_26mM}). 
    Use of $G(K)$ instead of $F(K)$ in analyzing synthetic data does not alter $K^*$ (Fig. \ref{fig_r_test_G}, Fig. \ref{fig_r_simul_K4L4}B). 
     
     Among the time traces of H-DNA, traces with more than 3 interconversions between distinct internal states, which enables us to estimate $\gamma^{(\mu)\rightarrow(\nu)}$, are rare, especially when [Na$^+$]=100 mM; thus it is not feasible to get a statistically meaningful scatter plot of ($D_{\text{conf}}$,$D_{\text{int}}$) (see Fig. \ref{fig_r_HDNA_phi_D}D, E, F); however, for those displayed in Fig. \ref{fig_r_HDNA_phi_D}D, E, and F, $\langle D_{\text{tot}}\rangle\approx 7$ suggests that $\chi\gtrsim 0.9$ (from Fig. \ref{fig_r_Dconf_Dint_simul}). 
     Therefore, at least the intra-basin rate constants extracted from H-DNA data using VB-DCMM are reliable.     
	Time traces that have $\tau_{int}$ comparable to experimental observation time ($\tau_{int}\approx\mathcal{T}_{obs}$) would exhibit on average no or only a single transition event between distinct internal states.
     	Indeed, we find that only a subset of total number of internal states is sampled by individual time traces due to the limited observation time. 
	For instance, at [NaCl]=100 mM, our analysis identified $K^* \leq 2$  in 265 out of 269 traces, and that  
     	only 4 time traces display $K^* > 2$ (Fig. \ref{fig_r_traces_100mM_multiple}).
     	Therefore, in order to identify the internal states present in the transition dynamics of H-DNA, clustering analysis is required against the whole ensemble of time trajectories. 
	We provide the procedure of clustering analysis and results in details in the following section.  
    \\

\subsection*{Clustering H-DNA data.}
\begin{figure}[ht!]
              	
              	\centering
              	\includegraphics[scale=0.44]{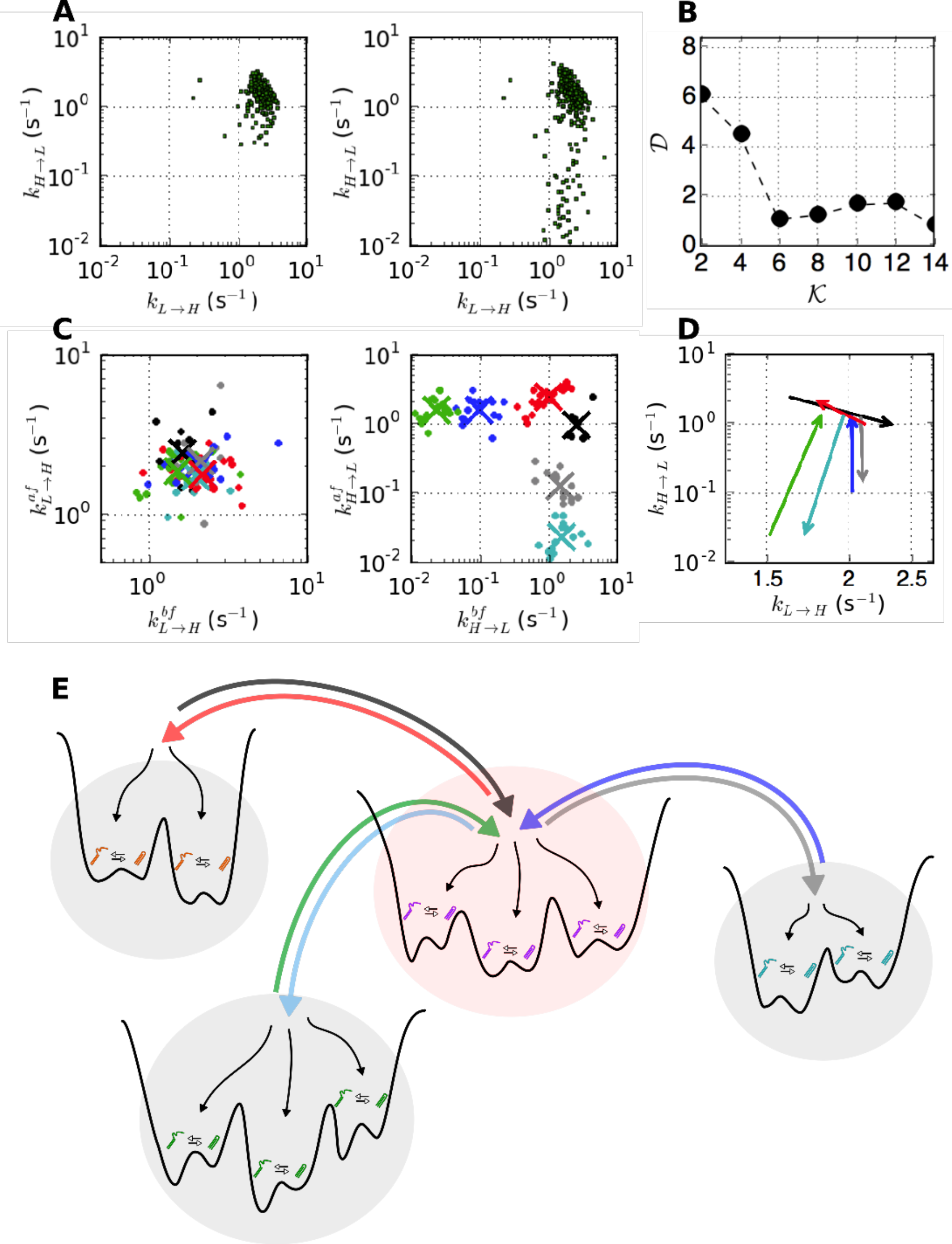}
              	
              	\caption{{\bf Clustering H-DNA data at [Na$^+$]=100 mM condition.}
              		(A) The scatter plots of ($k_{L \rightarrow H}$, $k_{H \rightarrow L}$) before (left) and after (right) applying VB-DCMM from [Na$^+$] =  100 mM data.
              		(B) The ``average pairing distance" $\mathcal{D}(\mathcal{K})$ as a function of the number of clusters ($\mathcal{K}$)(see {\bf Methods}). 
              		The minimum value of $\mathcal{D}(\mathcal{K})$ is found at $\mathcal{K}=6$. 
              		(C)  Left: the scatter plot of clustered data projected on $(k^{\text{bf}}_{L \rightarrow H},k^{\text{af}}_{L \rightarrow H})$ plane. 
              		Right: scatter plot of clustered data projected on $(k^{\text{bf}}_{H\rightarrow L},k^{\text{af}}_{H\rightarrow L})$ plane. 
              		The data belonging to different clusters are depicted in different colors, and the centroid of each cluster is marked with the $\times$ symbol.
              		Total 98 data points were used for analysis. 
              		(D) The result of the above clustering is represented using 6 kinetic arrows which 
              		represent the centroids of each cluster represented in $(k_{L\rightarrow H},k_{H\rightarrow L})$ plane.  
              		The starting point of the arrow is ($\langle k^{(\mu),\text{bf}}_{L \rightarrow H}\rangle$, $\langle k^{(\mu),\text{bf}}_{H \rightarrow L}\rangle$) whereas the ending point of the arrow is ($\langle k^{(\mu),\text{af}}_{L \rightarrow H}\rangle$, $\langle k^{(\mu),\text{af}}_{H \rightarrow L}\rangle$), where the superscript $\mu$ represents the index of each cluster $\mu=1,2,\ldots 6$.
              		The colors used for depicting kinetic arrows are consistent with the data points in (C). 
              		(E) A schematic of the conformational landscape of H-DNA. 
              	}
              	\label{fig_r_clustering_100mM_6}
              \end{figure}

    VB-DCMM algorithm allows us to decompose  
    individual H-DNA time traces with dynamic disorder into multiple ``components", each of which should satisfies the property of \emph{homogeneous Markov chain}. 
    In order to understand the structure of conformational space of H-DNA, the ensemble of components acquired from the VB-DCMM analysis should be clustered into the same kind. 
    To this end, 
    we produce scatter plots of $(k_{L\rightarrow H},k_{H\rightarrow L})$, representing the kinetic property of the ensemble of time traces, using the transition rates estimated for individual time traces. 
    The scatter plots of $(k_{L\rightarrow H},k_{H\rightarrow L})$ were calculated for the ensemble of H-DNA time traces (i) before (Fig. \ref{fig_r_clustering_100mM_6}A, left) and (ii) after decomposing the individual heterogeneous time traces retaining multiple components into the homogeneous ones (Fig. \ref{fig_r_clustering_100mM_6}A, right). 
    The scatter plot of $(k_{L\rightarrow H},k_{H\rightarrow L})$ after the decomposition has a greater dispersion, which is expected since a data point $(k_{L\rightarrow H},k_{H\rightarrow L})$ for a time trace with dynamic disorder is a mixture of $(k^{(\mu)}_{L\rightarrow H},k^{(\mu)}_{H\rightarrow L})$ with $\mu=1,2,\ldots K$. 
   In the presence of clear distinction 
   between internal states ($\mu\neq \nu$), the clustering of 
   $(k^{(\mu)}_{L\rightarrow H},k^{(\mu)}_{H\rightarrow L})$ would be straightforward, which is indeed the case for the synthetic data (Fig. \ref{fig_r_clustering_K3}A). 
  However, for the H-DNA data, even after the decomposition, the clustering of data on $(k_{L\rightarrow H},k_{H\rightarrow L})$ plane (Fig. \ref{fig_r_clustering_100mM_6}A) is not that clear. 

    To improve the quality of clustering,
    we extended the clustering of the kinetic data to a higher dimension by considering the kinetic information of internal states that are contiguous (kinetically linked) along time traces. 
 To be specific, 
 for a time trace exhibiting a transition from the $\mu$-th to $\nu$-th internal state ($\mu\neq\nu$), one can consider that the inter-basin transition has occurred from the time interval represented by its pair of kinetic rate $(k^{\text{bf}}_{L \rightarrow H}, k^{\text{bf}}_{H \rightarrow L})[=(k^{(\mu)}_{L \rightarrow H}, k^{(\mu)}_{H \rightarrow L})]$ to the next time interval represented by $(k^{\text{af}}_{L  \rightarrow H}, k^{\text{af}}_{H \rightarrow L})[=(k^{(\nu)}_{L \rightarrow H}, k^{(\nu)}_{H \rightarrow L})]$, where the superscripts, `bf' and `af' denotes `before' and `after' the transition, respectively.  
Thus, instead of $(k_{L\rightarrow H},k_{H\rightarrow L})$, a clustering at a higher dimension can be carried out by measuring the Euclidean distance between a pair of the four-dimensional (4-dim) arrays, $(\log{k^{\text{bf}}_{L \rightarrow H}}, \log{k^{\text{bf}}_{H \rightarrow L}}, \log{k^{\text{af}}_{L \rightarrow H}}, \log{k^{\text{af}}_{H \rightarrow L}})$. 
    
  In order to cluster the 4-dim arrays we used the k-means clustering algorithm.
 Application of the algorithm to the H-DNA data at [Na$^+$]=100 mM reveals that the average pairing distance, $\mathcal{D}(\mathcal{K})$ (see {\bf Methods}), is minimized when the number of clusters is 6 ($\mathcal{K}=6$), namely, the model with 6 clusters provides the best interpretation of the data (Fig. \ref{fig_r_clustering_100mM_6}B).
 	Although the model with 14 clusters shows a smaller $\mathcal{D}$, we selected 
	$\mathcal{K}=6$ as the best solution, since for $\mathcal{K}=14$ each of 12 clusters out of 14 has less than 10 data points, which makes the result of clustering statistically less significant (Fig. \ref{fig_r_clustering_100mM_2}).
	This results remain qualitatively identical when $L1$ distance (so called ``city block'' distance) was used instead of ``square-euclidean'' distance (Fig. \ref{fig_r_clustering_100mM_cityblock}). 
	Furthermore, the clustering algorithm using ``affinity propagation'' \cite{Frey:Science:2007}, which considers all the data points as possible exemplars (analogous to centroids in k-means clustering method) and iteratively exchanges messages between them, also gives qualitatively identical results, confirming the robustness of the  conclusion on H-DNA dynamics obtained from VB-DCMM and k-mean clustering (see Fig. \ref{fig_r_af_100mM}). 
	
	We present the result of clustering either (i) by projecting it on the two separate kinetic planes,   
    $(k_{L\rightarrow H}^{\text{bf}},k_{L\rightarrow H}^{\text{af}})$ and $(k_{H\rightarrow L}^{\text{bf}},k_{H\rightarrow L}^{\text{af}})$, which visualize the inter-basin transitions in terms of variable $L\rightarrow H$ and $H\rightarrow L$ transition rates (see Fig. \ref{fig_r_clustering_100mM_6}C), 
     or 
     (ii) by using ``interconversion arrows" linking the kinetic rates of two internal states, $\left[(\log{k^{\text{bf}}_{L \rightarrow H}},\log{k^{\text{bf}}_{H \rightarrow L}})\rightarrow (\log{k^{\text{af}}_{L \rightarrow H}},\log{k^{\text{af}}_{H \rightarrow L}})\right]$ on the $(k_{L\rightarrow H},k_{H\rightarrow L})$ plane (Fig. \ref{fig_r_clustering_100mM_6}D).    
Note that in the scatter plot visualized with $(k_{H\rightarrow L}^{\text{bf}},k_{H\rightarrow L}^{\text{af}})$, the distinction between different clusters is clear (the right panel of Fig. \ref{fig_r_clustering_100mM_6}C).
     Furthermore, for a system in equilibrium or at least near equilibrium,  
     the interconversion between two internal states, say $\mu$ and $\nu$, should occur in both directions, i.e., $\mu\rightarrow\nu$ and $\nu\rightarrow\mu$.  
     In the representation (i), a symmetry of $(k^{(\mu)}_{a\rightarrow b},k^{(\nu)}_{a\rightarrow b})=(k^{(\nu)}_{a\rightarrow b},k^{(\mu)}_{a\rightarrow b})$ is expected in the both panels of Fig. \ref{fig_r_clustering_100mM_6}C;  
    and in the representation (ii), the ``arrows", amounting to the kinetic connectivity between distinct internal states, should be bi-directional. 
    The symmetry of the data plotted in Fig.\ref{fig_r_clustering_100mM_6}C or the bidirectionality of the kinetic arrows confirms the condition of detailed balance being satisfied in the system in equilibrium.   
    Fig. \ref{fig_r_clustering_100mM_6}D depicts 6 kinetic arrows (3 pairs of reversible kinetic arrows) connecting the centroids of $(\log{k^{\text{bf}}_{L \rightarrow H}},\log{k^{\text{af}}_{L \rightarrow H}})$ or $(\log{k^{\text{bf}}_{H \rightarrow L}},\log{k^{\text{af}}_{H \rightarrow L}})$ data.

	Application of the above clustering method to synthetic data with $K=3$, $N=2$ (Fig. \ref{fig_r_clustering_K3}) is straightforward.
    To check the efficacy of clustering method for a more complicated case, we have tested with synthetic data generated with $K=4, N=4$, i.e. when there are as many as 4 observable states in each internal state (Fig. \ref{fig_r_simul_K4L4}, Fig. \ref{fig_r_clustering_K4L4_connected}).
    	In the case with 4 observable states, total 12 possible intrabasin transitions are conceivable. 
    	Thus, the dimension of the array associated with interbasin transition is 24.
    	As long as there is a clear time scale separation, it is expected that the pairing distance $\mathcal{D}(\mathcal{K}=12)$ shows  minimum as there are 12 connection paths between 4 internal states.
    	Indeed,	$\mathcal{D}(\mathcal{K})$ is minimized at $\mathcal{K} = 12$ (Fig. \ref{fig_r_clustering_K4L4_connected}A).

        Lastly, it is noteworthy that the clustering method presented here is not limited to data analysis for systems in equilibrium, but can be extended to systems in nonequilibrium steady state \cite{english2006NCB} where the individual state-to-state kinetic transition rate is well defined using the reversible Markov process although the condition of detailed balance is no longer anticipated \cite{QianBook,Hyeon09PCCP}. 
        The symmetry of data point and bidirectionality of kinetic arrows as in Fig.\ref{fig_r_clustering_100mM_6}C, D are still of use to cluster the kinetic information generated from a system in nonequilibrium steady states. 
\\
    
    \subsection*{Folding energy landscape of H-DNA.}
    We classified the ``components" of a similar kinetic pattern ($k_{L \rightarrow H}$, $k_{H \rightarrow L}$) obtained from VB-DCMM into a single cluster which  represents a kinetic path linking two independent basins of attraction (or internal states). 
    For example, the kinetic paths in Fig. \ref{fig_r_clustering_100mM_6}D can be best understood by hypothesizing 4 internal states (four basins) linked by 6 kinetic paths. 
    Thus, 
    the conformational transition landscape of H-DNA at [Na$^+$]=100 mM condition 
    consists of 4 internal states with 3 reversible kinetic paths being established as illustrated in Fig. \ref{fig_r_clustering_100mM_6}E. 
    At lower salt concentrations ([Na$^+$] = 50 mM (Fig. \ref{fig_r_clustering_50mM}) and [Na$^+$] = 26 mM (Fig. \ref{fig_r_clustering_26mM})),
    H-DNA transitions slow down and the dispersion of data also increases; however, the overall structure of conformational landscape of H-DNA remains unchanged from the picture suggested in Fig. \ref{fig_r_clustering_100mM_6}E; thus, there is a central superbasin to which three other superbasins are kinetically connected (Figs. \ref{fig_r_clustering_50mM} and \ref{fig_r_clustering_26mM}). 
    \\
    
    \subsection*{Contributions of our work.}
    In comparison to other pre-existing methods, the advantage of our VB-DCMM in decoding dynamic disorder from a given trajectory is highlighted as follows: 
   
    (1) Dynamic disorders in single molecule time trajectories are modeled using DCMM by assuming the presence of hidden internal states. 
    While Aggregated Markov Model (AMM), which has been adopted in ion-channel community for time trace analysis of varying current \cite{Fredkin_1986,Kienker_1989,Colquhoun_1981,Horn_1983,Qin_2004_BPJ,Qin_1997,Wagner_1999,Ball_1999,Rosales_2004,Rosales_2001,Gin_2009,Siekmann_2011_BPJ,Siekmann2012_BPJ,Hodgson_1999,Hines2015_BPJ}, can be employed to analyze our data with dynamic disorder, 
    DCMM is better in correctly decoding dynamic disorder than AMM. We found that AMM is prone to overpredict the transition between kinetic patterns (Fig. \ref{fig_r_iAMM}). 
    Our method is more suitable to the data showing persistent dynamic patterns by suppressing unwanted frequent transition between kinetic patterns.
    Detailed explanations of connection and quantitative comparison between DCMM and AMM are provided in SI and Fig.\ref{fig_r_iAMM}.

    (2) In this paper, Bayesian version of DCMM was developed by using variational Bayes (VB) method, which enabled us to determine the number of internal states straightforwardly. 
    Although Bayesian version of DCMM using Markov chain Monte Carlo (MCMC) method has previously been developed for the credit portfolio modeling \cite{Fitzpatrick:2013}, 
    the idea of Bayesian inference in ref. \cite{Fitzpatrick:2013} was used only for the purpose of calculating a posterior distribution of model parameters. 
    To determine the number of hidden states corresponding to the internal states in this study, the authors in ref. \cite{Fitzpatrick:2013} used the economic cycle fluctuation model, instead. 
    Our study combining VB with DCMM (i) can determine the number of internal states in a more objective fashion, (ii) offers intuitive way to incorporate prior knowledge, and (iii) is computationally more efficient than MCMC (See SI for details).
 
    (3) We tested VB-DCMM under various conditions, by varying the kinetic rates, the number of observables, the number of hidden states, and prior parameters.
    New metrics were also devised to quantify the performance of algorithm systematically.
    
    (4) Finally the connection paths (kinetic arrows) between internal states of H-DNA are clustered by using the kinetic components extracted from VB-DCMM and by applying k-means clustering algorithm to high dimensional arrays.

	To recapitulate, our entire process of analyzing single molecule data is composed of three stages: (i) noise-filtering using HMM; (ii) decomposition of heterogeneous time traces into the homogeneous components using VB-DCMM; (iii) clustering the decomposed components into the same cluster. 
	
		In principle, this three-stage analysis can be made more systematic by combining the noise-filtering and clustering procedure with VB-DCMM.
To be more specific,     
    	(1) The noise-filtering of observable trace ($\bm{o}_n$) is processed, independently from the main VB-DCMM algorithm, by using HMM, which has been proved to be reliable in noise-filtering \cite{McKinney2006}, 
	and the maximum number of observables ($N$) are predetermined as an input parameter.
	Current version of algorithm can be further automated by combining with the Bayesian version of HMM \cite{Bronson20093196}, which can determine the number of observables while filtering the noise in data 
	(See Fig. \ref{fig_r_intro}D). 
    	The resulting model will have a similar structure with the modified factorial HMM \cite{Ghahramani_1997_MachLearn,Jordan_1997_NIPS}.
    (2) The heterogeneous components identified from individual time traces are clustered separately from our main algorithm.
    	It would be also desirable to unify the post-processing step (clustering) 
	with VB-DCMM using 
	empirical Bayes method which has been applied recently to analyze single molecule data \cite{Meent2013,Meent20141327}.

	However, it should also be noted that a blind integration of noise-filtering and clustering steps inevitably complicates the implementation of VB-DCMM, as more number of prior parameters are ought to be decided by users. 
		For example, Bayesian implementation of HMM for noise filtering demands manual determination of additional $N (N+5) $ prior-parameters \cite{Bronson20093196}. 
		Compared to this, currently VB-DCMM requires users to pre-determine only one prior parameter which characterizes the final transition rate matrix, $\bm{A}$ (see the subsection: \emph{Selection of prior parameters} in SI).
		Moreover, the integration of other methods will obscure the flow of analysis, making it difficult to identify an error-causing step.
		Keeping each step in the algorithm separate makes the integration of VB-DCMM to other applications more transparent 
		(for example, if noise-filtering by HMM is unsuccessful, other advanced method can be employed \cite{Bronson20093196}).  
		We leave it as our future work to develop an algorithm that integrates the above-mentioned three procedures (noise-filtering, VB-DCMM, and clustering) without increasing complexity or obscuring the flow of analysis.
	
    In decoding SM FRET data, the most notable difference of our VB-DCMM from the previous studies employing the probabilistic models such as maximum likelihood and Bayesian statistics is that VB-DCMM 
    explicitly considers the situation that transition rates can change from one time interval to another within individual time traces. 
    The previous studies \cite{Schroder:2003:JCP,Antonik:2006:JPCB,Gopich:2009:JPCB,McKinney2006,Bronson20093196,Bronson2010Graphical,Okamoto2012} assumed that the transition rates were constant within individual time traces. 
    Also, currently, VB-DCMM is applicable to window-averaging FRET trajectories.
    It will be of great interest to extend VB-DCMM to analyzing time trajectories in which arrival times for individual photons are available.
	VB-DCMM is particularly powerful when there is a separation in time scales between $\tau_{int}$ and $\tau_{conf}$.

    \section*{Concluding Remarks}
    While the notion of dynamical heterogeneity or broken ergodicity seems better recognized in the research field of nucleic acids \cite{AlHashimi08COSB} than in proteins, which likely arises from more homopolymer-like nature of building block of nucleotides \cite{Thirum05Biochem}, biomolecules in general can have a rugged folding landscape with many local basins of attraction and kinetic barriers with varying heights \cite{Thirumalai10ARB}. 
    Conformational dynamics of biomolecules on rugged landscapes can be heterogeneous, which gives rise to static or dynamic disorder depending on the time scale of observation or the height distribution of kinetic barrier. 
    The presence of heterogeneity or disorder among individual molecules, unveiled by \emph{in vitro} SM experiments could be surprising at first sight; however, it is also important to note that the general hypotheses in the conventional molecular biology towards a single native state have been put forward based on the observations from ensemble experiments where the heterogeneity, if any, is usually masked by the process of ensemble averaging.  
    Given that the complexity of a molecular system increases with the system size ($N_{\text{sys}}$) as $\sim e^{N_{\text{sys}}}$ \cite{Palmer82AP}, it should not be too surprising to find such disorder in biomolecules in itself. 
    Cells are equipped with molecular chaperones that can tame misfolding-prone biomolecules with rugged landscapes \cite{Bhaskaran07Nature,Woodson10RNABiol,ThirumalaiARBBS01,Hyeon2013JCP}; thus the principle of optimization in biology, if it fails at the level of a molecule in isolation, can be extended further to the molecular system including its environmental factors.
    
    It is not easy to elucidate the molecular origin of disorder in a conclusive manner; yet, it has recently been suspected that interactions of biomolecules with cofactor such as ATP and multivalent metal-ions could be the microscopic causes for those molecules exhibiting dynamical heterogeneity  \cite{Solomatin2010,Hyeon:2012aa,Liu2013,Kowerko2015_PNAS,Segev2012_CurrOpinStructBiol}. 
    Modulating the concentration of Mg$^{2+}$ ions from high to low and again to high induced inter-conversions of dynamic patterns in equilibrium conformational fluctuations of \emph{T}. ribozyme \cite{Solomatin2010} and Holliday junctions \cite{Hyeon:2012aa}.  
    Distinct velocities of ATP-empowered individual RecBED helicase motors, which can move progressively along dsDNA by unwinding it into two separate strands, can be reset by introducing a long pause by halting the supply of ATP.   
    For the time trajectories of biomolecules displaying quenched disorder, a method to analyze such data was proposed using a concept from glass physics \cite{Hyeon:2012aa}. 
    Here, to deal with more general scenarios, we have developed a method to analyze single molecule time traces with dynamic disorder. 
    
    As demonstrated by testing the VB-DCMM algorithm on synthetic data, the algorithm is quite accurate in decoding dynamic disorder as long as a time trajectory of interest contains multiple time intervals, each of which display kinetic pattern distinct from others.
    When a clear separation in timescale is present between two distinct kinetic patterns, large value of $D_{\text{conf}}$, $D_{\text{int}}$, and $D_{\text{tot}}$ would be acquired.

    While we developed the VB-DCMM algorithm primarily to analyze dynamic disorder in duplex-triplex transitions of H-DNA, the method is applicable to any data in the form of one-dimensional time series with multiple transitions. 
    Together with a further technical advance in SM, which eliminates experimental artifacts as well as extends the measurement time, our algorithm developed here will contribute to better understanding of biomolecules that display heterogeneous dynamics.\\
    
\section*{Methods}  	   
\subsection*{Generation of synthetic data.}
Internal state sequence $\bm{x}$ was generated by using Monte Carlo method with a constant transition matrix (homogeneous Markov chain model). The observable sequence $\bm{o}$ was generated by using the same method but with the transition matrix that was defined at each time $t$ based on the internal state $x(t)$. Finally, Gaussian noise was added on $\bm{o}$ to produce $\bm{o}_n$. 
\\

\subsection*{Single-molecule FRET measurements to monitor duplex-triplex transitions of H-DNA}
We purchased triplex forming oligonucleotides from Integrated DNA Technologies (Coralville, IA, USA). 
The oligonucleotides were dissolved in T50 buffer solution (10 mM Tris-HCl, 50 mM NaCl, pH=7.5) and were heated beyond the melting temperature of DNA duplex ($\sim$ 90 $^o$C), and slowly cooled down on a heat block to room temperature over 8 hour to properly hybridize them.  
The DNA prepared as such is called ``H-DNA" here. 
The sequences of the triplex forming strands (purine-rich and pyrimidine-rich) are: 
Purine-rich strand: 5' AAG AAG AAG AAG AAG (Cy5) TGG CGA CGG CAG CGA (Biotin) 3', 
Pyrimidine-rich strand: 5' TCG CTG CCG TCG CCA CTT CTT CTT CTT CTT TTT TCT TCT TCT TCT TCT TC (Cy3) 3'.
In the purine-rich strand, the biotin at 3' terminus is used to attach the H-DNA molecule to a neutravidin-coated cover-glass. The Cy3 and Cy5 dyes in the H-DNA molecule correspond to a donor and an acceptor for FRET measurements, respectively.
In order to observe the transition between folded triplex and unfolded DNA, we used the reaction buffer containing 50 mM HEPES(Sigma-Aldrich) and various concentrations of Na$^+$ (26, 50, 100 mM). These buffer solutions also contained 2 mM trolox, 10 \% glucose and gloxy for single-molecule fluorescence experiments.
We utilized a home-made TIRF (Total Internal Reflection Fluorescence) microscope to measure the FRET efficiency between donor and acceptor dyes, which reveals the conformational state of the H-DNA molecule. A 532-nm laser (CrystaLaser DPSS, 10 mW) was used to excite donor molecules and fluorescence intensities of both dyes were measured by an EMCCD (Andor iXon DV887, Andor technology). To observe the change of FRET efficiency in real time, we measured the time-lapse FRET traces with the repetition rate of 10 Hz. 
To study kinetic features of the conformational transition with dynamic disorder, we acquired the FRET time traces for a long period ($> 100$ sec).
\\

\subsection*{Clustering at a higher dimension.} 
For given $N$ and $K$, total $N(N-1)$ intra-basin transition rates $k_{a,b}^{(\mu)}$ ($a, b \in \{1, 2, \cdots, N\}, a\neq b$) are defined in the $\mu$-th basin (or $\mu$-th internal state) and total $K(K-1)$ inter-basin transitions are conceivable. 
To cluster the kinetic information of H-DNA data obtained from VB-DCMM, 
we consider the \emph{kinetic arrow}, $2N(N-1)$-dimensional array of data, which has the structure of 
$\bm{C}_{i} \equiv ( \{\log{ k^{i, \text{bf}}_{a,b}} \}, \{ \log{ k^{i, \text{af}}_{a,b} } \})$ where the subscript $i$ denotes an index referring to one of $K(K-1)$ possible inter-basin transitions linking two internal states ($\mu\neq\nu$). 
For a kinetic scheme made of a network of reversible transitions between $K$ internal states, the transition between two internal states should be bidirectional; thus for a given inter-basin transition path $i$, there should be a kinetic path $j$ antiparallel to the path $i$, satisfying $\|\bm{C}_i-\tilde{\bm{C}}_j\|\approx 0$, where $\tilde{\bm{C}}_j\equiv (\{\log{k_{a,b}^{j,\text{af}}}\},\{\log{k_{a,b}^{j,\text{bf}}}\})$. 
In our problem, the set of all the data generated as an outcome of VB-DCMM can in principle be clustered into the disjoint subsets of size 2 partitioning the $\mathcal{K}$ transition paths, $\{\mathcal{K}|1\leq \mathcal{K}\leq K(K-1)\}$, and one realization of such disjoint subsets will minimize the pairwise sum of Euclidean distances $\|\bm{C}_{\alpha}-\tilde{\bm{C}}_{\beta}\|$ for all $\alpha$ and $\beta$; however, the method suffers from high computational cost as the possible number of clusters increases rapidly with $N$ and $K$.


To alleviate the computational cost for large $N$ and $K$, we modified the original method. 
We first searched the the best partitioning set of data $S^*(\mathcal{K})$ for a given $\mathcal{K}$ that minimizes the Euclidean distance between all the pairs of centroids, 
\begin{align}
\mathcal{D}^c(\mathcal{K})=\frac{2}{\mathcal{K}} \sum_{(i,j)} (d^c_{i j})^2
\end{align}
where $d^c_{i j}=\| \bm{C}^c_{i} - \tilde{\bm{C}}^{c}_{j}\|$ with $\bm{C}^c_{i} \equiv ( \{ \langle \log{ k^{i, \text{bf}}_{a,b}}\rangle \}, \{ \langle \log{ k^{i, \text{af}}_{a,b} } \rangle \})$, 
$\tilde{\bm{C}}^c_{j} \equiv ( \{ \langle \log{ k^{j, \text{af}}_{a,b}}\rangle \}, \{ \langle \log{ k^{j, \text{bf}}_{a,b} } \rangle \})$, 
and $\langle\ldots\rangle$ denotes the centroid of clustered data.
To obtain the best clustering result for a given $\mathcal{K}$, 
we conducted k-means clustering using $k\_means$ function from scikit-learn libraries \cite{scikit-learn} with 20,000 different random initial conditions in each analysis.
It is expected that 
$
\mathcal{D}^c(\mathcal{K})=\frac{2}{\mathcal{K}}\sum_{(i,j)} (d^c_{i j})^2\geq\frac{2}{\mathcal{K}}\sum_{S^*(\mathcal{K})}(d^c_{i j})^2
$. 
The summation, $\sum_{(i,j)}$, signifies that the sum is taken over the disjoint subsets of size 2 partitioning a set $\{1,\ldots,\mathcal{K}\}$ with $\mathcal{K}$ being an even number) 
and $S^*(\mathcal{K})$ is the best partitioning set that minimizes the value of $\mathcal{D}^c(\mathcal{K})$ for a given $\mathcal{K}$. 
For example, provided that there are 4 kinetic arrows made of centroids ($i=1,2,3,4$), which minimizes $\mathcal{D}^c$ at $\mathcal{K}=2$ when $i=1$ is paired with $i=3$ and $i=2$ with $i=4$, then $S^*(2)= \{ \{1,3\}, \{2,4\}\}$ and $\mathcal{D}^c(2) = d^c_{13} + d^c_{24}$.


Next, in order to decide the optimal $\mathcal{K}$, we calculated pairing distance between paired clusters in $S^*(\mathcal{K})$ again, but this time using all the elements in each cluster.
The total pairing score 
\begin{align}
\mathcal{D}(\mathcal{K})\equiv \frac{2}{\mathcal{K}}  \sum_{S^*(\mathcal{K})} \langle d_{ij} \rangle, 
\end{align} 
where the average pairing distance between two clusters $i$ and $j$ 
is defined as 
$\langle d_{i j}\rangle \equiv \frac{1}{M_{i} M_{j}} \sum_n^{M_i}\sum_{m}^{M_j} \| \bm{C}_{i_n} - \tilde{\bm{C}}_{j_m} \|$ 
where 
$\bm{C}_{i_n} = ( \{\log{ k^{i_n, \text{bf}}_{a,b} }\},\{\log{ k^{i_n, \text{af}}_{a,b} }\})$, 
$\tilde{\bm{C}}_{j_m} = ( \{\log{ k^{j_m, \text{af}}_{a,b} }\},\{\log{ k^{j_m, \text{bf}}_{a,b} }\})$, 
and $n$ refers to an index for the element in the $i$-th cluster and $m$ to an index for the elements in the $j$-th cluster.
$M_i$ is the total number of the elements in the $i$-th cluster.
Finally, the optimal $\mathcal{K^*}$, minimizing $\mathcal{D}(\mathcal{K})$, is selected, i.e., $\mathcal{K}^*=\arg{\min{\mathcal{D}(\mathcal{K})}}$, and the interpretation of data is conducted for the best partitioning set $S^*(\mathcal{K}=\mathcal{K}^*)$. 

For H-DNA data at three different Na$^+$ concentrations, 
the optimal $\mathcal{K}^*$ are determined at $\mathcal{K^*}=6$ for [Na$^+$]=100 mM (Fig. \ref{fig_r_clustering_100mM_6}A), $\mathcal{K^*}=10$ for [Na$^+$]=50 mM (Fig. \ref{fig_r_clustering_50mM}B), $\mathcal{K^*}=12$ for [Na$^+$] = 26 mM (Fig. \ref{fig_r_clustering_26mM}B).
This implies that 
the complexity of conformational space of H-DNA increases at low salt condition 
(also see the scatter plot of ($k_{L \rightarrow H}, k_{H \rightarrow L}$) in Fig. \ref{fig_r_clustering_100mM_6}A, Fig. \ref{fig_r_clustering_50mM}A, Fig. \ref{fig_r_clustering_26mM}A). 

The clustering results presented in this study remain robust regardless of the choice of distance metric. 
K-means clustering using $L1$ distance (``city block'') measure with 20,000 different random initial conditions also was led to qualitatively similar results (Fig. \ref{fig_r_clustering_100mM_cityblock}).  
Furthermore, as an alternative clustering algorithm, we also tested ``affinity propagation" \cite{Frey:Science:2007} on our data, and the results remain qualitatively identical (see Fig. \ref{fig_r_af_100mM}). 
In the affinity propagation method, negative square-euclidean distance was employed as a similarity metric ($s(i,j)=-||\bm{x}_i-\bm{x}_j||^2$) where $\bm{x}_i$ denotes the coordinate of the $i$-th data point. 
	The objective of the algorithm is to optimize the factorized probability distribution which approximates the net similarity $\mathcal{S}$, defined as $\mathcal{S}\sim \prod_{i=1}^Ne^{s(i,c_i)}$.  
	Here, $c_i$ is the index of the exemplar of $i$-th data point $\bm{x}_i$. For example, if $c_i = k$, $\bm{x}_k$ is an exemplar of $\bm{x}_i$ and $\bm{x}_i$ belongs to the cluster represented by $\bm{x}_k$.
	Multiple iterations of message passing are carried out until convergence is achieved in the result and the best result of clustering is acquired. 
For implementation, we used \emph{AffinityPropagation} class from \emph{scikit} \cite{scikit-learn} library with varying ``preference" as an input parameter, where the preference denotes the logarithm of probability that $i$-th data point $x_i$ selects itself as an exemplar. 
Further details of the algorithm are available in Ref.\cite{Frey:Science:2007}.
\\
    
\section*{Acknowledgements}
We thank the KIAS Center for Advanced Computation for providing computing resources.
This study was partly supported by National Research Foundation of Korea  NRF-2015R1D1A1A01060376.

\clearpage
\setcounter{figure}{0}  
\setcounter{equation}{0}
{\Large {\bf Supplementary Information} }

    \section{VB-DCMM algorithm: Backgrounds}
    We propose a new algorithm (Variational Bayes Double Chain Markov Model (VB-DCMM)) which combines three theoretical frameworks: Double Chain Markov Model (DCMM), maximum evidence, and Variational Bayes. 
    
    (1) DCMM consists of two layers of Markov chains.
    The elements of transition matrix in the Markov chain in the first layer are decided by the Markov chain in the second layer.
    In the light of analyzing single molecule time traces with dynamic disorder, the first and second layers of Markov chain are straightforwardly related to the transition dynamics along the sequences of hidden internal state ($\bm{x}$) and observable state ($\bm{o}$), respectively.
    While DCMM provides a straightforward conceptual framework to formulate the problem, the method itself, aiming to determine the best parameters for a given model, is not suitable for the best model selection (in our problem, the number of internal states, $K$).
    
    (2) The maximum evidence enables a comparison between models, allowing us to select the best model; however, its computational cost is too high because the method requires considering the entire parameter space. 
    
   (3) To circumvent this difficulty, we incorporated the Variational Bayes technique into the algorithm and calculated an approximate value of the maximum evidence.
    
    \subsection{Double Chain Markov Model}
     Double Chain Markov Model (DCMM), first formally introduced by Berchtold \cite{DCMM}, is defined with the following elements.
     \begin{itemize}
        \item $T$: The total length of data.
        \item $K$: The total number of internal states.
        \item $N$: The total number of observable states.
        \item $\bm{A}$: ($K\times K$)-transition matrix for $\bm{x}$.
        \item $\bm B = (\bm{B}^{1}, \bm{B}^{2}, ... \bm{B}^{K})$ where $\bm{B}^{\mu}$ denotes ($N\times N$)-transition matrix for $\bm{o}$ when internal state $x(t)=\mu$.
        \item $\bm{x}=(x(1), x(2), \cdots, x(t), \cdots,  x(T-1))$: The sequence of internal state. The transition,  $x(t-1)\xrightarrow{\bm{A}} x(t)$, is modeled as a homogeneous Markov process, the rate of which is determined by the transition matrix $\bm{A}$. 
	The value of the internal state at time $t$, $x(t)\in\{1,2,\ldots,\mu,\ldots,K\}$, set the transition rate matrix $\bm{B}^{x(t)}$ which determines the transition of observable state from $o(t)$ to $o(t+1)$. 
        \item $\bm{o}=(o(1), o(2), \cdots, o(t), \cdots, o(T))$: The sequence of observable state. 
        The observable state denotes an index assigned to the value of data after filtering noises from experimental data, such that $o(t)\in\{1,2,\ldots,N\}$.  
        The transition, $o(t)\xrightarrow{\bm{B}^{\mu}} o(t+1)$, is modeled as non-homogeneous Markov chain with a transition matrix $\bm{B}^{\mu}$, whose elements are decided by the internal state of $x$ at time $t$ ($x(t)=\mu$).
        	\item ${\bm \pi} = (\pi_1, \pi_2, ..., \pi_K)$ where $\pi_{\mu} = P(x(1)=\mu | \bm{o, A, B})$ is the conditional probability of having $x(1)=\mu$ for a given $\bm{o}$, $\bm{A}$, and $\bm{B}$.

    \end{itemize}
    The probability of observing $\bm{o}$ and $\bm{x}$ with a given set of  parameters $\bm{\lambda = (\pi, A, B)}$ can be written as 
    \begin{widetext}
    \begin{equation}\label{eq:p_ox_dcmm}
    P( \bm{o, x | \lambda}) = \pi_{x(1)} B_{x(1), o(1),o(2)} \prod_{t=1}^{T-2}A_{x(t), x(t+1)} B_{x(t+1), o(t+1),o(t+2)},
    \renewcommand{\theequation}{S\arabic{equation}} \end{equation}
    \end{widetext}
    where $A_{i,j}\equiv (\bm{A})_{ij}$ denotes the (i,j) element of the transition matrix $\bm{A}$, and $B_{\mu,i,j}\equiv (\bm{B}^{\mu})_{ij}$ denotes the $(i,j)$ element of the transition matrix $\bm{B}^{\mu}$.  
    By using Eq. \ref{eq:p_ox_dcmm} and adapting a similar procedure in Hidden Markov Model (HMM) (Forward-Backward algorithm and Baum-Welch algorithm), it is possible to determine the optimal parameter $\bm \lambda^*$ that (locally) maximizes $P(\bm{o}| \bm{\lambda})\left(=\sum_{\bm{x}}P(\bm{o},\bm{x}|\bm{\lambda})\right)$ \cite{DCMM}. For given $\bm o$ and $\bm \lambda^*$, the sequence $\bm{x}$ for the internal state is determined (Viterbi algorithm) \cite{DCMM}.

    \subsection{Maximum Evidence}
    In contrast to the maximum likelihood method used to identify optimal $\boldsymbol{\lambda}$ maximizing $P(\bm{o} | \boldsymbol{\lambda})$, the maximum evidence method selects the optimal model (in our case, optimal number of internal states $K$) maximizing $P(\bm{o} | \boldsymbol{K})$.
    \begin{equation}
   	P(\bm{o} | \boldsymbol{K}) = \int P( \bm{o}| \boldsymbol{\lambda'} ) P( \boldsymbol{\lambda'} | \boldsymbol{K} ) d \boldsymbol{\lambda'}. 
   	\renewcommand{\theequation}{S\arabic{equation}} 
   	\end{equation}
   	In the maximum evidence, the likelihood value ($P( \bm{o}| \boldsymbol{\lambda} )$) from an optimal $\boldsymbol{\lambda}$ is reduced by the factor $P( \boldsymbol{\lambda} | \boldsymbol{K} )$, which could be smaller in more complex model since there are more freedom in choosing $\boldsymbol{\lambda}$.
   	For $P( \boldsymbol{\lambda'} | \boldsymbol{K} )= \delta(\lambda - \lambda') $, the evidence becomes the likelihood.

    \subsection{Variational Bayes}
    The maximum evidence is formally suited for model selection, but the computational cost of the method, which requires integrating over the entire parameter space, is too large. 
    To circumvent this difficulty, we combine the variational Bayes method with DCMM.
    
    Let $q(\bm{Z})$ be an arbitrary probability distribution of a set of variable $\bm Z$ consisting of parameters and hidden variables of model (In DCMM, $\bm {Z = (x, \lambda)}$).  
    Then, from $\int q(\bm{Z}) d\bm{Z} = 1$, the logarithm of the evidence, i.e., 
    $\log{ P(\bm{o } | \bm{K}) }$ can be written as \cite{Bishop:2006}
\begin{widetext}
    \begin{equation}\label{eq: p_ox_K}
    \begin{aligned}
    \log (P(\bm{o } | \bm{K}))&= \int  q(\bm{Z})\  \log (P(\bm{o } | \bm{K})) d\bm{Z} \\
    &= \int  q(\bm{Z})\  \log\left( P(\bm{o } | \bm{K}) \frac{ P(\bm{o ,Z}|\bf{K}) }{q(\bm{Z})} \frac{ q(\bm{Z}) }{ P(\bm{o ,Z}|\bf{K}) }\right) d\bm{Z} \\
    &=  \int  q(\bm{Z})\ \log\left( \frac{P(\bm{o ,Z} | \bm{K})}{q(\bm{Z})}\right)  d\bm{Z}
    + \int q(\bm{Z})\ \log\left(\frac{q(\bm{Z}) }{ P(\bm{Z}|\bf{o , K})} \right) d\bm{Z} \\
    &= F[q] + D_{KL}(q || p )
    \end{aligned}
    \renewcommand{\theequation}{S\arabic{equation}} \end{equation}
    \end{widetext}
    where $p$ denotes $P(\bm{Z}|\bm{o},\bm{K})$,  
    \begin{equation}
    P(\bm{Z} | \bm{o, K}) = P( \bm{o, Z} | \bm{K} ) / P( \bm{o} | \bm{K}), 
    \renewcommand{\theequation}{S\arabic{equation}} \end{equation}
    \begin{equation}
    F[q] \equiv  \int  q(\bm{Z}) \ \log\left( \frac{P(\bm{o,Z} | \bm{K})}{q(\bm{Z})}\right)  d\bm{Z}, 
    \label{eq:Fq}
    \renewcommand{\theequation}{S\arabic{equation}} \end{equation}
     and 
     \begin{equation} \label{eq: p_q_KL}
     D_{KL}(q || p) \equiv \int q(\bm{Z})\ \log \left(\frac{q(\bm{Z}) }{ P(\bm{Z}|\bf{o, K})} \right) d\bm{Z}.
     \renewcommand{\theequation}{S\arabic{equation}} \end{equation}
    As Kullback-Leibler divergence always satisfies $D_{KL}(q||p) \geq 0$, the following inequality holds.
    \begin{equation}
    \log( P(\bm{o} | \bm{K} ) ) = F[q] + D_{KL}(q||p) \geq F[q]
    \renewcommand{\theequation}{S\arabic{equation}} \end{equation}
    
    The Variational Bayes method aims to maximize the lower bound of $F[q]$ (and thus the lower bound of $  \log (P(\bm{o} | \bm{K}))$ by refining $q(\bm Z)$ iteratively, anticipating that $F[q]$ converges to $  \log (P(\bm{o} | \bm{K}))$. 
    When $F[q]$ converges to $  \log (P(\bm{o} | \bm{K}))$, $D_{KL}(q||p)$ converges to 0, indicating that $q(\bm Z)$ converges to $P(\bm{Z}|\bf{o, K})$. 
    Thus the variational method simultaneously find the approximate values of the evidence and $P(\bm{Z}|\bf{o, K})$, the probability distribution of model parameters and hidden variables of each model for given data.
    In the light of DCMM, $P( \bm{o, Z} | \bm{K} )$ is written as
    \begin{equation} \label{eq: P_factor}
    \begin{aligned}
    P( &\bm{o, Z} | \bm{K}) = P( \bm{o, x, \pi, A, B} | \bm{K}) \\
    &=P( \bm{o, x | \pi, A, B}) P(\bm{\pi, A, B} | \bm{K})  \\
    &=P( \bm{o, x | \pi, A, B}) P( \bm{\pi}|\bm{K})P( \bm{A} | \bm{K}) P(\bm{B}|\bm{K}) 
    \end{aligned}
    \renewcommand{\theequation}{S\arabic{equation}} \end{equation}
    Dirichlet distributions are used for prior distributions $P( \bm \pi | \bm K )$, $P( \bm A | \bm K )$, and $P( \bm B | \bm K )$ to render $q(\bm Z)$ into the same type of function (Dirichlet distribution) as well \cite{vbhmm}.
    \begin{equation} \label{eq: prior_pi}
    \begin{aligned}
    P( \bm{\pi}|\bm{K}) & = Dir( \pi_1, \pi_2, ..., \pi_K | u^{\pi}_1, u^{\pi}_2, ..., u^{\pi}_K) \\
    &= \frac{\Gamma( u^{\pi}_0)}{\prod_{\mu=1}^{K} \Gamma(u^{\pi}_\mu)}\prod_{\mu=1}^{K}\pi_\mu^{u ^{\pi}_\mu - 1} \\
    \end{aligned}
    \renewcommand{\theequation}{S\arabic{equation}} \end{equation}
    where $u^{\pi}_{\mu}$ ($\mu \ge 1$) refers to a parameter of Dirichlet distribution with $u^{\pi}_0 = \sum_{\mu=1}^{K} u^{\pi}_\mu$, $\sum_{\mu=1}^{K}\pi_\mu = 1$, and $\Gamma(\cdot)$ denotes the gamma function. 
    The superscript $\pi$ in $u_{\mu}^{\pi}$ implies that $u_{\mu}^{\pi}$ is the parameter involving the probability $\pi$.  
 \begin{widetext}
    \begin{equation} \label{eq: prior_A}
    \begin{aligned}
    P( \bm{A}|\bm{K}) & = \prod_{\mu=1}^{K} Dir( A_{\mu,1}, A_{\mu,2}, ..., A_{\mu,K} | u^A_{\mu,1}, u^A_{\mu,2}, ..., u^A_{\mu,K}) \\
    &= \prod_{\mu=1}^{K}\frac{\Gamma(u^A_{\mu,0})}{\prod_{\nu=1}^{K} \Gamma(u^A_{\mu,\nu})}\prod_{\nu=1}^{K}A_{\mu,\nu}^{u^A_{\mu,\nu} - 1} \\
    \end{aligned}
    \renewcommand{\theequation}{S\arabic{equation}} \end{equation}
    where $u^{A}_{\mu,0} = \sum_{\nu=1}^{K} u^{A}_{\mu,\nu}$, $\sum_{\nu=1}^{K}A_{\mu,\nu} = 1$, and $u^{A}_{\mu,\nu}$ ($ \nu \geq 1$) again refers to a parameter of Dirichlet distribution with the superscript $A$ implying that the parameter is involved with the transition matrix $\bm{A}$. 
    
    \begin{equation} \label{eq: prior_B}
    \begin{aligned}
    P( \bm{B}|\bm{K}) & = \prod_{\mu=1}^{K}\prod_{i=1}^{N} Dir( B_{\mu, i,1}, B_{\mu, i,2}, ..., B_{\mu, i, N} | u^B_{\mu,i,1}, u^B_{\mu,i,2}, ..., u^B_{\mu,i,N}) \\
    &= \prod_{\mu=1}^{K}\prod_{i=1}^{N}\frac{\Gamma(u^B_{\mu,i,0})}{\prod_{j=1}^{N} \Gamma(u^B_{\mu,i,j})}\prod_{j=1}^{N}B_{\mu,i,j}^{u^B_{\mu,i,j} - 1} \\
    \end{aligned}
    \renewcommand{\theequation}{S\arabic{equation}} \end{equation}
       where $u^{B}_{\mu,i,0} = \sum_{j=1}^{N} u^{B}_{\mu,i,j} $, $\sum_{j=1}^NB_{\mu,i,j} = 1$, and $u^{B}_{\mu, i, j}$ ($j \geq 1$) again stands for a parameter of Dirichlet distribution involving the transition matrix $\bm{B}^{\mu}$.
        \end{widetext}

    \section{VB-DCMM: Implementation}
    \subsection{Derivations}
A factorized form of $q(\bm Z) = q(\bm \pi) q(\bm A) q(\bm B) q(\bm x)$ was assumed to find an approximate $q(\bm Z)$. 
    As a result, $F[q]$ (Eq.\ref{eq:Fq}) can be expanded in term by term as
    \begin{widetext}
    \begin{equation} \label{eq: Fq_1}
    \begin{aligned}
    F[q] &= \int  q(\bm{Z}) \ \log\left( \frac{P(\bm{o,Z} | \bm{K})}{q(\bm{Z})}\right)  d\bm{Z} \\
    &=\int q(\bm \lambda) q(\bm x)  
			  \bigg(  \log \frac{P( \bm{\pi}|\bm{K})}{q(\bm \pi)} 
					  + \log \frac{P( \bm{A} | \bm{K})}{q(\bm A)}
					  + \log \frac{P(\bm{B}|\bm{K})} {q(\bm B) }+ \log \frac{   P( \bm{o, x} |\bm{ \pi, A, B})  }{   q(\bm x)   } \bigg) d\bm{\lambda} d\bm{x}
    \end{aligned}
    \renewcommand{\theequation}{S\arabic{equation}} \end{equation}  
    By substituting Eq.(\ref{eq:p_ox_dcmm}) to Eq.(\ref{eq: Fq_1}) we obtain
    \begin{equation} \label{eq: F_piABx}
    \begin{aligned}
    F[q] = F[q(\bm \pi)] + F[q(\bm A)] + F[q(\bm B)] + F[q(\bm x)]
    \end{aligned}
    \renewcommand{\theequation}{S\arabic{equation}} \end{equation}    
    where 
    \begin{equation}
    \begin{aligned}
    F[q(\bm \pi)] &= \int q(\bm \lambda) q(\bm x)  \bigg(\log \frac{P( \bm{\pi}|\bm{K})}{q(\bm \pi)} 
								  											+ \log( \pi_{x(1)})  
								  										 \bigg) d\bm{\lambda} d\bm{x},\nonumber\\
    F[q(\bm A)] &= \int q(\bm \lambda) q(\bm x)  \bigg(
																		    \log \frac{P( \bm{A} | \bm{K})}{q(\bm A)}
																		   + \sum_{t=1}^{T-2}\log( A_{x(t), x(t+1) } )  
																		\bigg) d\bm{\lambda} d\bm{x},\nonumber\\
    F[q(\bm B)] &= \int q(\bm \lambda) q(\bm x)  \bigg(
																			    \log \frac{P( \bm{B} | \bm{K})}{q(\bm B)}
																			    + \sum_{t=1}^{T-1}\log( B_{x(t), o(t), o(t+1) } )  
																	   \bigg) d\bm{\lambda} d\bm{x},\nonumber\\
    F[q(\bm x)] &= -\int q(\bm x) 
											   \log ( q(\bm x) )
											    d\bm{x}.\nonumber
  \renewcommand{\theequation}{S\arabic{equation}} 
  \end{aligned}
  \end{equation}
     \end{widetext}    
     
    \subsubsection{Updating $q(\bm {\lambda})$ ($ q(\bm {\pi} )$, $q(\bm {A} )$, and $ q(\bm {B} ) $).}
    We first set $q(\bm x) = P( \bm{x} | \bm{ o, \lambda}' )$ with given initial values of $\bm{\lambda}' = $ ($\bm{\pi}'$, $\bm A'$, and $\bm B'$). 
    Substitution of Eqs.(\ref{eq: prior_pi}),(\ref{eq: prior_A}),(\ref{eq: prior_B}) to $F[q]$ (Eq.(\ref{eq: F_piABx})) and integration over $\bm x$ lead to 
\begin{widetext}
    \begin{equation}   
    \begin{aligned}
     F[q(\bm \pi)]&= \int q(\bm \pi) q(\bm A) q(\bm B) q(\bm x)  \bigg(
																			      \log \frac{P( \bm{\pi}|\bm{K})}{q(\bm \pi)} 
																			      + \log( \pi_{x(1)})  
																	      \bigg) d\bm{\pi} d\bm{A} d\bm{B} d\bm{x} \\
						&= \int q(\bm \pi) q(\bm x)  \bigg(
																					\log \frac{P( \bm{\pi}|\bm{K})}{q(\bm \pi)} 
																			 		+ \log( \pi_{x(1)})  
												 					  		\bigg) d\bm{\pi} d\bm{x}	\\										      
						&= \int q(\bm \pi) \bigg(  
															\log \frac{   \prod_{\mu=1}^{K}\pi_\mu^{u ^{\pi}_\mu - 1}   }{q(\bm \pi)} 
															+ \sum_{\mu=1}^{K} P(x(1)=\mu | \bm{o, \lambda'}) \log( \pi_{\mu} )
													\bigg) d\bm{\pi} + const.\\
						&= \int q(\bm \pi) \bigg(  
															\log \frac{   \prod_{\mu=1}^{K}\pi_\mu^{u ^{\pi}_\mu - 1} 	}{q(\bm \pi)} 
																+ \sum_{\mu=1}^{K}  \log( \pi_{\mu}^{  P(x(1)=\mu | \bm{o, \lambda'})  } )
													\bigg) d\bm{\pi} + const.\\
						&= \int q(\bm \pi) \bigg(  
															\log \frac{   \prod_{\mu=1}^{K}\pi_\mu^{u ^{\pi}_\mu + P(x(1)=\mu | \bm{o, \lambda'})-1} }{q(\bm \pi)} 
													\bigg) d\bm{\pi} + const. \\						
    \end{aligned}
    \renewcommand{\theequation}{S\arabic{equation}} 
    \end{equation}

    To derive the equations above, we first use $\int q( \bm A ) d\bm{A} = \int q( \bm B ) d\bm{B} = 1$, and then replace the $ \int q(\bm{x}) \log{\pi_{x(1)}}d\bm{x}$ with $\sum_{x(1), x(2), ..., x(T-1)} P( x(1), x(2), ..., x(T-1) | \bm{o}, \bm{ \lambda' } ) \log{\pi_{x(1)}}$ in the second to the third line.  
    The normalization factor of $P( \bm{\pi}|\bm{K} )$ (Eq.(\ref{eq: prior_pi})) is added as a constant term. Finally, by changing the sum of $\log$ to multiplication of its arguments and combine all the integrands together, the final result is obtained.
    By a similar procedure, $F[q(\bm A)]$ and $F[q(\bm B)]$ can be written as 

    \begin{equation} \label{eq:F_A2}
    \begin{aligned}
        F[q(\bm A)] &=  \int q(\bm \pi) q(\bm A) q(\bm B) q(\bm x) 
																			        \log \frac{	P( \bm{A} | \bm{K})	  }{	q(\bm A)	}
																			        + \sum_{t=1}^{T-2}\log( A_{x(t), x(t+1) } )  
										 							        \bigg) d\bm{\pi} d\bm{A} d\bm{B} d\bm{x} \\
    					   &= \int q(\bm A)  \bigg(
																	   \log \frac{	\prod_{\mu, \nu=1}^{K} A_{\mu, \nu}^{u^A_{\mu, \nu} - 1}  }{	q(\bm A)	}+ \sum_{\mu, \nu=1}^{K} \sum_{t=1}^{T-2} P( x(t) = \mu, x(t+1) = \nu | \bm{o, \lambda'}  )~ \log( A_{\mu, \nu} )  
												   \bigg) d\bm{ A }  + const. \\
					   &= \int q(\bm A)  \bigg(	
																	   \log \frac{	\prod_{\mu, \nu=1}^{K} A_{\mu, \nu}^{u^A_{\mu, \nu} + \sum_{t=1}^{T-2} P(x(t)=\mu, x(t+1)=\nu | \bm{o, \lambda'}) - 1}  }{	q(\bm A)	}
												   	\bigg) d\bm{A}  + const. 
    \end{aligned}
    \renewcommand{\theequation}{S\arabic{equation}} \end{equation}          
    \begin{equation} \label{eq:F_B2}
    \begin{aligned}
        F[q(\bm B)] &=  \int q(\bm \pi) q(\bm A) q(\bm B) q(\bm x)
																			         \log \frac{P( \bm{B} | \bm{K})}{q(\bm B)}
																			        + \sum_{t=1}^{T-1}\log( B_{x(t), o(t), o(t+1) } )  
																	        \bigg) d\bm{\pi} d\bm{A} d\bm{B} d\bm{x} \\
				         &= \int q(\bm B) \bigg(
																	         \log \frac{		\prod_{\mu=1}^{K} \prod_{i, j=1}^{N}  B_{\mu,i,j}^{u^B_{\mu,i,j} - 1} 	}{q(\bm B)} + \sum_{\mu=1}^{K} \sum_{t=1}^{T-1} P( x(t) = \mu | \bm{o, \lambda'}  )~
																	           \log( B_{\mu, o(t), o(t+1) } )  
													              \bigg) d\bm{ B }  + const.\\
						 &= \int q(\bm B)  \bigg(
																	        \log \frac{		\prod_{\mu=1}^{K} \prod_{i, j=1}^{N}  B_{\mu,i,j}^{u^B_{\mu,i,j} - 1} 	}{q(\bm B)} + \sum_{\mu=1}^{K} \sum_{i,j=1}^{N} \sum_{\substack{t=1 \\o(t)=i, o(t+1)=j} }^{T-1} P( x(t) = \mu | \bm{o, \lambda'}  )~ \log( B_{\mu, o(t), o(t+1) } )  
															        \bigg) d\bm{ B } + const.\\								
 						 &= \int q(\bm B)  \bigg(
									        						 \log \frac{		\prod_{\mu=1}^{K} \prod_{i, j=1}^{N}  B_{\mu,i,j}^{u^B_{\mu,i,j}+\sum_{\substack{t=1, o(t)=i, o(t+1)=j} }^{T-1} P( x(t) = \mu | \bm{o, \lambda'} ) - 1} 	}{q(\bm B)}
									        	   \bigg) d\bm{ B } + const.										              					        
    \end{aligned}
    \renewcommand{\theequation}{S\arabic{equation}} \end{equation}    
    After combining the above three equations together we get
    \begin{equation}
    \begin{aligned}
    F[q] = & \int q(\bm \pi) \log\bigg(
													   \frac{  \prod_{\mu=1}^{K} \pi_\mu^{W_\mu^\pi - 1 }  } {   q(\bm \pi)   } 
											  \bigg) d\bm{\pi} \\
							   			   &+ \int q(\bm A) \log\bigg(
																					  \frac{  \prod_{\mu=1,\nu=1}^{K} A_{\mu,\nu}^{W_{\mu,l}^A - 1 }  } {   q(\bm A)   } 
																					  \bigg) d\bm{A} \\
								         &+ \int q(\bm B) \log\bigg(
																		  \frac{  \prod_{\mu=1}^{K} \prod_{i=1,j=1}^{N} (B_{\mu, i, j})^{W_{\mu,i,j}^B - 1 }  } {   q(\bm B)   } 
								`										  \bigg) d\bm{B}\\
						   				 & + const. \\
									 =  & - D_{KL}( q(\bm \pi) || Dir( \pi_1, \pi_2, ..., \pi_K | W^{\pi}_1, W^{\pi}_2, ..., W^{\pi}_K) )\\
									   & - D_{KL}( q(\bm A) || \prod_{\mu=1}^{K} Dir( A_{\mu,1}, A_{\mu,2}, ..., A_{\mu,K} | W^A_{\mu,1}, W^A_{\mu,2}, ..., W^A_{\mu,K}) )\\
									   & - D_{KL}( q(\bm B) ||  \prod_{\mu=1}^{K}\prod_{i=1}^{N} Dir( B_{\mu,i,1}, B_{\mu,i,2},  ..., B_{\mu,i,L} | W^B_{\mu,i,1}, W^B_{\mu,i,2}, ..., W^B_{\mu,i,L}) ) \\
													 & + const.
    \end{aligned}
    \renewcommand{\theequation}{S\arabic{equation}} \end{equation}  
    \end{widetext}   
    where 
    \begin{equation}
    \begin{aligned}
    W^\pi_\mu &= u^\pi_\mu + P(x(1)=\mu | \bm{o, \lambda'} ) \textnormal{,}\nonumber\\
    W^A_{\mu, \nu} &= u^A_{\mu,\nu} + \sum_{t=1}^{T-2}P(x(t)=\mu, x(t+1)=\nu | \bm{o, \lambda'}) \textnormal{,}\nonumber\\
    W^B_{\mu.i,j} &= u^B_{\mu,i,j} + \sum_{\substack{  t=1 \\ o(t)=i, o(t+1)=j } }^{T-1}  P(x(t)=\mu | \bm{o, \lambda'}) \textnormal{.}\nonumber
    \renewcommand{\theequation}{S\arabic{equation}} 
    \end{aligned}
    \end{equation}
    Now by setting $q(\bm \pi), q(\bm A)$ and $q(\bm B)$ equal to Dirichlet distributions with new parameter $W$, we can increase $F[q]$ as $ -D_{KL}( \cdot ) \leq 0 $. $ P(x(1)=\mu | \bm{o, \lambda'} )$ and $P(x(t)=\mu, x(t+1)=\nu | \bm{o, \lambda'})$ can be calculated efficiently by using Forward-Backward algorithm \cite{DCMM}.

    \subsubsection{Updating $q(\bm x)$.} 
    Now we integrate $F[q]$ over $\bm \pi$, $\bm A $, and $\bm B$ with fixed (and updated) $q(\bm \pi)$, $q(\bm A)$, and $ q(\bm B)$ to optimize $q(\bm x)$. From Eq.(\ref{eq: Fq_1}), $F[q(\bm \pi)]$ can be written as 
    \begin{widetext}
    \begin{equation}  \label{eq: F_pi3}
    \begin{aligned}
    F[q(\bm \pi)]&= \int q(\bm \pi) q(\bm A) q(\bm B) q(\bm x)
      \bigg(
		    \log \frac{P( \bm{\pi}|\bm{K})}{q(\bm \pi)} 
		    + \log( \pi_{x(1)})  
    \bigg) d\bm{\pi}d\bm{A}d\bm{B} d\bm{x} \\
    &= \int q(\bm \pi) q(\bm x)  \bigg(
		    \log\frac{P( \bm{\pi}|\bm{K})}{q(\bm \pi)} 
		    + \log( \pi_{x(1)})  
    \bigg) d\bm{\pi} d\bm{x}	\\										      
 &= \int q(\bm \pi) q(\bm x)  \bigg(
	 	   \log( \pi_{x(1)})  
	 \bigg) d\bm{\pi} d\bm{x} + const.	\\			
    \end{aligned}
    \renewcommand{\theequation}{S\arabic{equation}} \end{equation}
    \end{widetext}
    We first use $\int q( \bm A ) d\bm{A} = \int q( \bm B ) d\bm{B} = 1$ as the integrand does not depend on $\bm A$ and $\bm B$.
    As $\log \frac{ P(   \bm{\pi}  | \bm{K}  )  }{q( \bm \pi)} $ does not depend on $\bm x$, the result of integration of this term can be written as a constant ($const.$).
    By similar procedure, $F[q(\bm A)]$ and $F[q(\bm B)]$ are written as 
    \begin{widetext}
    \begin{equation} \label{eq:F_A3}
    \begin{aligned}
    F[q(\bm A)] &= \int \int q(\bm \pi) q(\bm A) q(\bm B) q(\bm x) 
     \bigg(
		    \log \frac{	P( \bm{A} | \bm{K})	  }{	q(\bm A)	}
		    + \sum_{t=1}^{T-2}\log( A_{x(t), x(t+1) } )  
    \bigg)d\bm{\pi}d\bm{A}d\bm{B} d\bm{x} \\
	&= \int q(\bm A) q(\bm x)  
	\bigg(
	 \sum_{t=1}^{T-2}  \log( A_{x(t), x(t+1) } )  
	\bigg) d\bm{ A } d\bm{x} + const.\\
    \end{aligned}
    \renewcommand{\theequation}{S\arabic{equation}} \end{equation}          
    \begin{equation} \label{eq:F_B3}
    \begin{aligned}
    F[q(\bm B)] &= \int q(\bm \pi) q(\bm A) q(\bm B) q(\bm x)  
    \bigg(
	    \log \frac{P( \bm{B} | \bm{K})}{q(\bm B)}
	    + \sum_{t=1}^{T-1}\log( B_{x(t), o(t), o(t+1) } )  
    \bigg) d\bm{\pi}d\bm{A}d\bm{B} d\bm{x} \\
    &= \int q(\bm B) q(\bm x)  
    \bigg(
         \sum_{t=1}^{T-1}\log( B_{x(t), o(t), o(t+1) } )  
    \bigg) d\bm{ B } d\bm{x} + const.								              					        
    \end{aligned}
    \renewcommand{\theequation}{S\arabic{equation}} \end{equation}   
    By combining Eq.(\ref{eq: F_pi3}-\ref{eq:F_B3}), we get
    \begin{equation}
    \begin{aligned}
    F[q] = &\int q(\bm x) \bigg(
	\int q(\bm \pi)  \log( \pi_{x(1)})  d\bm{\pi} + \int q(\bm A) \sum_{t=1}^{T-2}\log(A_{x(t), x(t+1)}) d\bm{A}\\
	&+ \int q(\bm B) \sum_{t=1}^{T-1}\log(B_{x(t), o_{t},o(t+1)}) d\bm{B} - \log(q(\bm x)) \bigg)d\bm{x} \\
			&\hspace{1.5cm}+  const. \\
			= &\int q(\bm x) \log\bigg( \frac{ \pi''_{x(1)} B''_{x(1),o(1),o(2)} \prod_{t=1}^{T-2}A''_{x(t), x(t+1)} B''_{x(t+1),o(t+1),o(t+2)} }
																{ q(\bm x)} 
										\bigg) d\bm{x}+ const.
    \end{aligned}
    \renewcommand{\theequation}{S\arabic{equation}} \end{equation}         
    where    
    \begin{equation}
    \begin{aligned}
    \log{  \pi''_{x(1)}  } &= \int q(\bm \pi) \log( \pi_{x(1)}) d\bm{\pi} = \psi(W^\pi_{x(1)}) - \psi(\sum_{k=1}^{K}W^\pi_k) \textnormal{,}\nonumber\\
    \log{  A''_{x(t), x(t+1)  }  } &= \int q(\bm A) \log( A_{x(t), x(t+1)} ) d\bm{A} = \psi(W^A_{x(t), x(t+1)}) - \psi( \sum_{k=1}^{K} W^A_{x(t), k})\textnormal{,}\nonumber\\
    \log{  B''_{x(t),o(t), o(t+1)}  } &=\int q(\bm B) \log( B_{x(t), o(t), o(t+1)}) d\bm{B} =\psi(W^B_{x(t), o(t), o(t+1)}) - \psi( \sum_{j=1}^{N} W^B_{x(t), o(t), j})\nonumber. 
   \renewcommand{\theequation}{S\arabic{equation}} 
   \end{aligned}
   \end{equation}
    \end{widetext} 
    Here, $\psi(\cdot)$ denotes the digamma function ($\psi(x) = \frac{d}{dx} \log{ \Gamma(x) } $).
    Now that $F[q]$ again has a form of $-D_{KL}(\cdot) + const.$, 
    $F[q]$ can be maximized by minimizing the $D_{KL}(\cdot)$ term, which is achieved by setting 
    \begin{widetext}
    \begin{equation} \label{eq: qstar}
    q''(\bm x) = \frac{ \pi''_{x(1)} B''_{x(1),o(1),o(2)} \prod_{t=1}^{T-2}A''_{x(t), x(t+1)} B''_{x(t+1), o(t+1),o(t+2)} }
							    { P(\bm o | \bm{ \pi}'', \bm{A}'', \bm{B}'') } 
	\renewcommand{\theequation}{S\arabic{equation}} \end{equation}
	\end{widetext}
	Note that, the numerator of the equation above is equal to $P(\bm{o, x} | \bm{\pi}'', \bm{A}'', \bm{B}'')$
  implying $q''(\bm{x}) = P( \bm{x} | \bm{o}, \bm{\pi}'', \bm{A}'', \bm{B}'')$.

 With $q(\bm x)$ and by replacing $\bm{ \lambda}' = (\bm{\pi}', \bm{A}', \bm{B}')$ with $\bm{ \lambda}'' =(\bm{\pi}'',\bm{A}'', \bm{B}'')$, one can further update $q(\bm \pi), q(\bm A)$, and $q(\bm B)$. 
    These procedures are iterated until the value of $F[q]$ converges to a desired precision. 
    
    Finally, the converged $F[q]$ can be calculated by substituting the converged argument $q=q^*$ and parameters $\bm{\pi}^*, \bm{A}^*, \bm{B}^*$ into Eq.(\ref{eq: Fq_1}). 
    \begin{equation} \label{eq: F_final}
    \begin{aligned}
    F[q^*] =& -D_{KL}( Dir(W^{\bm{\pi}*}) || Dir(u^{\bm{\pi}} ) ) \\
			    &- D_{KL}( Dir(W^{\bm{A}*}) || Dir(u^{\bm{A}} ) ) \\
			    &-  D_{KL}( Dir(W^{\bm{B}*}) || Dir(u^{\bm{B}} ) ) \\
			    &+ \log{ \ P(\bm o | \bm{ \pi^*, A^*, B^* })} \\
			    &+ \log{K!}
	\end{aligned}
    \renewcommand{\theequation}{S\arabic{equation}} \end{equation}
    The first three terms, $- D_{KL}(\cdot)$, correspond to penalties against the model complexity. 
    The fourth term corresponds to the likelihood, which generally increases with $K$. 
    The final $\log{K!}$ term is added to account for the symmetry of model \cite{Bishop:2006}.
    $K$ is the number of possible internal states in the model.  
Degeneracy arises from the freedom of permutating the labels. 
For example, if two internal states $x=1,2$ are found from VB-DCMM, a new model with $x=2,4$ and $\bm{B}^{x_{new}=2}$ ($ = \bm{B}^{x=1}$), $\bm{B}^{x_{new}=4}$ ($= \bm{B}^{x=2}$) can also be a possible solution with an equal probability. Thus, overall evidence should be calculated with the sum of all possible cases that can be obtained from the permutation of labels for internal states. Thus a corrected evidence should be multiplied by $K!$, which results in introducing the additional factor $\log {K!}$ to $\log{evidence}$. 
In the analysis of real single molecule data, 
the number of observed internal states $K_{obs}$ is not generally identical to the parameter $K$. 
In this case, the actual number of degenercy in labeling internal states should be $_KC_{K_{obs}}\times K_{obs}!$ instead of $K!$. 
To take this effect into account in calculating evidence function, we modified the original evidence function into the following form: 
\begin{equation} \label{eq: G(K)}
G(K) \equiv F(K) - \log{(K-K_{obs})!}. 
\renewcommand{\theequation}{S\arabic{equation}} \end{equation} 

  According to Eq. (\ref{eq: p_ox_K}), the increase of lower bound of $F[q]$ accompanies the decrease of $D_{KL}( q || p )$. 
 	Thus, it is expected that after multiple iterations,  	
 	$F[q]$ (or $G[q]$) converges to $F[q^*]$ which satisfies $F[q] < F[q^*] \simeq \log{ P(\bm{o} | \bm{K}) }$).
 	This implies that $D_{KL}( q^* || p ) \simeq 0$.
 	From 
	 	\begin{equation*}
 	P(\bm{Z} | \bm{o}, \bm{K}) = P(\bm{\pi} | \bm{K})P(\bm{A} | \bm{K})P(\bm{B} | \bm{K}) P(\bm{x} | \bm{o}, \bm{\pi}, \bm{A}, \bm{B}),
 	\end{equation*}
 	and $q(\bm{Z}) = q(\bm{x}) q(\bm{\pi}) q(\bm{A}) q(\bm{B})$,
	it follows that 
 	\begin{equation}
 	\begin{aligned}
 	D_{KL}(q^* || p )  =&  D_{KL}( q^*(\bm{\pi}) || P( \bm{\pi} | \bm{K})) \\
							 	 &+ D_{KL}( q^*(\bm{A}) || P( \bm{A} | \bm{K}))  \\
							 	 &+D_{KL}( q^*(\bm{B}) || P( \bm{B} | \bm{K})) \\
							 	 &+D_{KL}( q^*(\bm{x}) || P( \bm{x} |\bm{o}, \bm{\pi},\bm{A},\bm{B})).
	\end{aligned}
 	\renewcommand{\theequation}{S\arabic{equation}} \end{equation}
 	Thus, $D_{KL}( q^* || p ) \simeq 0$ implies 
 	$q^*(\bm{\pi}) \simeq P (\bm{\pi}| \bm{K}) $, 
 	$q^*(\bm{A}) \simeq P (\bm{A}| \bm{K}) $, 
 	$q^*(\bm{B}) \simeq P (\bm{B}| \bm{K}) $, and 
 	 \begin{widetext}
 	\begin{equation} \label{eq: KL_qx}
 	D_{KL}( q^*(\bm{x}) || P( \bm{x} |\bm{o}, \bm{\pi},\bm{A},\bm{B}))=\int d\bm{x} d\bm{\pi}d\bm{A}d\bm{B} ~ q^*(\bm{x}) \log{ \left( \frac{ q^*(\bm{x}) }{ P( \bm{x} |\bm{o}, \bm{\pi},\bm{A},\bm{B})}  \right) } \simeq 0.
	\renewcommand{\theequation}{S\arabic{equation}} 
	\end{equation}
 	\end{widetext}
	Eq. \ref{eq: KL_qx} also implies that, 
 	$q^*(\bm{x})=P(\bm{x}  | \bm{o}, \bm{\pi}^* \bm{A}^* \bm{B}^*) \simeq P( \bm{x} |\bm{o}, \bm{\pi},\bm{A},\bm{B}) $.
 	Finally, $\bm{\pi}^*, \bm{A}^*, \bm{B}^*$, which provide us with a set of rate constants (e.g. $\{k^{(\mu)}_{a\rightarrow b}\}$, $\{\gamma^{(\mu)\rightarrow(\nu)}\}$), are interpreted as the estimated model parameters.  

  
  \subsection{Implementation.}  
  \subsubsection{Selection of prior parameters}     	
    The likelihood, $\log{P(\bm o | \bm{ \pi^*, A^*, B^* })}$ in Eq.(\ref{eq: F_final}) generally increases with $K$. Other terms, $- D_{KL}(\cdot)$, are always negative, which imposes a penalty against the model with a higher $K$. 
    As the difference between two Dirichlet distributions vanishes when the posterior value $W$ is equal to the prior parameter $u$ and is minimized when the ratios between the element of $W$ and that of $u$ are identical (for example when $W^A_{i,j}/W^A_{i,k} = u^A_{i,j}/u^A_{i,k}$), Eq.(\ref{eq: F_final}) provides a natural guideline for selecting the prior parameters. 
    We have selected the prior parameters using the following rule. 
    \begin{itemize}
    	\item $u^\pi_\mu = 1$
    	\item $u^A_{\mu,\nu} = 1$ for $\mu \neq \nu$
    	\item $u^A_{\mu,\nu} = $ (transition rate (with $\Delta t=1$) using a visual estimation)$^{-1}$. 
    	\item Perform Hidden Markov Analysis assuming $K=1$ to construct a transition matrix $\bm B^{h}$ of homogeneous Markov process.
    	\item Set $u^{B}_{\mu,i,j} = B^h_{i,j} / \min(\{B^{h}_{i,1}, B^{h}_{i,2}, ..., B^{h}_{i,N}\})$ for all $\mu$.
    \end{itemize}
    For example, when roughly one internal-state transition is observed in the  trace with $T_{obs} / \Delta t$= 2000, we set $u^A_{\mu,\nu} = 1/0.001 = 1000$. 
    If 
    $\bm B^h = 
    \begin{pmatrix}
    0.93 & 0.07 \\
    0.05 & 0.95
    \end{pmatrix}
    $, then
	$\bm (\bm u^B)^\mu = 
	\begin{pmatrix}
	0.93/0.07 & 1 \\
	1    		  & 0.95 / 0.05
	\end{pmatrix}
	=
	\begin{pmatrix}
	13			 & 1 \\
	1    		  & 19
	\end{pmatrix}$ 
	for all $\mu $. The results do not depend critically on the choice of prior parameters as long as they are in a reasonable range (Fig \ref{fig_r_simul_ub}, \ref{fig_r_simul_ua}).
	
	\subsubsection{Avoiding local minima}
	To avoid local minimum, the evidence was calculated 20 times for each model with random initial parameters and the result with a larger evidence was selected. Initial values for transition matrices were generated by using Dirichlet distribution:  $\bm{A}$ with parameters $u_a = 0.3, u_{ad}=200$; $\bm{B}$ with parameters $u_b = 1, u_{ad}=20$. $u_{ad}, u_{bd}$ are used to generate the diagonal elements of transition matrices.
	
	\subsubsection{Computation time}
	Computation time depends on the length of data, the number of models to be tested, and the number of repeat (to avoid local minimum). For example, the analysis of one time trace with $T_{obs} / \Delta t$= 4400, K=1, 2, and 3, and 20 repeats takes $\sim$ 3 min whereas the same test but with $T_{obs} / \Delta t$= 8800 takes $\sim$6 min on Macbook pro 13 (3 GHz intel core i7). 
	Linear dependence of analysis time on $T_{obs}/\Delta t$ is expected because each implementation requires execution of DCMM. The running time scales linearly with the length of data as it involves a similar procedure of parameter estimation as HMM \cite{DCMM}.
	$F$ converged usually after ~10 iterations in our test conditions except the case when poor guess for $u_a, u_{ad}, u_b, u_{bd}$ was used on purpose while testing the algorithm (Fig. \ref{fig_r_simul_ub}, \ref{fig_r_simul_ua}). 
	All the implementations of algorithm and data analysis were conducted by using our custom-code written in python with the following libraries: Matplotlib \cite{Matplotlib}, Numpy \cite{Numpy}, Scipy \cite{Scipy}, IPython \cite{Ipython}, Scikit-learn \cite{scikit-learn} and Cython \cite{Cython}.

 \section{Efficacy of VB-DCMM assessed by the law of large number}
\begin{figure*}[h]
		\centering \renewcommand{\thefigure}{S\arabic{figure}} 
		\includegraphics[scale=0.5]{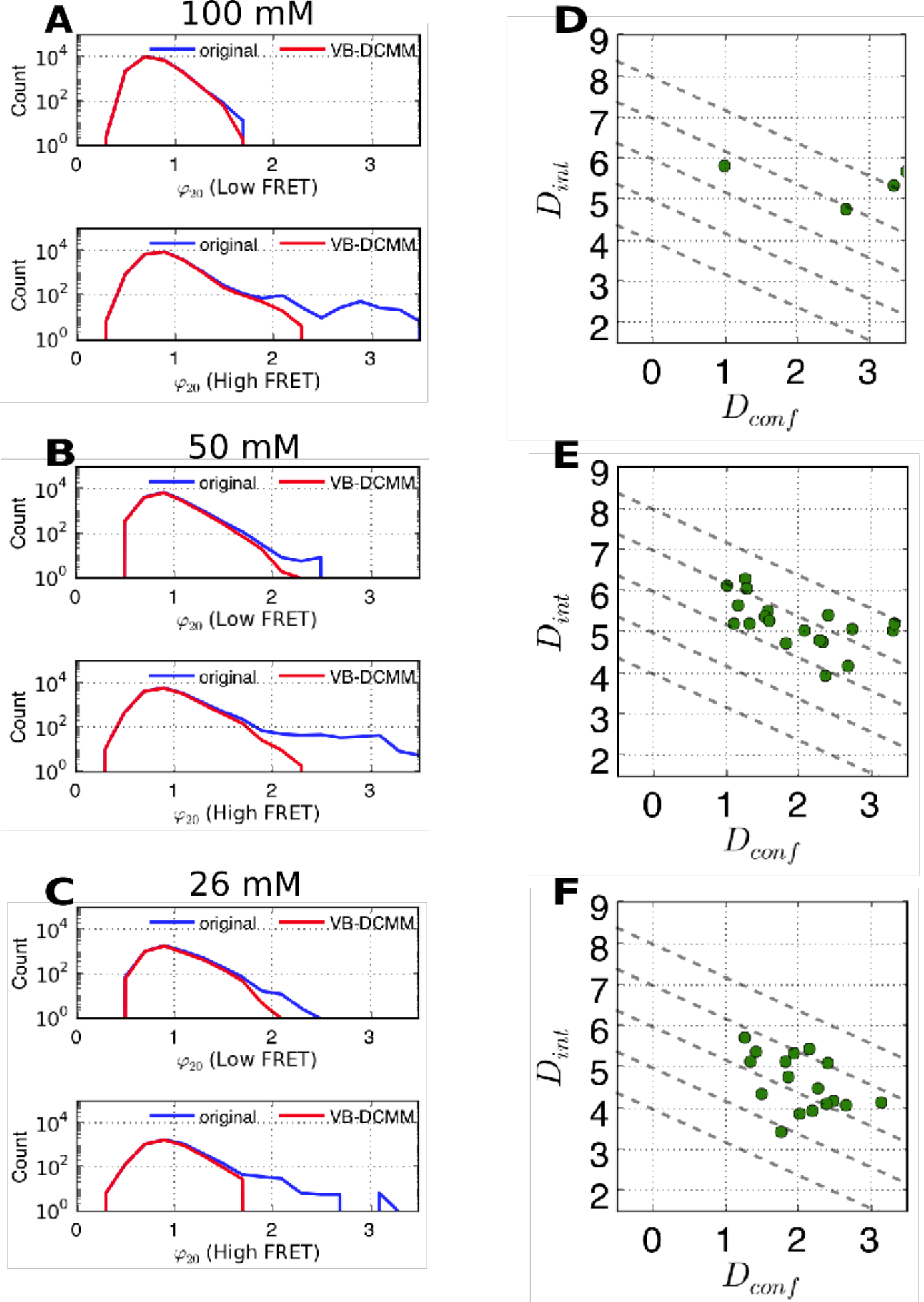}
		
		\caption{
			Effect of decomposing the original H-DNA time traces into its homogeneous Markov components. 
			(A) (Top): Comparison of $\varphi_{20} = \sigma_{20} / \mu_{20}$ histograms from low FRET dwell time data before (blue) and after removing dynamic disorder (red) by VB-DCMM.
			(Bottom): Comparison of $\varphi_{20} = \sigma_{20} / \mu_{20}$ histograms for high FRET dwell time data. [Na$^+$]=100 mM data were used.
			Same analyses for [Na$^+$]=50 mM and [Na$^+$]=26 mM are shown in (B) and (C) respectively.
			(D-F) $D_{\text{conf}}$ and $D_{\text{int}}$ of (D) [Na$^+$]=100 mM data, (E) [Na$^+$]=50 mM data, and (F) [Na$^+$]=26 mM data.
			Each data point denotes the values of $D_{\text{conf}}$ and $D_{\text{int}}$ of individual time traces. Only the time traces exhibiting more than three transition events between internal states are depicted.
		}
		\label{fig_r_HDNA_phi_D}
	\end{figure*}

    To assess the efficacy of VB-DCMM in identifying 
    dynamic disorder (hidden internal state) of a given time trace, we divided an ensemble of heterogeneous time traces into shorter homogeneous traces by using the information of internal states in $x^{\text{model}}(t)$, and calculated the distribution of $\varphi_{20} \equiv \sigma_{20} / \mu_{20}$ of dwell times, where the subscript 20 means that 20 consecutive data of dwell times along the time traces are used in evaluating the standard deviation 
    ($\sigma_{20}^2 =  \frac{1}{20} \sum_{i=1}^{20} ( \tau_i - \mu_{20} )^2$) and the mean ($\mu_{20} = \frac{1}{20} \sum_{i=1}^{20} \tau_{i}$).
    It is expected that $\varphi_{20}=1$ for the time traces generated from a completely homogeneous Markov process; however, $\varphi_{20} > 1 $ when it is evaluated at the boundaries where different internal states coexist. Thus the distribution of $\varphi_{20}$ will be sharply defined as $P(\varphi_{20}) \sim \delta( \varphi_{20} - 1)$ if a heterogeneous trace is correctly decomposed into several pieces of homogeneous traces, so that each piece contains only one internal state.
    Indeed, after the decomposition of original time trace the histogram of $\varphi_{20}$ become narrower and more Gaussian like (Fig. \ref{fig_r_HDNA_phi_D}A--C).
    Test on synthetic data generated using $K=3$ 
    also shows a similar trend (Fig. \ref{fig_r_simul_phi}).
    Next we analyzed H-DNA traces with more than 3 interconversion events between internal states (Fig. \ref{fig_r_HDNA_phi_D}D--F). ($D_{\text{conf}}$, $D_{\text{int}}$) values of these traces are in the region where the synthetic traces displaying $\langle \chi \rangle  \sim 0.9 $.

\begin{figure*}[htbp]
       	\centering \renewcommand{\thefigure}{S\arabic{figure}} 
       	\includegraphics[scale=0.45]{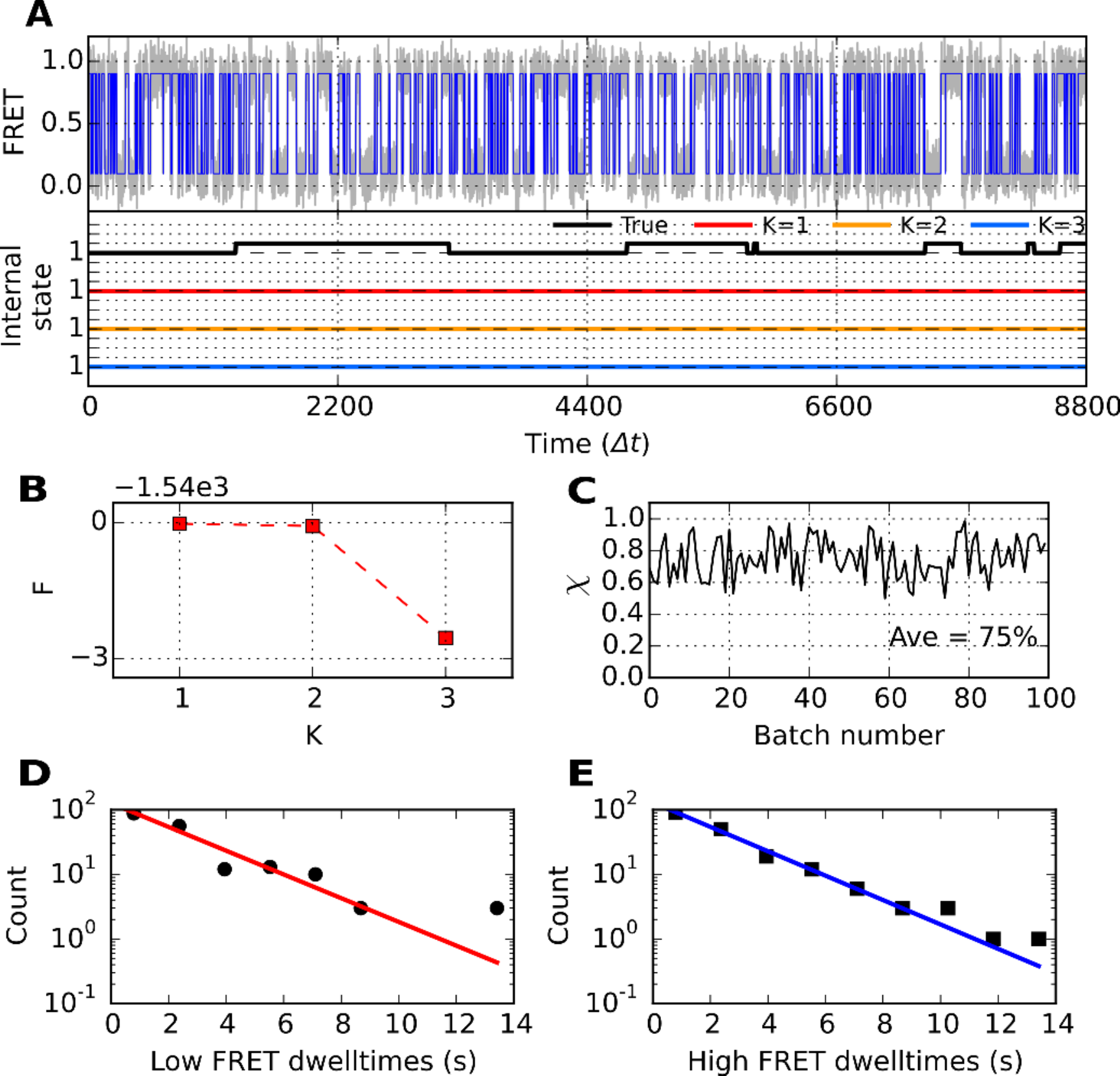}
       	
       	\caption{
       		VB-DCMM analysis on synthetic data generated with the following parameters: $K^{\text{\text{true}}}=2$, $\gamma^{(1) \rightarrow (2)} \Delta t=\gamma^{(2) \rightarrow (1)} \Delta t=0.001, 	k^{(1)}_{L \rightarrow H} \Delta t=k^{(1)}_{H \rightarrow L} \Delta t=0.05, k^{(2)}_{L \rightarrow H} \Delta t =k^{(2)}_{H \rightarrow L}  \Delta t =0.0025$. 
       		(A) (Top) : Gray line is FRET trace and blue line is noise-filtered FRET obtained by using HMM. (Bottom) : True internal state trace (black) and estimated internal state traces by assuming the model with different $K$ (red: $K=1$, orange: $K=2$, blue: $K=3)$.
       		(B) $F(K)$ from the result of VB-DCMM analysis.
       		(C) The accuracy of the model prediction on 100 traces generated under identical condition with the FRET trace shown in (A).
       		(D) Low FRET dwell time histogram and (E) high FRET dwell time histogram obtained from the FRET trace in (A). 
		The solid line denotes a single exponential fit.
	       	}
       	\label{fig_r_simul_nondistinguishable}
	\end{figure*}

 In Markov model, the transition probability from an observable state $a$ to $b$ is estimated as $w_{a \rightarrow b} = k_{a \rightarrow b} ~ \Delta t = n_{a \rightarrow b} / \sum_{b} n_{a \rightarrow b}$ ($n_{a \rightarrow b}$ is the actual number of transitions from $a$ to $b$ observed from a given trace) and the ratio between the standard deviation ($\sigma_{n_{a \rightarrow b} } = \sqrt{\langle (\delta n_{a \rightarrow b})^2} \rangle = \sqrt{ \langle n_{a \rightarrow b} \rangle }~$) and mean ($\mu_{a \rightarrow b} = \langle n_{a \rightarrow b} \rangle $) of the number of transitions $n_{a \rightarrow b}$ satisfies 
    $\varphi_{n_{a \rightarrow b} }=\sigma_{n_{a \rightarrow b} } / \mu_{n_{a \rightarrow b}} = 1/\sqrt{ \langle n_{a \rightarrow b} \rangle } \sim 1/\sqrt{n_{a \rightarrow b}}$. 
    Thus, we expect $\varphi_{k_{a \rightarrow b} } \sim \varphi_{n_{a \rightarrow b}} \sim 1/\sqrt{n_{a \rightarrow b}} \sim 1 /   \sqrt{   \tau_{int} / \tau_{conf}   }$.
    Since $\sim$4 fold difference in $k^{ (\mu) }_{a \rightarrow b}$ and $k^{ (\nu) }_{a \rightarrow b}$ (with $\mu \neq \nu$) is sufficient for the reliable detection of internal states (Fig. \ref{fig_r_test_k_gamma1}A, Fig. \ref{fig_r_simul_nondistinguishable}),
   	VB-DCMM is expected to work for $\varphi_{k_{a \rightarrow b} } \sim \varphi_{n_{a \rightarrow b}} \lesssim 1/4$ which leads to a requirement of time scale separation between $\tau_{int}$ and $\tau_{conf}$ as $\tau_{int} / \tau_{conf} \gtrsim 16$ (or $D_{\text{int}} \gtrsim 4$ (Eq. (\ref{eq: D_int}))). Indeed when all synthetic data were plotted with two metrics $D_{\text{conf}}$ and $D_{\text{int}}$, all the data with $ D_{\text{int}} \gtrsim  4 $  show high $\langle \chi \rangle$ for $D_{\text{conf}} \gtrsim 2$ (Eq. (\ref{eq: D_conf})) (Fig. \ref{fig_r_Dconf_Dint_simul}). 
   	Large $D_{\text{conf}}$ is important for the internal states to be discernible, whereas large $D_{\text{int}}$ is required for accurate estimation of $k$. The performance of algorithm relies on these two factors.

   \section{Other approaches}
   \subsection{Markov Chain Monte Carlo (MCMC) technique}
   As an alternative way of calculating the evidence, Bayesian version of DCMM using MCMC method has previously been developed for credit portfolio modeling \cite{Fitzpatrick:2013}. 
   They, however, used Bayesian inference to calculate posterior distribution of model parameters, instead of selecting a model with optimal number of internal states, and determined the number of internal states based on well-accepted economic cycle fluctuation model. This approach is not applicable when solid knowledge on internal states is not available. Also, they have used constant value for all prior parameters without investigating the effect of prior parameters on the analysis. 
    Furthermore, unlike VB-DCMM, MCMC method does not offer analytical expression for the evidence, 
    which makes it difficult to select prior parameters (or to incorporate prior information).
    
     \subsection{Comparison with Infinite Aggregated Markov Model (iAMM)}
\begin{figure*}[h]
    	\centering \renewcommand{\thefigure}{S\arabic{figure}} 
    	\includegraphics[scale=0.65]{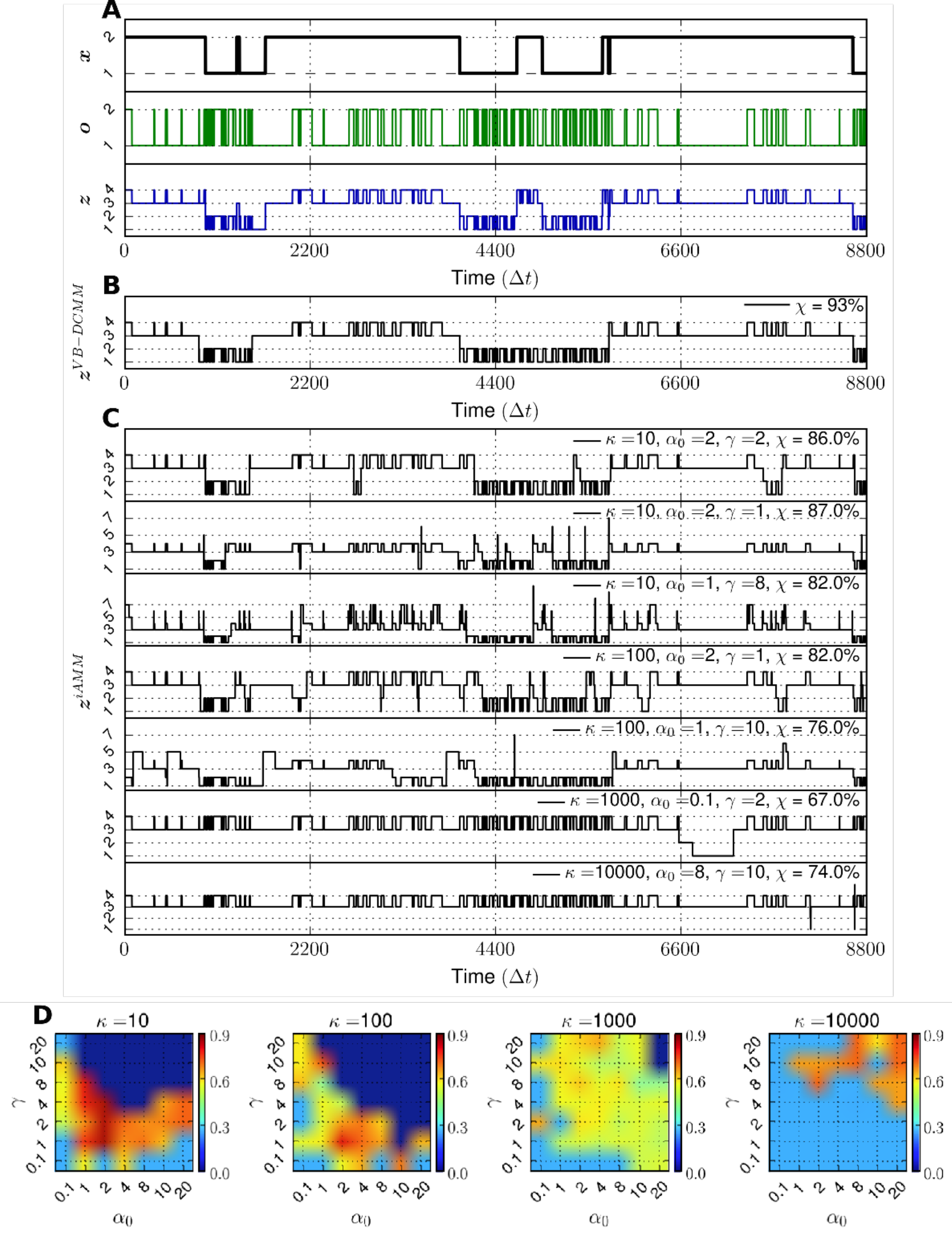}
    	\caption{Comparison between VB-DCMM and sticky-iAMM.
	    	(A) Top, middle: Sequence of internal states ($\bm{x}$) and corresponding observable sequence ($\bm{o}$) from the same synthetic data from Fig. \ref{fig_r_demonstration}A-(i), (ii).
	    	Bottom: Flattened version of $\bm{x}$ and $\bm{o}$ ($\bm{z}$).
	    	(B) Estimated $\bm{z}^{\text{VB-DCMM}}$ using VB-DCMM.
	    	$\chi = \frac{1}{T} \sum_{t=1}^{T} \delta_{z(t), z^{\text{\text{model}}}(t)}=0.93$.
	    	(C) Examples of estimated $\bm{z}^{\text{iAMM}}$ using sticky-iAMM (the code in Ref. \cite{Hines2015_BPJ} was used) after 2000 iterations under various prior parameters ($\kappa$, $\alpha_0$, and $\gamma$).
	    In all the results from iAMM, $\chi$ values are lower than the one obtained from VB-DCMM in (B). 
		   	(D) Result of sticky-iAMM analysis by varying the prior parameters ($\kappa$, $\alpha_0$, $\gamma$) against the time trace ($\bm{o}$) shown in the middle panel in (A).
	    	The values of $\chi$ between $\bm{z}$ and $\bm{z}^{\text{iAMM}}$ under various conditions are color-coded.
	    	When the number of states identified from iAMM is greater than 10, we set $\chi=0$ because in this case the agreement between $\bm{z}^{\text{\text{iAMM}}}$ and $\bm{z}$ is practically very low.     	
		}
    		\label{fig_r_iAMM}
    \end{figure*}

There has been a study developing a method that can detect the presence of hidden states by analyzing the dynamic pattern of  single ion channel data 
using (sticky) infinite aggregated Markov model (iAMM) with nonparametric Bayesian method \cite{Fox:ICML08,Hines2015_BPJ}. 
     VB-DCMM differs from iAMM in several ways and sometimes can be more advantageous: 
     (1) iAMM uses Markov chain Monte Carlo method whereas VB-DCMM employs the variational Bayes method which is computationally less expensive.
     (2) In iAMM, the aggregated Markov model (AMM) is the basic structure in which only a single Markov Chain exists.  
     The model aims to detect distinct transition rates from a  signal layer of signal. 
       In fact, one can map DCMM onto the structure of AMM by \emph{flattening} the two layers of states in DCMM (internal and observable states) into a sequence of one state.
For example, a data structure of DCMM retaining two internal states $X_1, X_2$ and two observables $O_1, O_2$ can be mapped onto four states in AMM as follows: $Z_1 = (X_1, O_1), Z_2 = (X_1, O_2), Z_3 = (X_2, O_1)$, and $Z_4 = (X_2, O_2)$ (Fig. \ref{fig_r_iAMM}A).
     While the transition between two dynamic patterns in DCMM are more strictly regulated, so that the transition rates $k_{Z_1 \rightarrow Z_4}, k_{Z_4 \rightarrow Z_1}, k_{Z_2 \rightarrow Z_3}$, and $k_{Z_3 \rightarrow Z_2}$ are practically zero as the transitions of observables are slaved to the internal state, 
     AMM does not impose such condition. 
     Although AMM could be more flexible in accommodating possible transitions, and could accurately predict the sequence of internal states under very carefully selected the prior parameters (see Fig. \ref{fig_r_iAMM}C), we found that the results obtained from iAMM analysis against our synthetic data was highly sensitive to the prior parameters being selected (Fig.\ref{fig_r_iAMM}D), and that in the most of prior parameters, the traces predicted by iAMM ($z^{\text{iAMM}}(t)$), predicting unwanted frequent transitions between the states, do not match with the synthetic data ($z(t)$), which gives rise to a low $\chi$ value ($\chi=\frac{1}{T}\sum_{t=1}^T\delta_{z(t),z^{\text{iAMM}}(t)}$). 
     
\begin{figure*}[h]
	\centering \renewcommand{\thefigure}{S\arabic{figure}} 
	\includegraphics[scale=0.42]{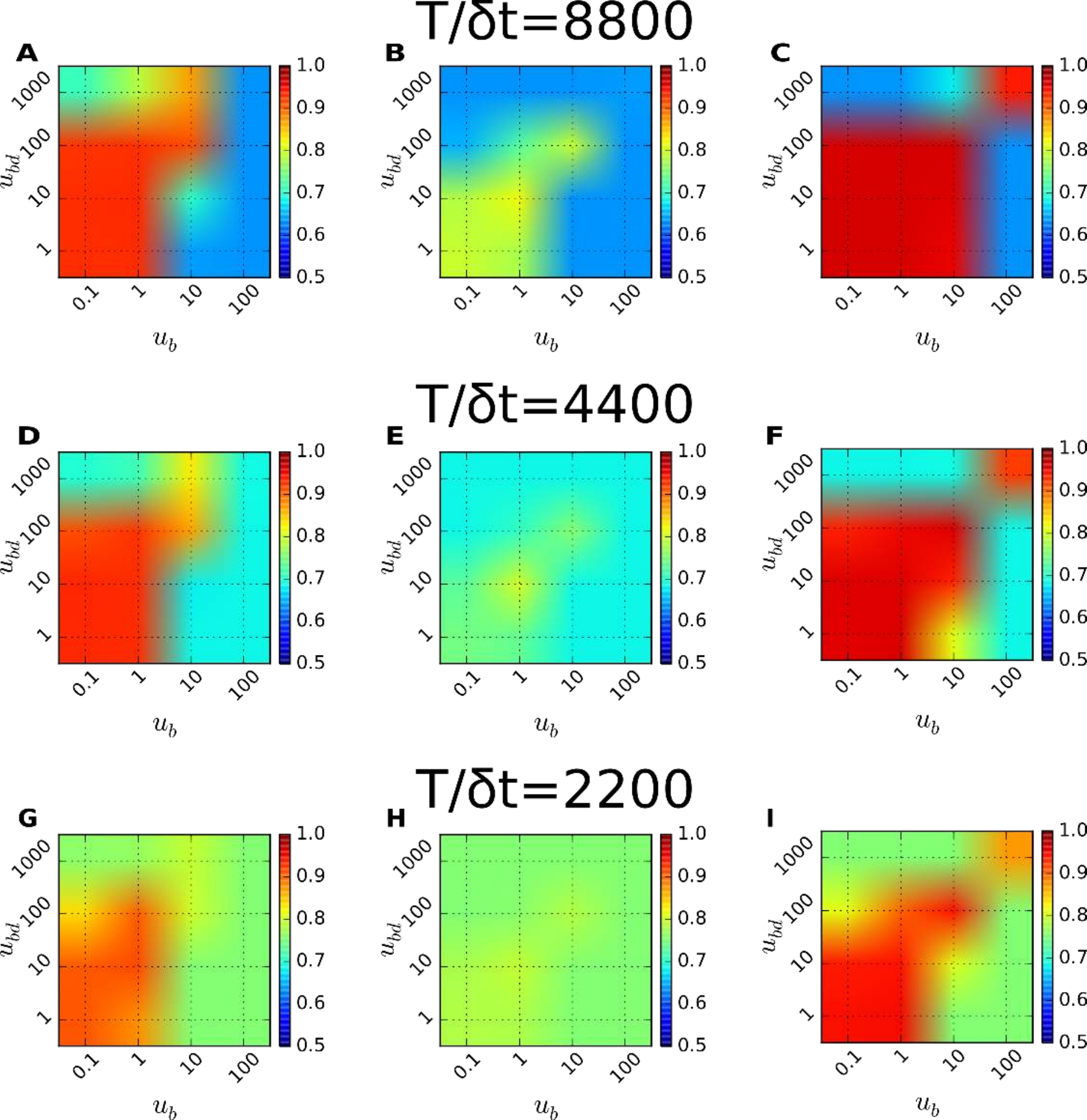}
	
	\caption{
		Accuracy of the model prediction in terms of $\langle \chi \rangle $ under varying prior parameters $u_b$, and $u_{bd}$.
		To calculate the diagram 100 time traces were analyzed at each condition.
		$\langle \chi \rangle$ of each graph was evaluated for the data generated with 
		the fixed parameters $\gamma^{(1) \rightarrow (2)} \Delta t=\gamma^{(2) \rightarrow (2)} \Delta t =0.001, k^{(1)}_{L \rightarrow H} \Delta t =k^{(1)}_{H \rightarrow L} \Delta t = 0.05$, and  
		(A) $k^{(2)}_{L \rightarrow H} \Delta t = 0.00625, k^{(2)}_{H \rightarrow L} \Delta t = 0.0125$, $T_{obs} / \Delta t$= 8800.
		(B) $k^{(2)}_{L \rightarrow H} \Delta t = 0.025, k^{(2)}_{H \rightarrow L} \Delta t = 0.1$, $T_{obs} / \Delta t$= 8800.
		(C) $k^{(2)}_{L \rightarrow H} \Delta t = 0.1, k^{(2)}_{H \rightarrow L} \Delta t = 0.2$, $T_{obs} / \Delta t$= 8800.
		(D-F) Identical conditions with (A-C) except $T_{obs} / \Delta t$= 4400.
		(G-I) Identical conditions with (A-C) except $T_{obs} / \Delta t$= 2200.
    	}
	\label{fig_r_simul_ub}
\end{figure*}

\begin{figure*}[h]
	\centering \renewcommand{\thefigure}{S\arabic{figure}} 
	\includegraphics[scale=0.40]{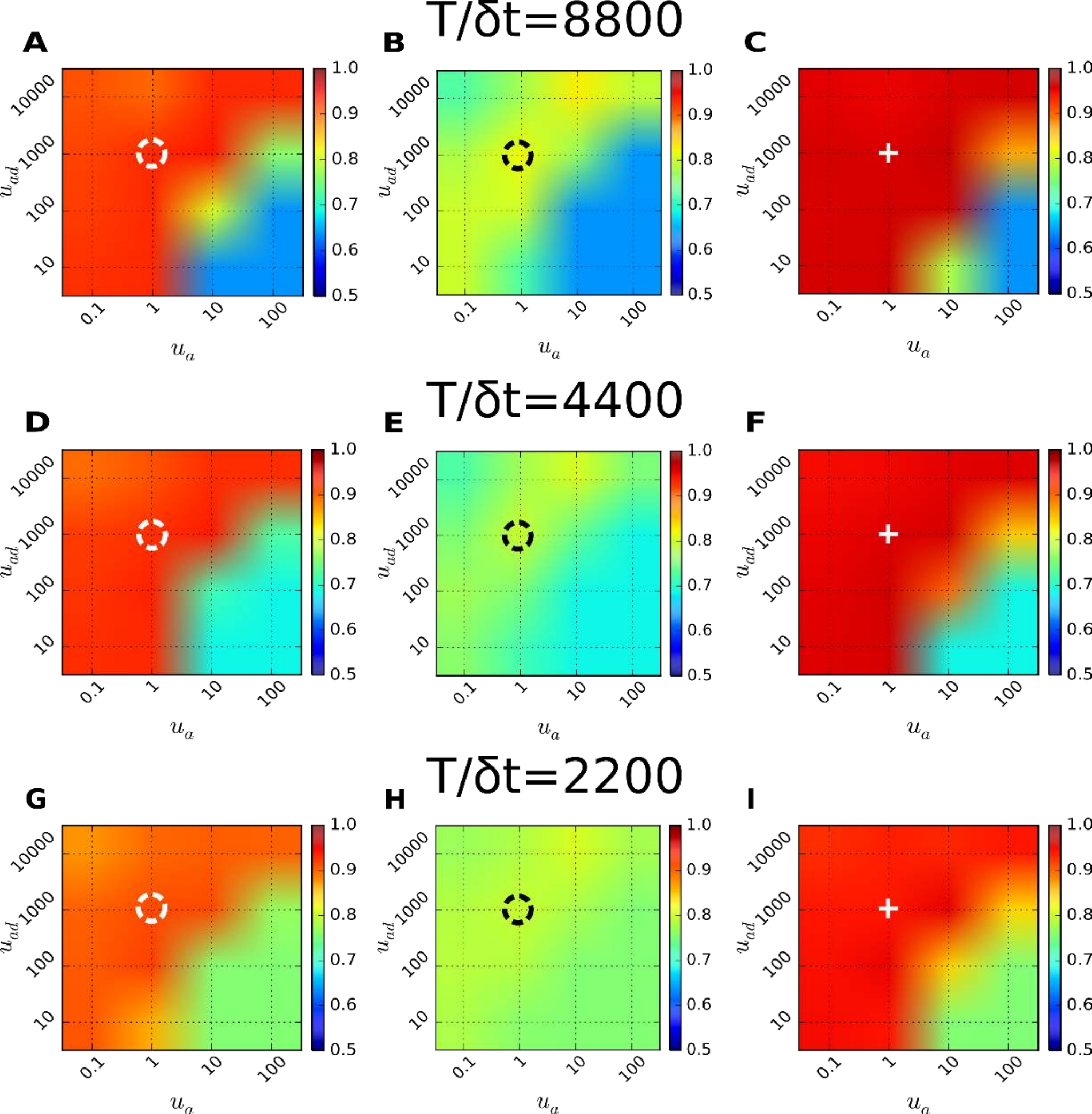}
	
	\caption{
		Accuracy of the model prediction in terms of $\langle \chi \rangle $ under varying prior parameters $u_a$, and $u_{ad}$.
		To calculate the diagram 100 time traces were analyzed at each condition.
		$\langle \chi \rangle$ of each graph was evaluated for the data generated with 
		the fixed parameters 
		$\gamma^{(1) \rightarrow (2)} \Delta t=\gamma^{(2) \rightarrow (2)} \Delta t =0.001, k^{(1)}_{L \rightarrow H} \Delta t =k^{(1)}_{H \rightarrow L} \Delta t = 0.05$, and  
		(A) $k^{(2)}_{L \rightarrow H} \Delta t = 0.00625, k^{(2)}_{H \rightarrow L} \Delta t = 0.0125$, $T_{obs} / \Delta t$= 8800.
		(B) $k^{(2)}_{L \rightarrow H} \Delta t = 0.025, k^{(2)}_{H \rightarrow L} \Delta t = 0.1$, $T_{obs} / \Delta t$= 8800.
		(C) $k^{(2)}_{L \rightarrow H} \Delta t = 0.1, k^{(2)}_{H \rightarrow L} \Delta t = 0.2$, $T_{obs} / \Delta t$= 8800.
		(D-F) Identical conditions with (A-C) except $T_{obs} / \Delta t$= 4400.
		(G-I) Identical conditions with (A-C) except $T_{obs} / \Delta t$= 2200.
		Black and white circles, and white cross in figure indicate the standard choice of $u_a=1$ and $u_{ad}=1000$ used for other analyses.
	}
	\label{fig_r_simul_ua}
\end{figure*}

     The persistent dynamic pattern as shown in H-DNA and preQ$_1$-riboswitch \cite{Rinaldi:2016iw} dynamics can be better modeled with VB-DCMM whose result is not sensitive to the choice of prior parameters (Fig. \ref{fig_r_simul_ub}, \ref{fig_r_simul_ua}).
      \\


	\begin{figure*}[h!]
		\centering \renewcommand{\thefigure}{S\arabic{figure}}  \renewcommand{\thefigure}{S\arabic{figure}} 
		\includegraphics[scale=0.65]{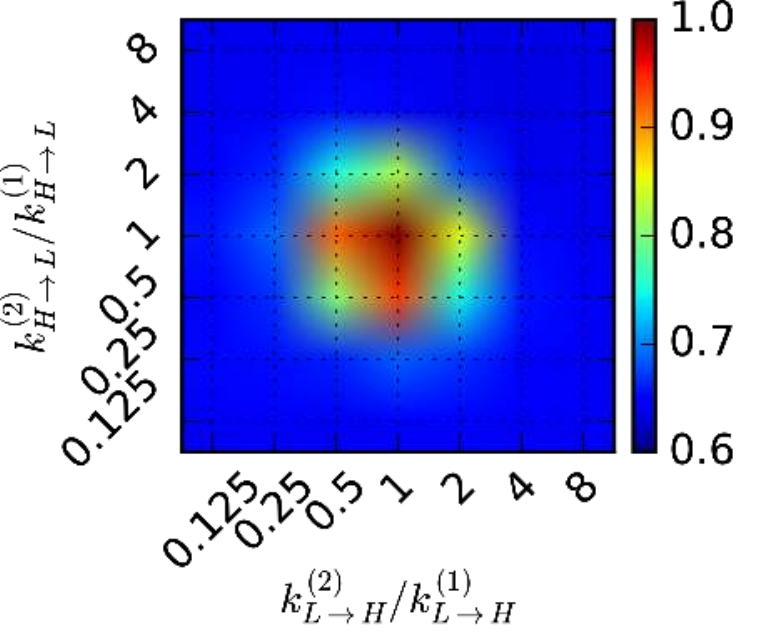}
		\caption{
		Re-calculated accuracy of model prediction ($\langle \chi \rangle$) by assuming ``$K=1$" (i.e., assuming $x(t)^{\text{true}} =1$ for all $t$ in Eq.(\ref{eq: chi})) for the same set of parameters used to calculate Fig. \ref{fig_r_test_k_gamma1}A.
		}
		\label{fig_r_test_k_K1}
	\end{figure*} 
	\begin{figure*}[h!]
		\centering \renewcommand{\thefigure}{S\arabic{figure}} 
		\includegraphics[scale=0.78]{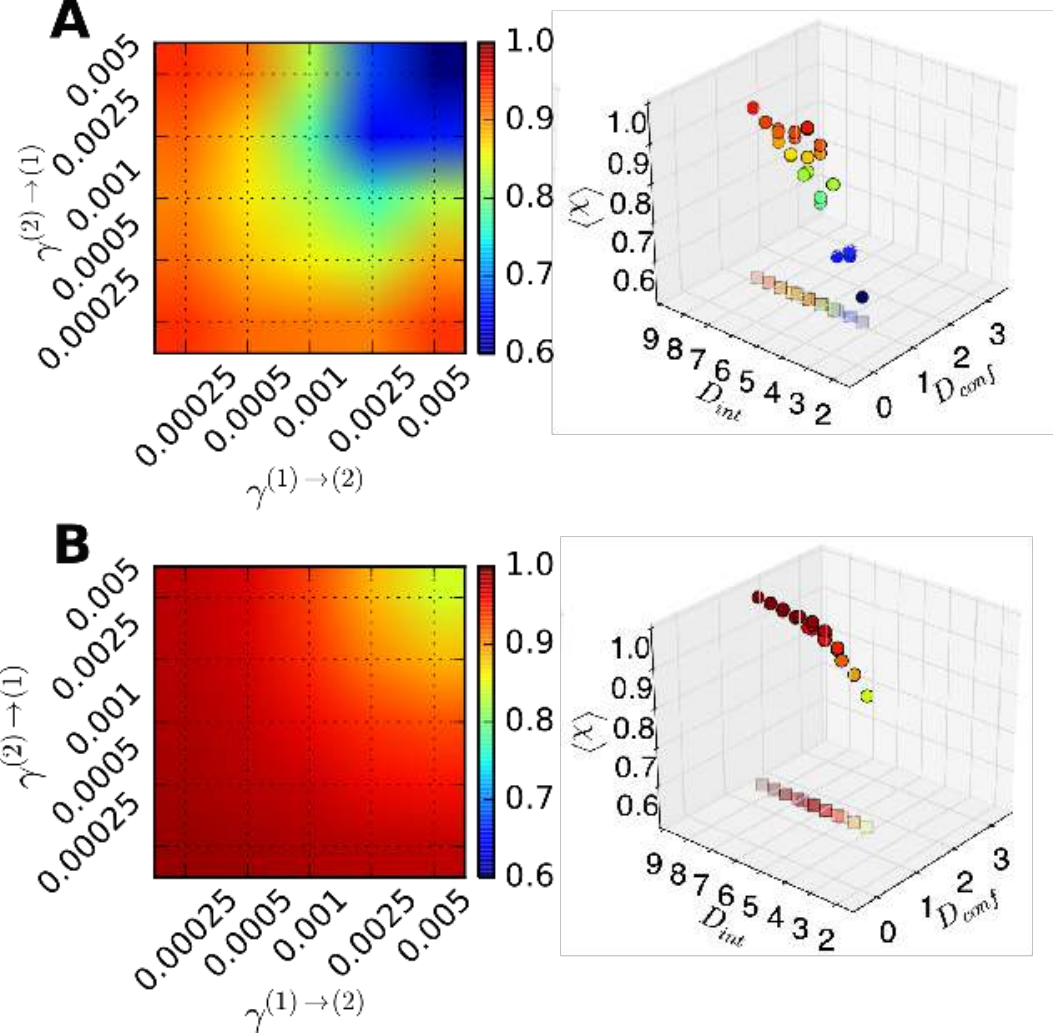}
		\caption{
			Systematic validation of VB-DCMM on synthetic data generated under various conditions with $T_{obs} / \Delta t = 8800$.
			Color code denotes the accuracy of the model prediction in terms of $\langle \chi \rangle$, averaged over 100 traces for each condition. 
			Same analysis were performed with Fig. \ref{fig_r_test_k_gamma1}B under different conditions.
    		(A) $\langle \chi \rangle$ under varying $\gamma^{(1) \rightarrow (2)}$ and $\gamma^{(2) \rightarrow (1)}$ with $K^{\text{\text{true}}}=2$, $k^{(1)}_{L \rightarrow H}  \Delta t = k^{(1)}_{H \rightarrow L}  \Delta t =0.05$, $k^{(2)}_{L \rightarrow H}  \Delta t = 0.025$, $k^{(2)}_{H \rightarrow L}  \Delta t = 0.1$.
    		(B) $\langle \chi \rangle$ under varying $\gamma^{(1) \rightarrow (2)}$ and $\gamma^{(2) \rightarrow (1)}$ with $K^{\text{\text{true}}\text{\text{true}}}=2$, $k^{(1)}_{L \rightarrow H}  \Delta t = k^{(1)}_{H \rightarrow L}  \Delta t =0.05$, $k^{(2)}_{L \rightarrow H}  \Delta t = 0.1$, $k^{(2)}_{H \rightarrow L}  \Delta t = 0.2$.
		}
		\label{fig_r_test_gamma2_gamma3}
	\end{figure*} 
    
	\begin{figure*}[h]
       	\centering \renewcommand{\thefigure}{S\arabic{figure}} 
       	\includegraphics[scale=0.6]{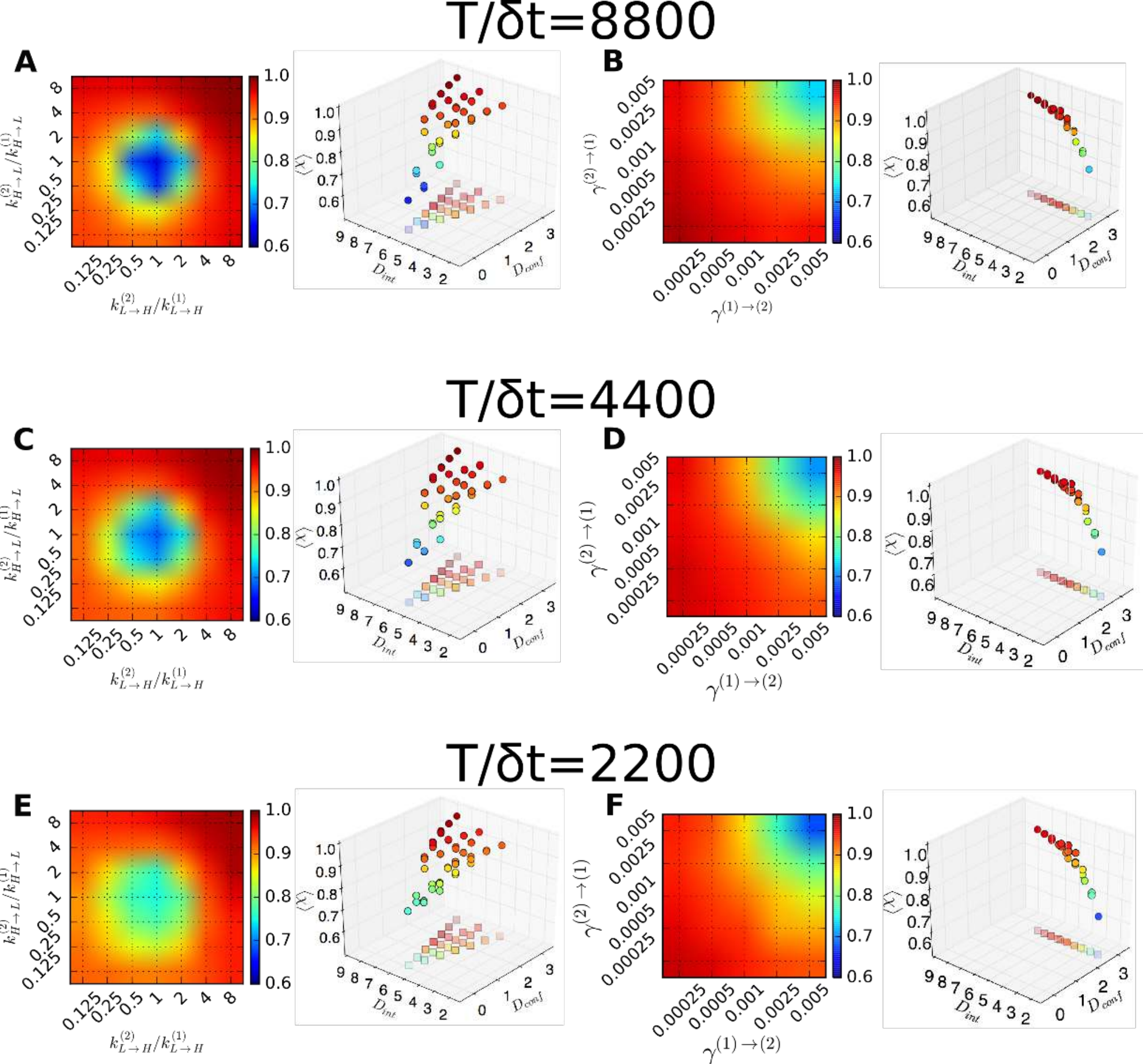}
       	\caption{
       		Accuracy of VB-DCMM on synthetic data with $T_{obs} / \Delta t = 4400$ and  2200.
       		Color code denotes the accuracy of the model prediction in terms of $\langle \chi \rangle$, averaged over 100 traces for each condition, under varying $k^{(2)}_{L \rightarrow H}, k^{(2)}_{H \rightarrow L}$.
       		(A-B) Results with $T_{obs} / \Delta t = 8800$. Same graphs from Fig. \ref{fig_r_test_k_gamma1} are showed again for clarity.
       		(C-D) Results with $T_{obs} / \Delta t = 4400$ and (E-F) Results with $T_{obs} / \Delta t = 2200$. Same analysis with Fig. \ref{fig_r_test_k_gamma1} were performed.
			}
       	\label{fig_r_test_T}
     \end{figure*} 
        
    \begin{figure*}[h!]
       	\centering \renewcommand{\thefigure}{S\arabic{figure}} 
       	\includegraphics[scale=0.65]{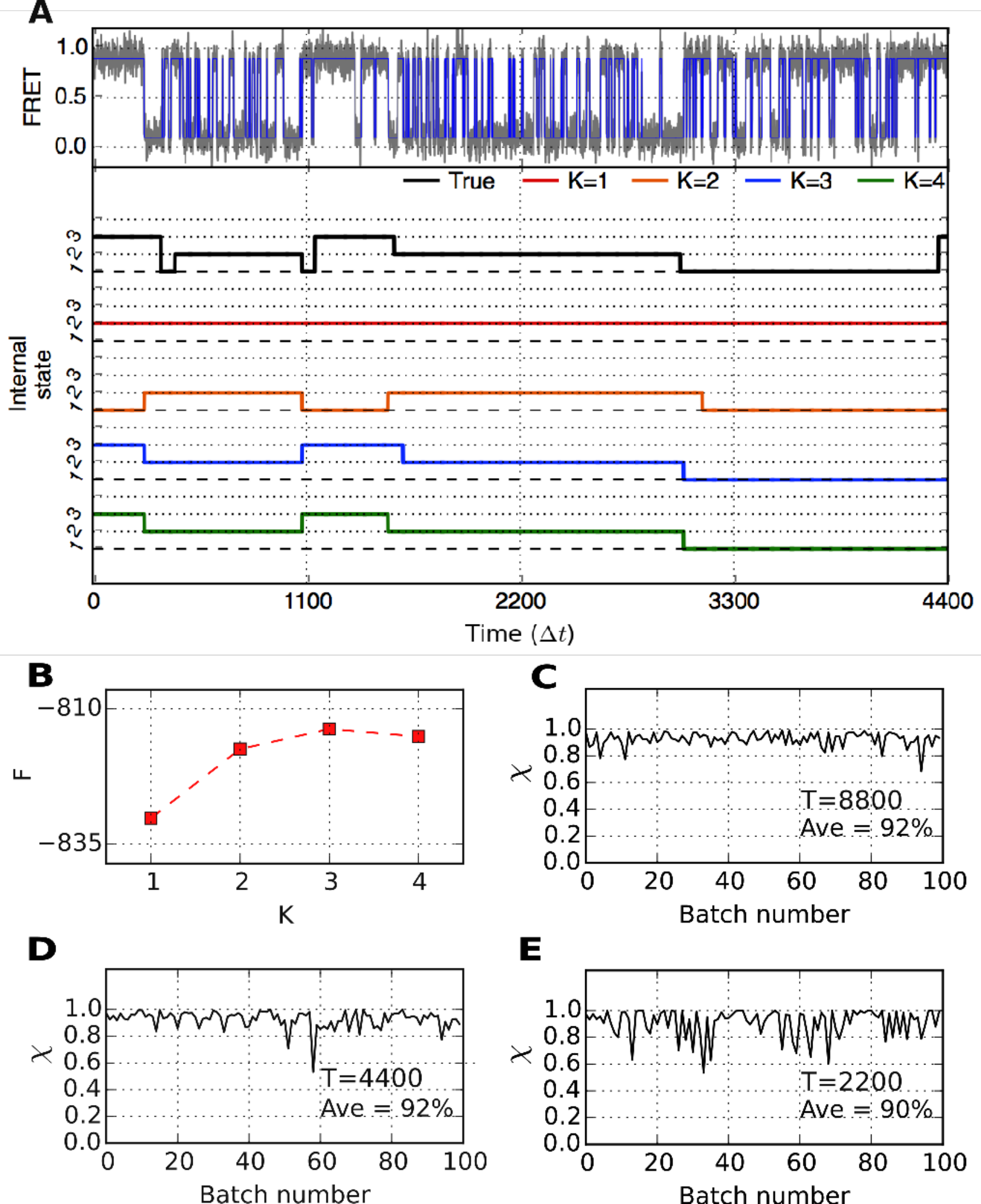}
       	\caption{
       		VB-DCMM analysis on synthetic data having three internal states generated with the following parameters: $K^{\text{\text{true}}}=3$, $\gamma^{(1) \rightarrow (2)} \Delta t=\gamma^{ (1) \rightarrow (3) }\Delta t=\gamma^{ (2) \rightarrow (3) }\Delta t=\gamma^{ (3) \rightarrow (2) }\Delta t=\gamma^{ (3) \rightarrow (1) }\Delta t=\gamma^{ (2) \rightarrow (1) }\Delta t=0.0005, 	k^{(1)}_{L \rightarrow H}\Delta t=0.1, 
       		k^{(1)}_{H \rightarrow L}\Delta t=0.04, 
       		k^{(2)}_{L \rightarrow H}\Delta t=0.04,
       		k^{(2)}_{H \rightarrow L}\Delta t=0.1,
       		k^{(3)}_{L \rightarrow H}\Delta t=0.008,
       		k^{(3)}_{H \rightarrow L}\Delta t=0.012$.
       		(A) (Top) : Gray line indicates FRET trace and blue line is noise-filtered FRET obtained by using HMM. (Bottom) : True internal state trace (Black) and estimated internal state traces (red: $K=1$, orange: $K=2$, blue: $K=3$, green: $K=4$).
       		(B) $F(K)$ from VB-DCMM analysis.
       		(C) $\chi$ on 100 traces with $T_{obs} / \Delta t$= 8800,
       		(D) with $T_{obs} / \Delta t$= 4400, and 
       		(E) with $T_{obs} / \Delta t$= 2200.
       		}
       	\label{fig_r_simul_K3}
    \end{figure*}  
    
    \begin{figure*}[h!]
       	\centering \renewcommand{\thefigure}{S\arabic{figure}} 
       	\includegraphics[scale=0.65]{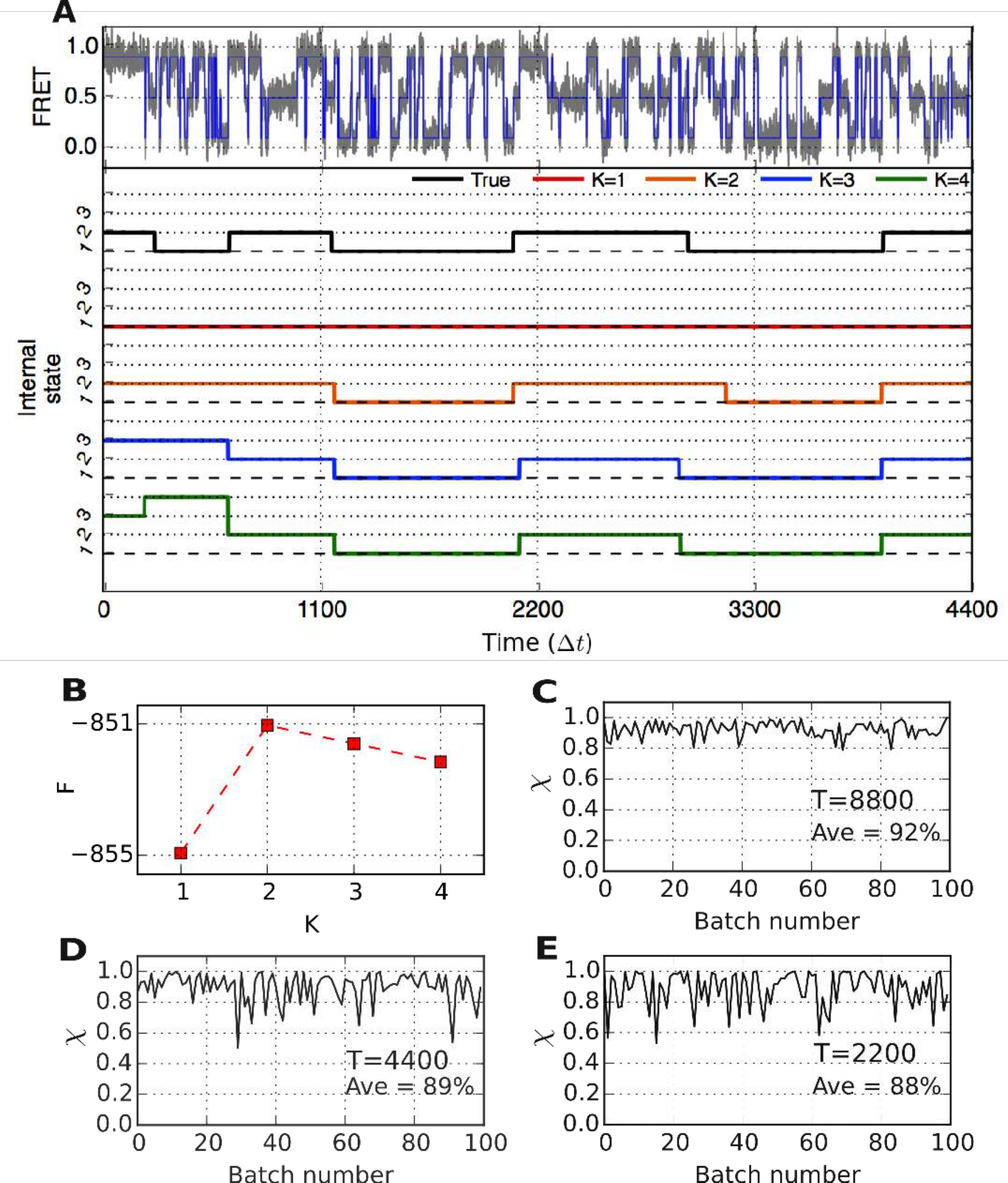}
       	\caption{
       		VB-DCMM analysis on synthetic data having 3 observables generated with following parameters: 
       		$K^{\text{\text{true}}}=2$,  
       		$\gamma^{(1) \rightarrow (2)} \Delta t=\gamma^{2 \rightarrow 2} \Delta t =0.001, 
       		k^{(1)}_{L \rightarrow M}\Delta t=0.015,
       		k^{(1)}_{L \rightarrow H}\Delta t=0.023,
       		k^{(1)}_{M \rightarrow L}\Delta t=0.032,
       		k^{(1)}_{M \rightarrow H}\Delta t=0.05, 
       		k^{(1)}_{H \rightarrow L}\Delta t=0.03,
       		k^{(1)}_{H \rightarrow M}\Delta t=0.014,
       		k^{(2)}_{L \rightarrow M}\Delta t=0.058,
       		k^{(2)}_{L \rightarrow H}\Delta t=0.065,
       		k^{(2)}_{M \rightarrow L}\Delta t=0.021,
       		k^{(2)}_{M \rightarrow H}\Delta t=0.004, 
       		k^{(2)}_{H \rightarrow L}\Delta t=0.0093,
       		k^{(2)}_{H \rightarrow M}\Delta t=0.014$.
       		(A) (Top) : Gray line indicates FRET trace and blue line is noise-filtered FRET obtained by using HMM. (Bottom) : True internal state trace (Black) and estimated internal state traces (red: $K=1$, orange: $K=2$, blue: $K=3$, green: $K=4$).
       		(B) $F(K)$ from VB-DCMM analysis.
       		(C) $\chi$ on 100 traces with $T_{obs} / \Delta t$= 8800,
       		(D) with $T_{obs} / \Delta t$= 4400, and 
       		(E) with $T_{obs} / \Delta t$= 2200.
			}
       	\label{fig_r_simul_L3}
    \end{figure*}   
    
    \begin{figure*}[h!]
    	\centering \renewcommand{\thefigure}{S\arabic{figure}} 
    	\includegraphics[scale=0.65]{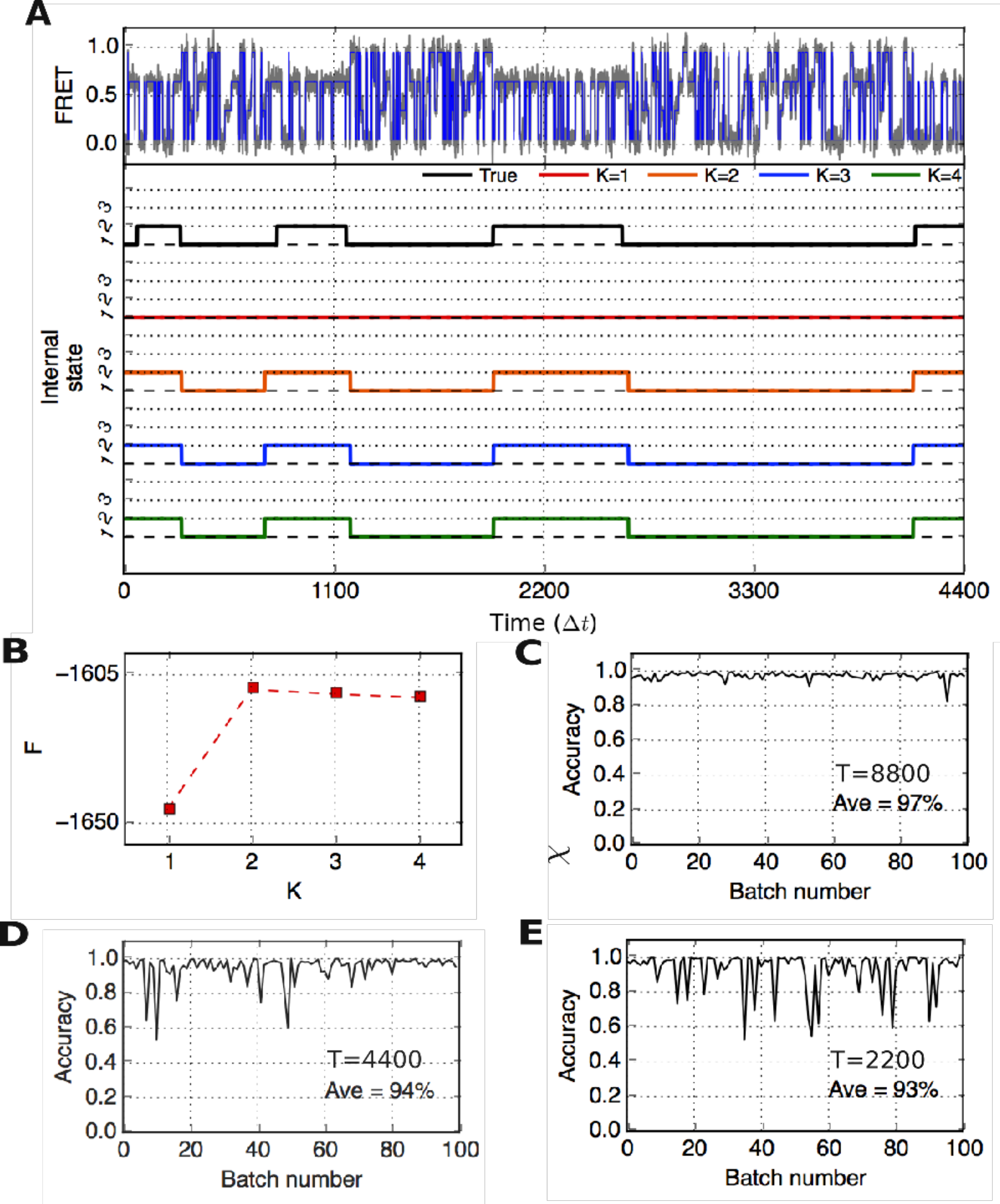}
    	\caption{
    		VB-DCMM analysis on synthetic data having 4 observables ($o$=1, 2, 3, and 4) when internal state $x$=1, or having 2 ($o$=1, 3) observables when $x$=2. Following parameters are used to generate synthetic data: 
    		$K^{\text{\text{true}}}=2$,  
    		$\gamma^{(1) \rightarrow (2)} \Delta t=\gamma^{2 \rightarrow 2} \Delta t =0.001, 
    		k^{(1)}_{1 \rightarrow 2}\Delta t=0.015,
    		k^{(1)}_{1 \rightarrow 3}\Delta t=0.023,
    		k^{(1)}_{1 \rightarrow 4}\Delta t=0.05,	
    		k^{(1)}_{2 \rightarrow 1}\Delta t=0.032,
    		k^{(1)}_{2 \rightarrow 3}\Delta t=0.05, 
  			k^{(1)}_{2 \rightarrow 4}\Delta t=0.01, 
    		k^{(1)}_{3 \rightarrow 1}\Delta t=0.03,
    		k^{(1)}_{3 \rightarrow 2}\Delta t=0.014,
		    k^{(1)}_{3 \rightarrow 4}\Delta t=0.03,
    		k^{(1)}_{4 \rightarrow 1}\Delta t=0.1,
    		k^{(1)}_{4 \rightarrow 2}\Delta t=0.02,
    		k^{(1)}_{4 \rightarrow 3}\Delta t=0.01,
    		k^{(2)}_{1 \rightarrow 3}\Delta t=0.085,
    		k^{(2)}_{3 \rightarrow 1}\Delta t=0.063$.
    		To make only $o$=1, 3 appears when $x=2$, the transition rates $k^{(2)}_{i \rightarrow j}$ set to zero when $i$ or $j$ is 2 or 4.
    		(A) (Top) : Gray line indicates FRET trace and blue line is noise-filtered FRET obtained by using HMM. (Bottom) : True internal state trace (Black) and estimated internal state traces (red: $K=1$, orange: $K=2$, blue: $K=3$, green: $K=4$).
    		(B) $F(K)$ from VB-DCMM analysis.
    		(C) $\chi$ on 100 traces with $T_{obs} / \Delta t$= 8800,
    		(D) with $T_{obs} / \Delta t$= 4400, and 
    		(E) with $T_{obs} / \Delta t$= 2200.}
    	\label{fig_r_simul_L24}
    \end{figure*}   

    \begin{figure*}[h!]
    	\centering \renewcommand{\thefigure}{S\arabic{figure}} 
    	\includegraphics[scale=0.70]{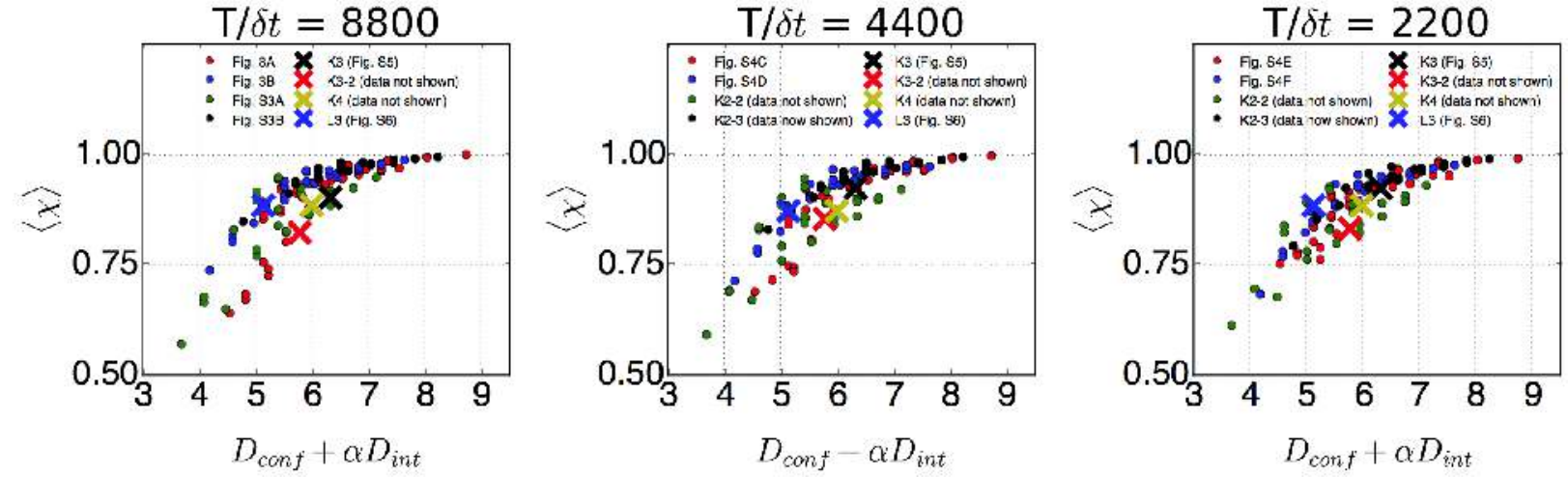}
    	
    	\caption{
    		Average accuracy of the model prediction $\langle \chi\rangle$ versus $D_{\text{tot}}=D_{\text{conf}}+0.8D_{\text{int}}$ from various synthetic data, where results are shown in  Fig. \ref{fig_r_test_k_gamma1}, \ref{fig_r_test_gamma2_gamma3} ($T_{obs} / \Delta t$=8800), Fig. \ref{fig_r_test_T} ($T_{obs} / \Delta t$=4400, 2200).
    	}
       	\label{fig_r_simul_D}
    \end{figure*} 
   
   \begin{figure*}[h]
   	\centering \renewcommand{\thefigure}{S\arabic{figure}} 
   	\includegraphics[scale=0.7]{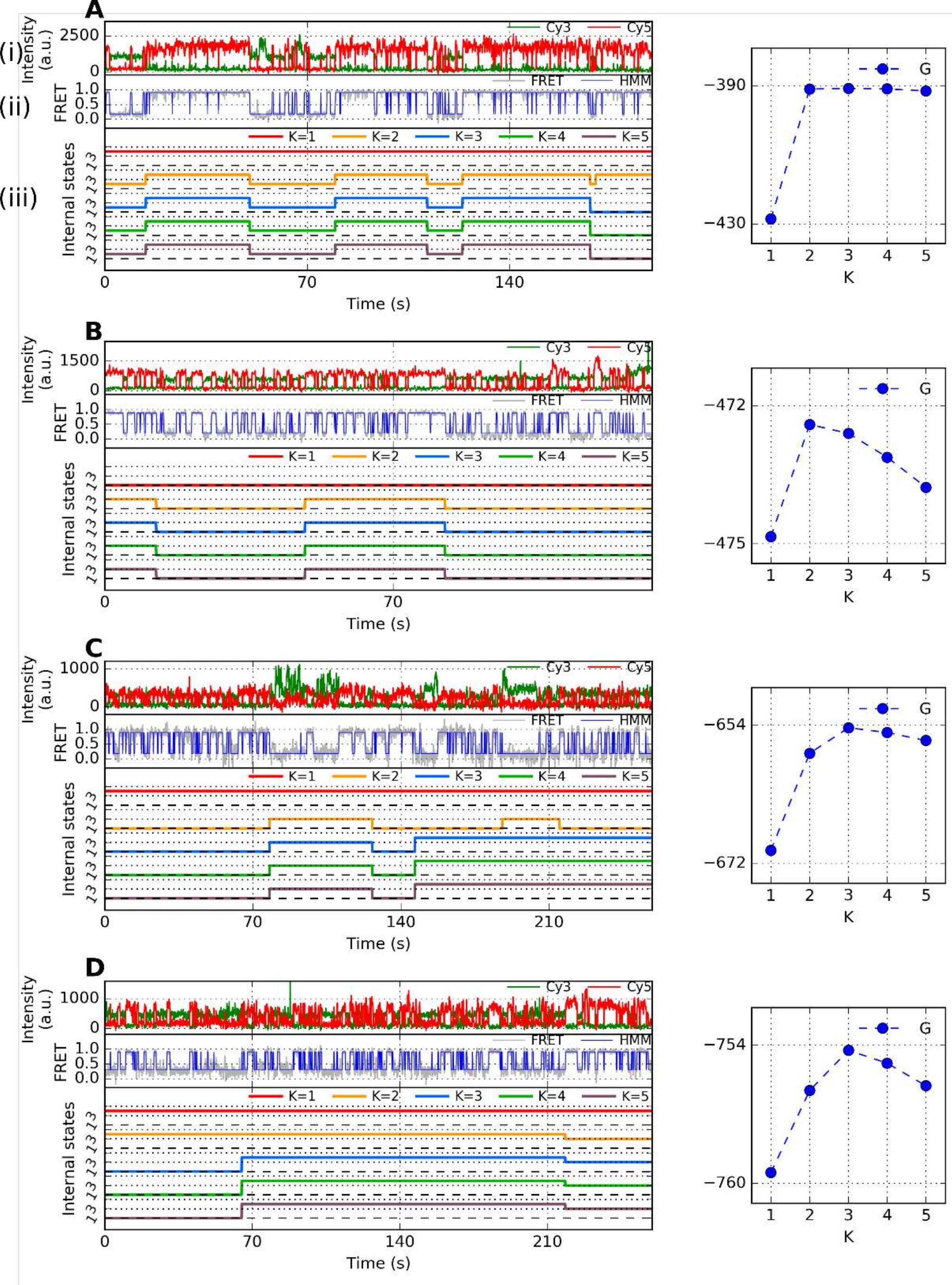}
   	
   	\caption{Representative time traces of H-DNA dynamics and their analysis using VB-DCMM at [Na$^+$]= 50 mM.
   		(A) (i) Representative fluorescence signal and (ii) their FRET state. (iii) Internal states estimated for $K=1, 2, \ldots 5$.
   		Right panel shows $G(K)$ (blue circle).
   		(B, C, D) Other representative time traces and their lower bound obtained under the same experimental condition.
   	}
   	\label{fig_r_traces_50mM}
   \end{figure*}    
   
   \begin{figure*}[h]
   	\centering \renewcommand{\thefigure}{S\arabic{figure}} 
   	\includegraphics[scale=0.7]{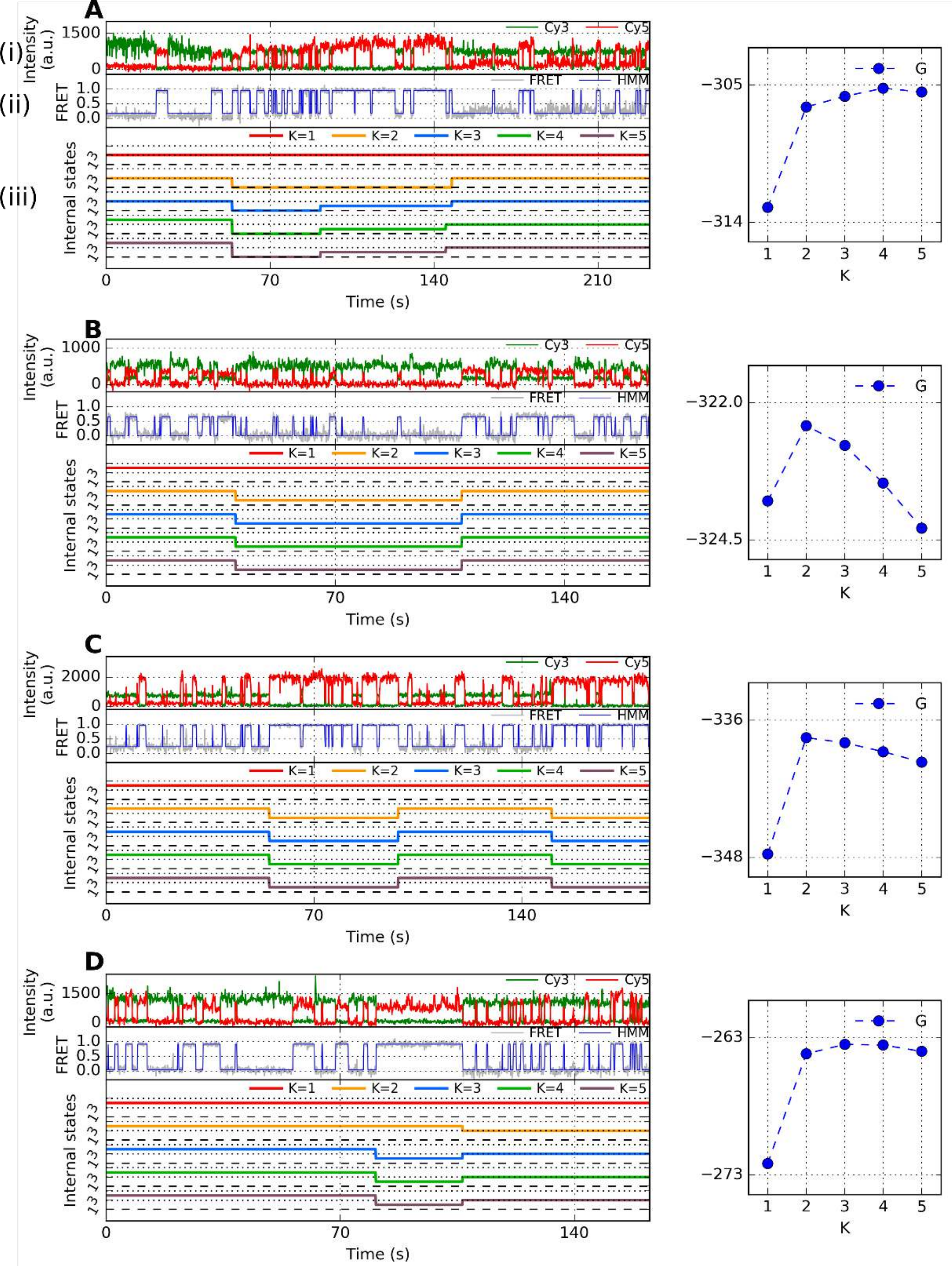}
   	
   	\caption{Representative time traces of H-DNA dynamics and their analysis using VB-DCMM at [Na$^+$]= 26 mM.
   		(A) (i) Representative fluorescence signal and (ii) their FRET state. (iii) Internal states estimated for $K=1, 2, \ldots 5$.
   		Right panel shows $G(K)$ (blue circle).. 
   		(B, C, D) Other representative time traces and their lower bound obtained under the same experimental condition.
   	}
   	\label{fig_r_traces_26mM}
   \end{figure*}    

	\begin{figure*}[h]
		\centering \renewcommand{\thefigure}{S\arabic{figure}} 
		\includegraphics[scale=0.65]{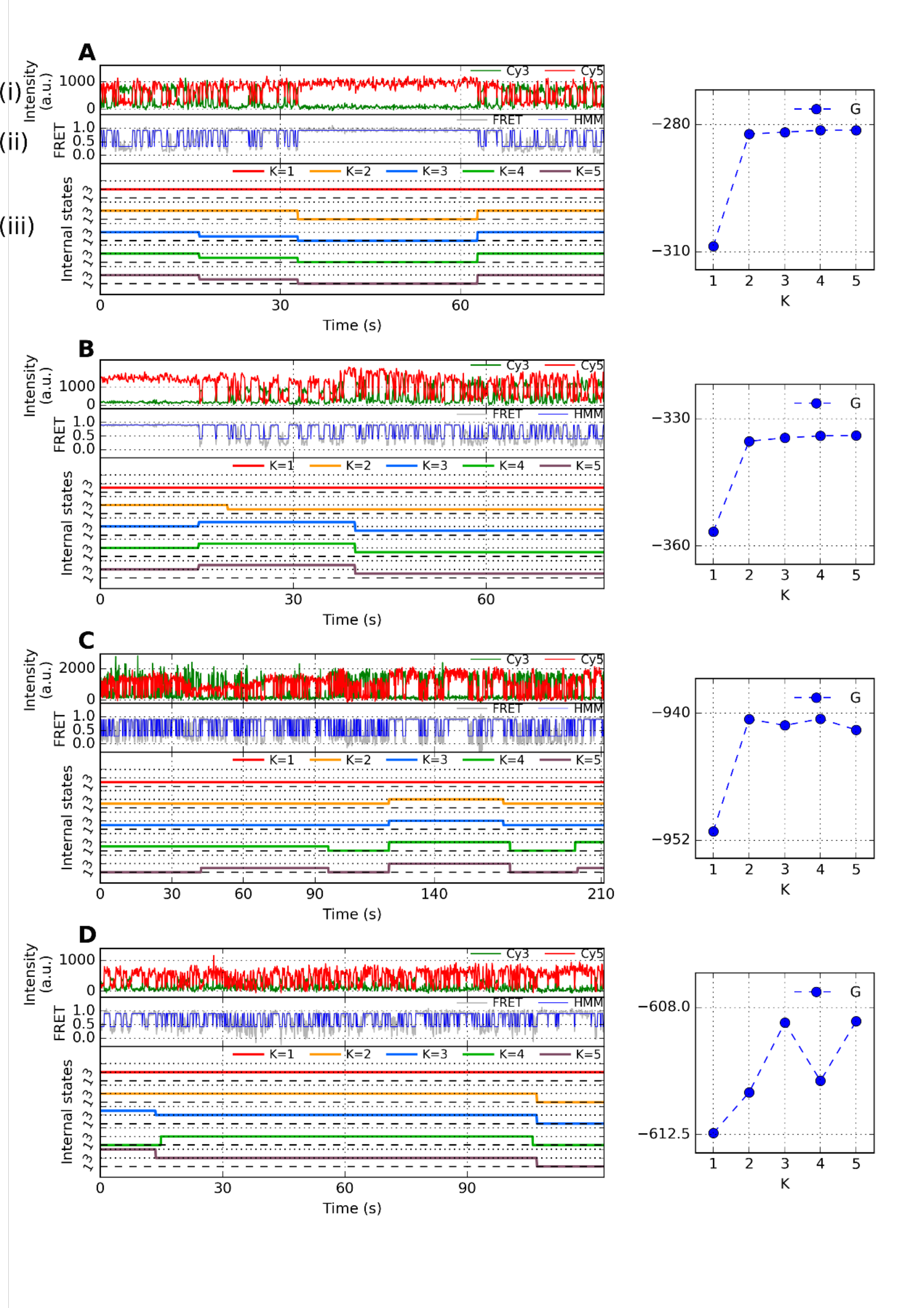}
		
		\caption{
			Representative time traces of H-DNA which display more than two internal states within the trace ($K^* > 2$) and their analysis using VB-DCMM at [Na$^+$]= 100 mM.
			(A) (i) Representative fluorescence signals and (ii) their FRET state. (iii) Internal states estimated for $K=1, 2, \ldots 5$.
			Right panel shows $G(K)$ (blue circle).
			(B-D) Other representative time trace and their lower bound obtained under the same experimental condition.
			Decrease of $G(K)$ at $K=3$ in (C) and at $K=4$ in (D) is due to trapping in local minimum (In the analysis of H-DNA, the best solution was selected after applying VB-DCMM 20 times with random initial conditions for each $K$).
		}
		\label{fig_r_traces_100mM_multiple}
	\end{figure*}

    \begin{figure*}[h]
    	\centering \renewcommand{\thefigure}{S\arabic{figure}} 
    	\includegraphics[scale=0.6]{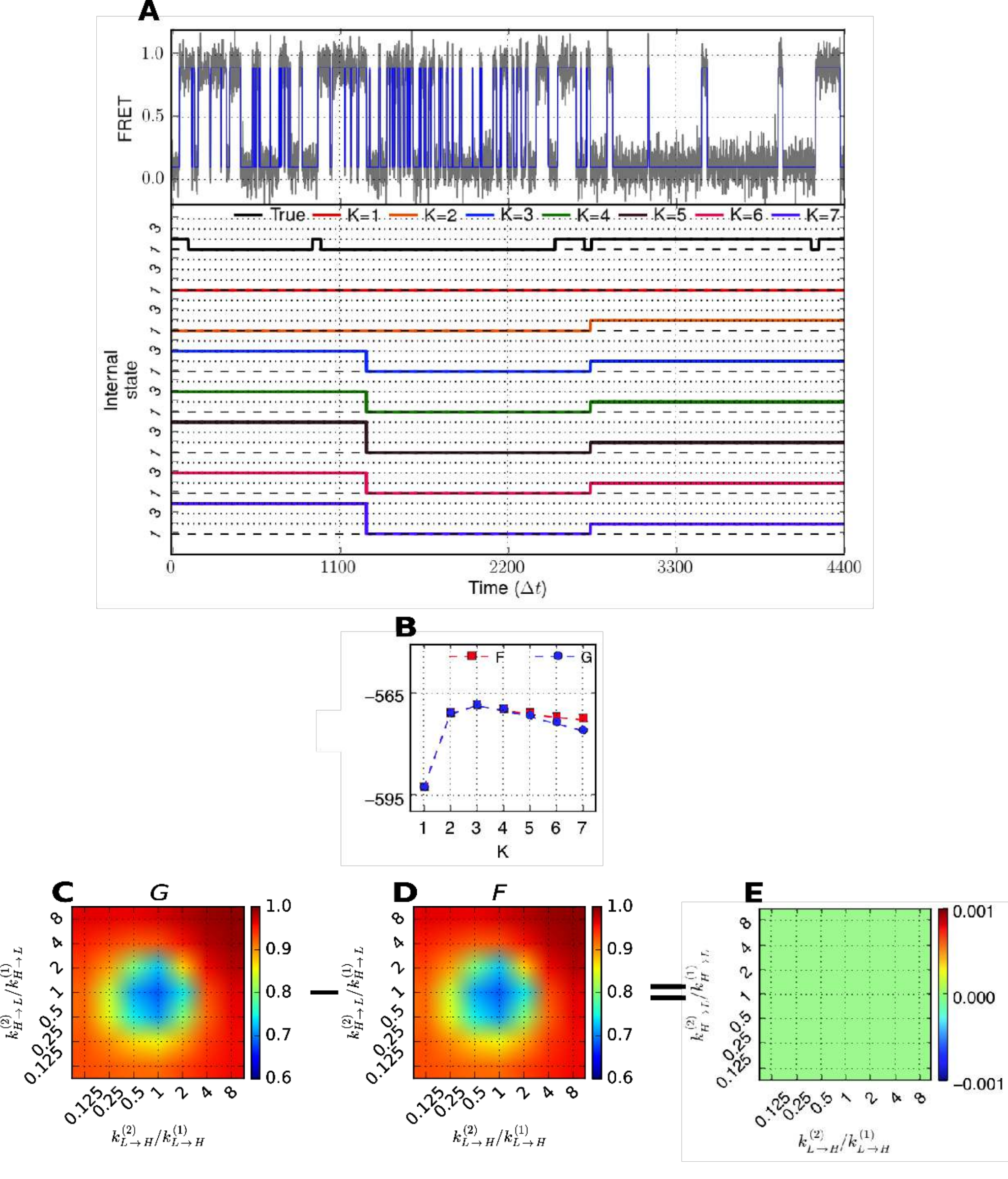}
    	
    	\caption{
    		VB-DCMM analysis on synthetic data generated with following parameters: $T_{obs}  / \Delta t$= 4400, $K^{\text{\text{true}}}=2$, $\gamma^{(1) \rightarrow (2)}\Delta t=\gamma^{(2) \rightarrow (1)}\Delta t=0.001, k^{(1)}_{L \rightarrow H}\Delta t=k^{(1)}_{H \rightarrow L}\Delta t=0.05, k^{(2)}_{L \rightarrow H}\Delta t=0.00625, k^{(2)}_{H \rightarrow L}\Delta t=0.0125$. 
    		(A) (Top): Gray line indicates FRET trace and blue line is noise-filtered FRET obtained after HMM analysis. (Bottom) : The traces of true internal state (Black) and estimated internal state (red: $K=1$, orange: $K=2$, blue: $K=3)$, green: $K=4$, brown: $K=5$, pink: $K=6$, and purple: $K=7$).
    		(B) $F(K)$ (red, square) and $G(K)$ (blue, circle) from the results of VB-DCMM analysis.
    		Accuracy of the model prediction using $\langle \chi \rangle$. 
    		In (C), the final model is selected using $G(K)$ whereas $F(K)$ is used in (D).
    		(E) The difference between $\langle \chi \rangle$s from (C) and (D) is practically zero for all parameter range.
    		}
	\label{fig_r_test_G}
    \end{figure*}

    \begin{figure*}[h]
    	\centering \renewcommand{\thefigure}{S\arabic{figure}} 
    	\includegraphics[scale=0.65]{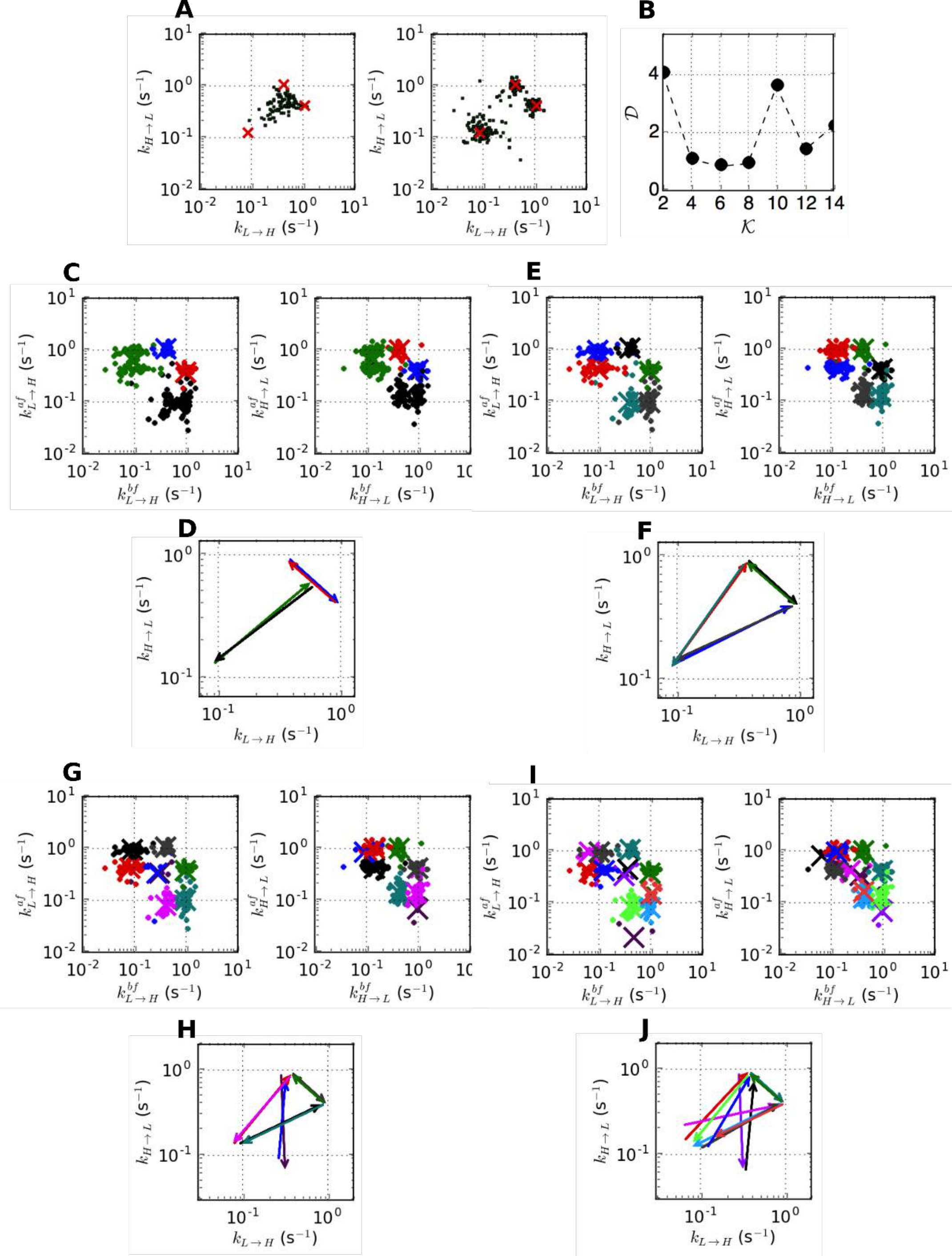}
    	\caption{Clustering synthetic data with three internal states ($K=3$) with two observable states ($N=2$).
    			(A) The scatter plots of ($k_{L \rightarrow H}$, $k_{H \rightarrow L}$) before (left) and after (right) applying VB-DCMM from synthetic data generated with $K=3$ (data from Fig. \ref{fig_r_simul_K3}A, D).
    			Red crosses are the observable transition rates used to generate synthetic data which demonstrates how reliably VB-DCMM can recover the input transition rates.
	    		(B) The sum of pairing distances as a function of the number of centroids, $\mathcal{K}$ (See Methods).
    			The clustering analysis was performed for $\mathcal{K}=4$ (C-D), or $\mathcal{K}=6$ (E-F).
  }      		\label{fig_r_clustering_K3}
    \end{figure*}

   \begin{figure*}[h!]
   	\centering \renewcommand{\thefigure}{S\arabic{figure}} 
   	\includegraphics[scale=0.65]{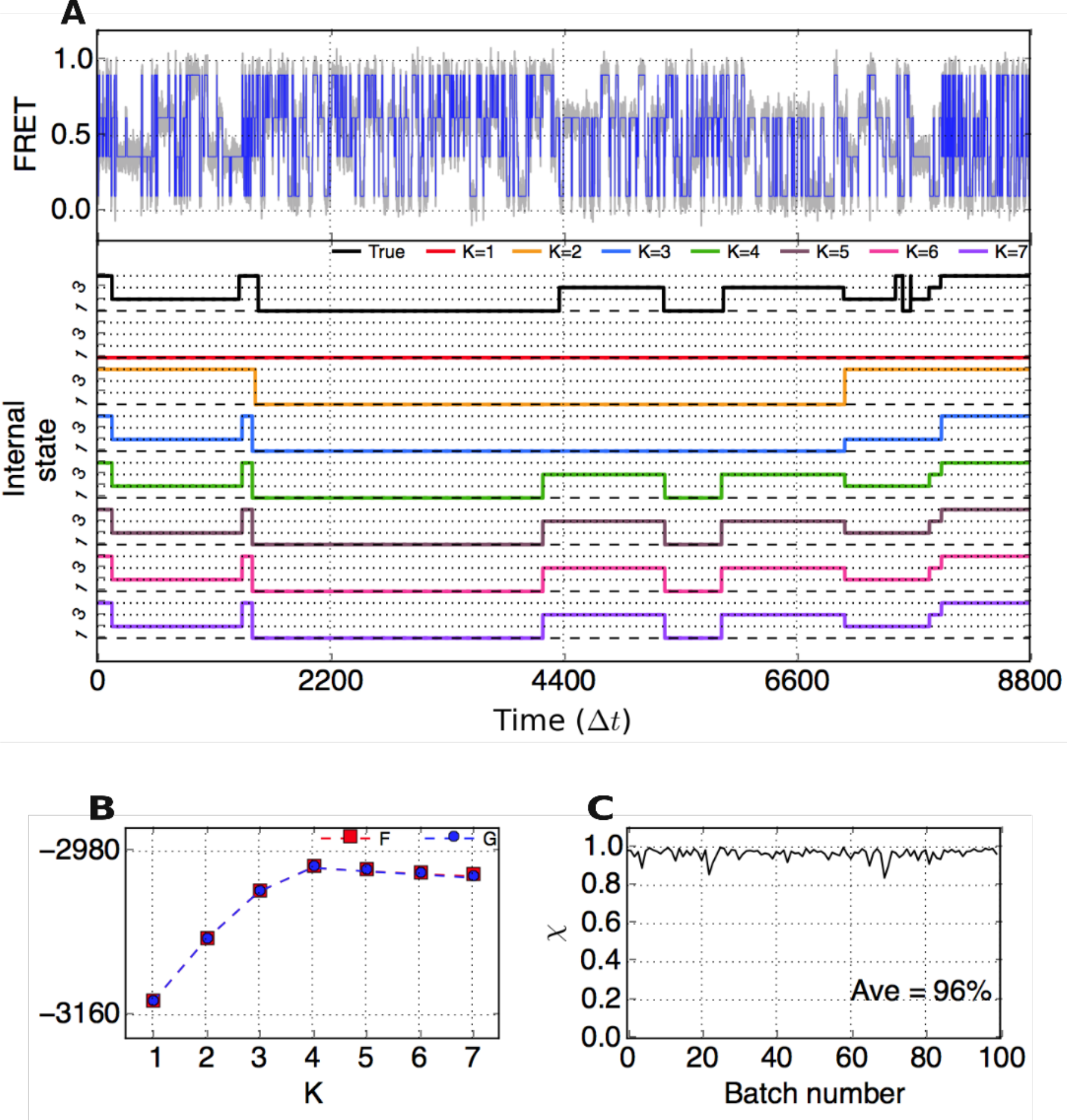}
   	\caption{
   		VB-DCMM analysis on synthetic data having 4 internal states generated with the following parameters: $K^{\text{\text{true}}}=4$, $\gamma^{(1) \rightarrow (2)} \Delta t=\gamma^{ (1) \rightarrow (3) }\Delta t=\gamma^{ (2) \rightarrow (3) }\Delta t=\gamma^{ (3) \rightarrow (2) }\Delta t=\gamma^{ (3) \rightarrow (1) }\Delta t=\gamma^{ (2) \rightarrow (1) }\Delta t=0.00033,
   		k^{(1)}_{1 \rightarrow 2}\Delta t=0.015, 
   		k^{(1)}_{1 \rightarrow 3}\Delta t=0.023,
   		k^{(1)}_{1 \rightarrow 4}\Delta t=0.015,
   		k^{(1)}_{2 \rightarrow 1}\Delta t=0.032, 
   		k^{(1)}_{2 \rightarrow 3}\Delta t=0.05,
   		k^{(1)}_{2 \rightarrow 4}\Delta t=0.025,
   		k^{(1)}_{3 \rightarrow 1}\Delta t=0.03, 
   		k^{(1)}_{3 \rightarrow 2}\Delta t=0.014,
   		k^{(1)}_{3 \rightarrow 4}\Delta t=0.06,
   		k^{(1)}_{4 \rightarrow 1}\Delta t=0.058, 
   		k^{(1)}_{4 \rightarrow 2}\Delta t=0.065,
   		k^{(1)}_{4 \rightarrow 3}\Delta t=0.058,
   		k^{(2)}_{1 \rightarrow 2}\Delta t=0.058, 
   		k^{(2)}_{1 \rightarrow 3}\Delta t=0.065,
   		k^{(2)}_{1 \rightarrow 4}\Delta t=0.058,
   		k^{(2)}_{2 \rightarrow 1}\Delta t=0.011, 
   		k^{(2)}_{2 \rightarrow 3}\Delta t=0.004,
   		k^{(2)}_{2 \rightarrow 4}\Delta t=0.004,
   		k^{(2)}_{3 \rightarrow 1}\Delta t=0.0093, 
   		k^{(2)}_{3 \rightarrow 2}\Delta t=0.014,
   		k^{(2)}_{3 \rightarrow 4}\Delta t=0.003,
   		k^{(2)}_{4 \rightarrow 1}\Delta t=0.06, 
   		k^{(2)}_{4 \rightarrow 2}\Delta t=0.002,
   		k^{(2)}_{4 \rightarrow 3}\Delta t=0.02,
   		k^{(3)}_{1 \rightarrow 2}\Delta t=0.001, 
   		k^{(3)}_{1 \rightarrow 3}\Delta t=0.05,
   		k^{(3)}_{1 \rightarrow 4}\Delta t=0.003,
   		k^{(3)}_{2 \rightarrow 1}\Delta t=0.07, 
   		k^{(3)}_{2 \rightarrow 3}\Delta t=0.08,
   		k^{(3)}_{2 \rightarrow 4}\Delta t=0.01,
   		k^{(3)}_{3 \rightarrow 1}\Delta t=0.03, 
   		k^{(3)}_{3 \rightarrow 2}\Delta t=0.01,
   		k^{(3)}_{3 \rightarrow 4}\Delta t=0.005,
   		k^{(3)}_{4 \rightarrow 1}\Delta t=0.004, 
   		k^{(3)}_{4 \rightarrow 2}\Delta t=0.04,
   		k^{(3)}_{4 \rightarrow 3}\Delta t=0.002,
   		k^{(4)}_{1 \rightarrow 2}\Delta t=0.08, 
   		k^{(4)}_{1 \rightarrow 3}\Delta t=0.002,
   		k^{(4)}_{1 \rightarrow 4}\Delta t=0.02,
   		k^{(4)}_{2 \rightarrow 1}\Delta t=0.011, 
   		k^{(4)}_{2 \rightarrow 3}\Delta t=0.043,
   		k^{(4)}_{2 \rightarrow 4}\Delta t=0.11,
   		k^{(4)}_{3 \rightarrow 1}\Delta t=0.026, 
   		k^{(4)}_{3 \rightarrow 2}\Delta t=0.07,
   		k^{(4)}_{3 \rightarrow 4}\Delta t=0.045,
   		k^{(4)}_{4 \rightarrow 1}\Delta t=0.1, 
   		k^{(4)}_{4 \rightarrow 2}\Delta t=0.07,
   		k^{(4)}_{4 \rightarrow 3}\Delta t=0.03$,
   		Here, sub-indexes $i$, $j$ in $k^{(\mu)}_{i \rightarrow j}$ indicate observables $o$'s with following FRET values:
   		$o=1$, FRET=0.1; $o=2$, FRET=0.36; $o=3$, FRET=0.62; $o=4$, FRET=0.9. 
   		(A) (Top) : Gray line depicts FRET trace, and blue line is noise-filtered FRET obtained by using HMM. (Bottom) : True internal state trace (Black) and estimated internal state traces (red: $K=1$, orange: $K=2$, blue: $K=3$, green: $K=4$, brown: $K=5$, pink: $K=6$, and purple: $k=7$).
   		(B) $F(K)$ and $G(K)$ (Eq. \ref{eq: G(K)}) from VB-DCMM analysis.
   		(C) $\chi$ on 100 traces with $T_{obs} / \Delta t$= 8800,
   	}
   	\label{fig_r_simul_K4L4}
   \end{figure*}  
     \begin{figure*}[h]
     	\centering \renewcommand{\thefigure}{S\arabic{figure}} 
     	\includegraphics[scale=0.65]{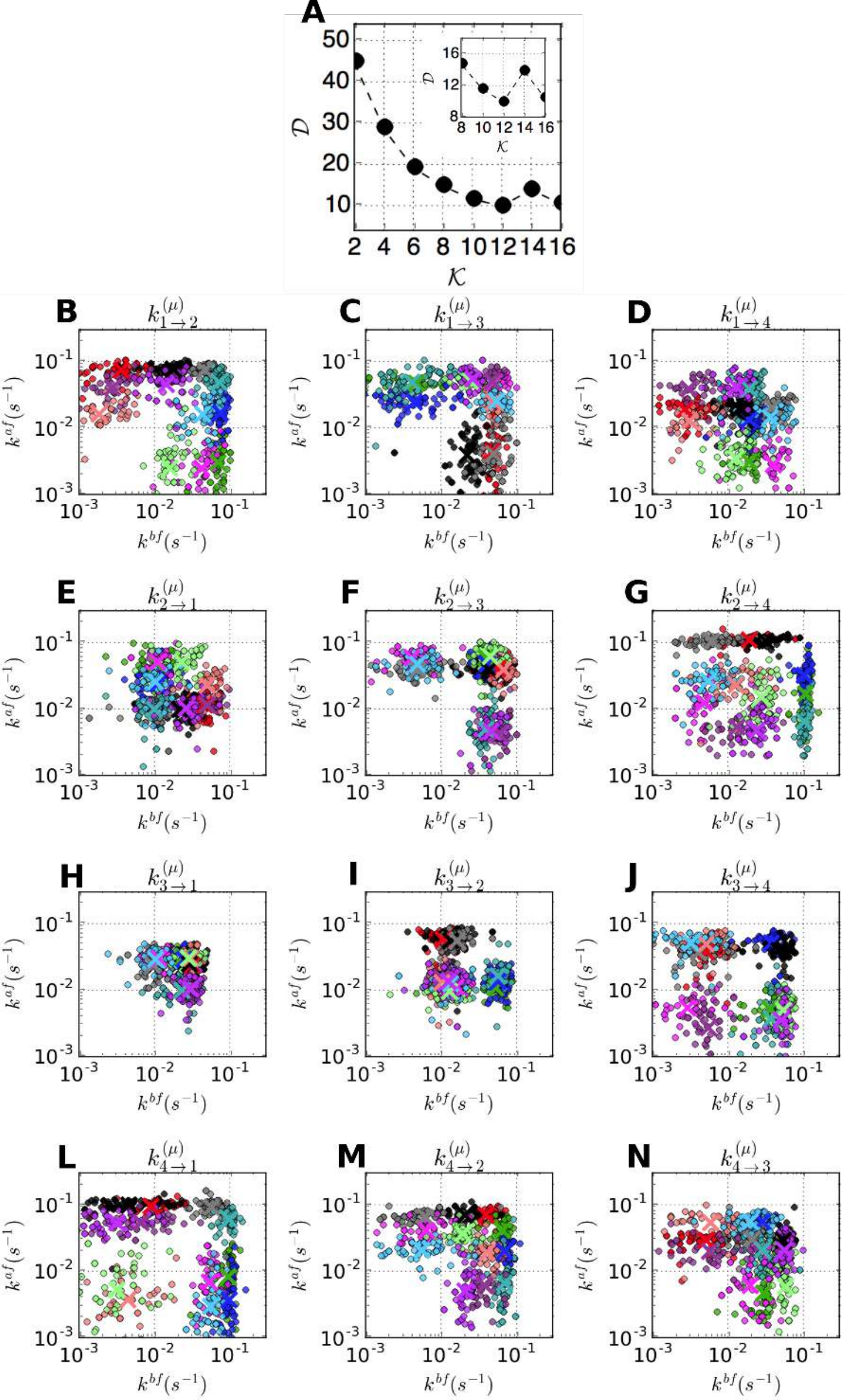}
     	\caption{Clustering synthetic data with four internal states ($K=4$), and four observable states ($N=4$).
		   (A) The sum of pairing distances as a function of $\mathcal{K}$ the number of centroids, $\mathcal{K}$ (See Methods). 
		   The inset shows the region around $\mathcal{K}=12$ of the same graph.
     	   (B) The result of clustering analysis performed when $\mathcal{K}=12$ (B-N).
     	}
     	\label{fig_r_clustering_K4L4_connected}
     \end{figure*}        

    \begin{figure*}[h]
    	\centering \renewcommand{\thefigure}{S\arabic{figure}} 
    	\includegraphics[scale=0.65]{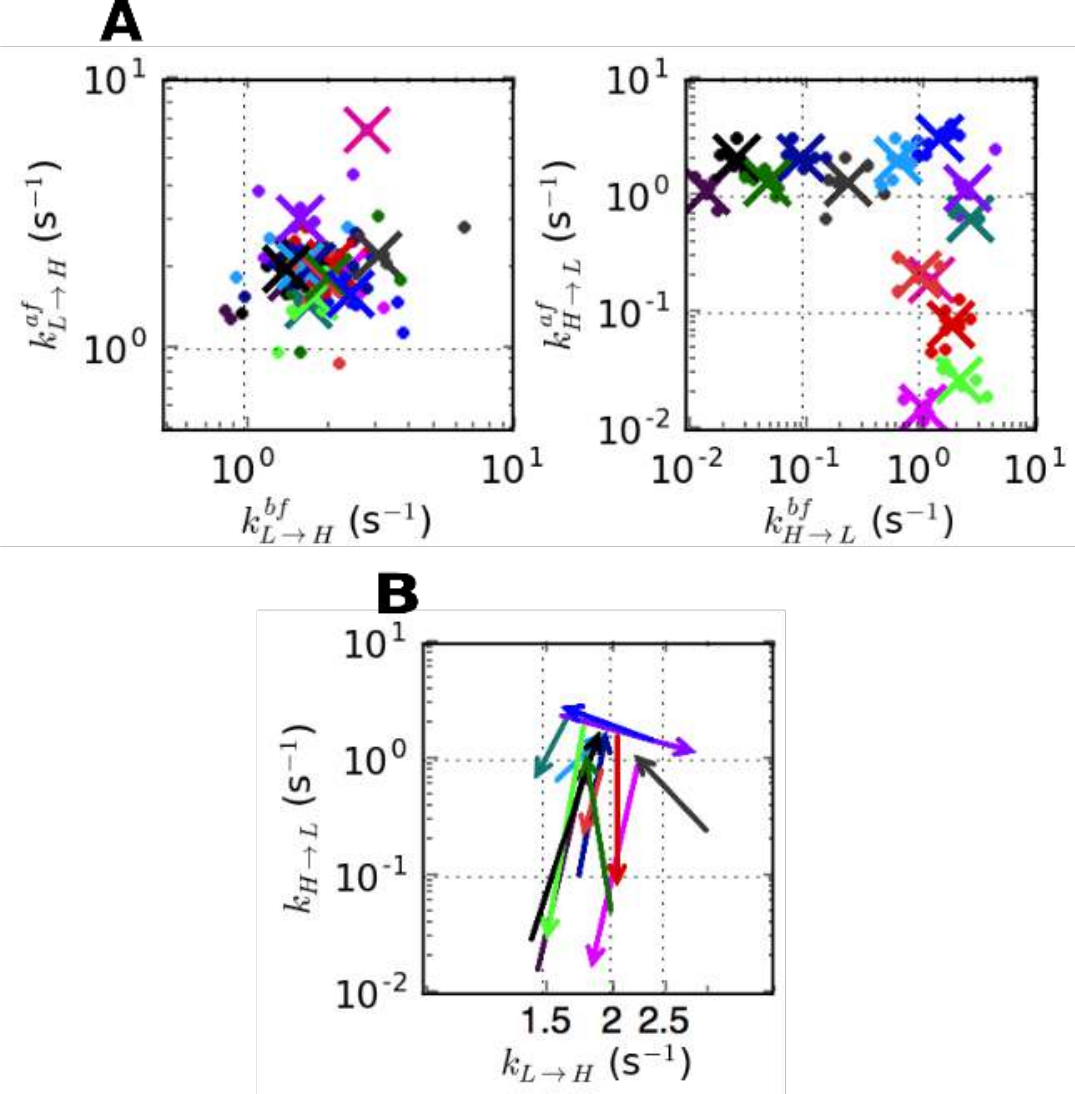}
    	\caption{Clustering H-DNA data ([Na$^+$] =  100 mM).
    		(A-B) The clustering analysis used in Fig. \ref{fig_r_clustering_100mM_6}C, D was performed for $\mathcal{K}=14$.
    		       }
    		\label{fig_r_clustering_100mM_2}
    \end{figure*}
     
     \begin{figure*}[h]
        	\centering \renewcommand{\thefigure}{S\arabic{figure}} 
        	\includegraphics[scale=0.6]{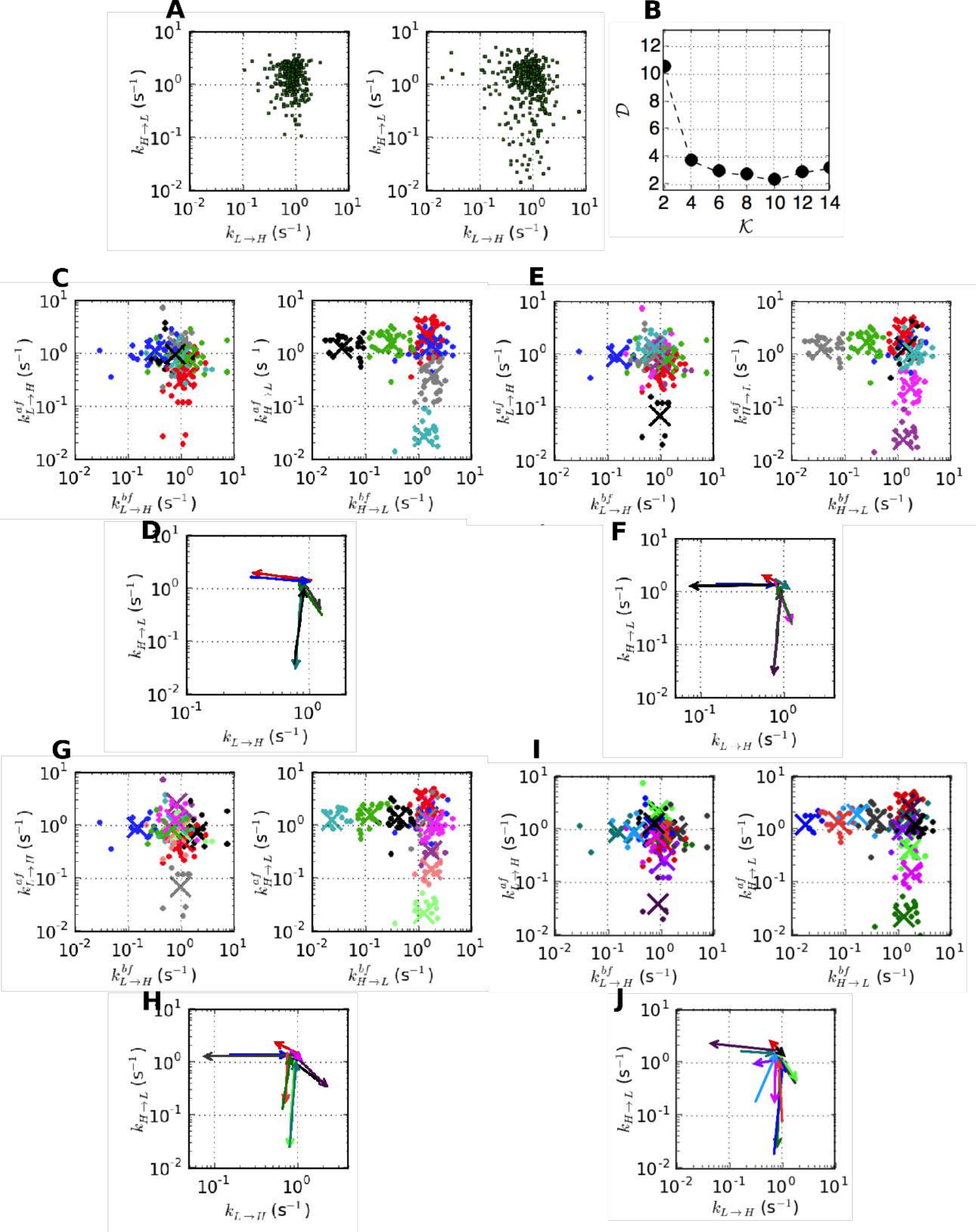}
        	\caption{Clustering H-DNA data ([Na$^+$] =  50 mM).
    			(A) The scatter plots of ($k_{L \rightarrow H}$, $k_{H \rightarrow L}$) before (left) and after (right) applying VB-DCMM from [Na$^+$] =  50 mM H-DNA data.
    			(B) The sum of pairing distances as a function of the number of centroids (See Methods).
    			The clustering analysis used in Fig. \ref{fig_r_clustering_100mM_6}C, D was performed for $\mathcal{K}=6$ (C-D), $\mathcal{K}=8$ (E-F), $\mathcal{K}=10$ (G-H), or $\mathcal{K}=12$ (I-J).
    			Total 186 data points were used in each clustering analysis.
        	}
        	\label{fig_r_clustering_50mM}
     \end{figure*} 
     
        \begin{figure*}[h]
        	\centering \renewcommand{\thefigure}{S\arabic{figure}} 
        	\includegraphics[scale=0.6]{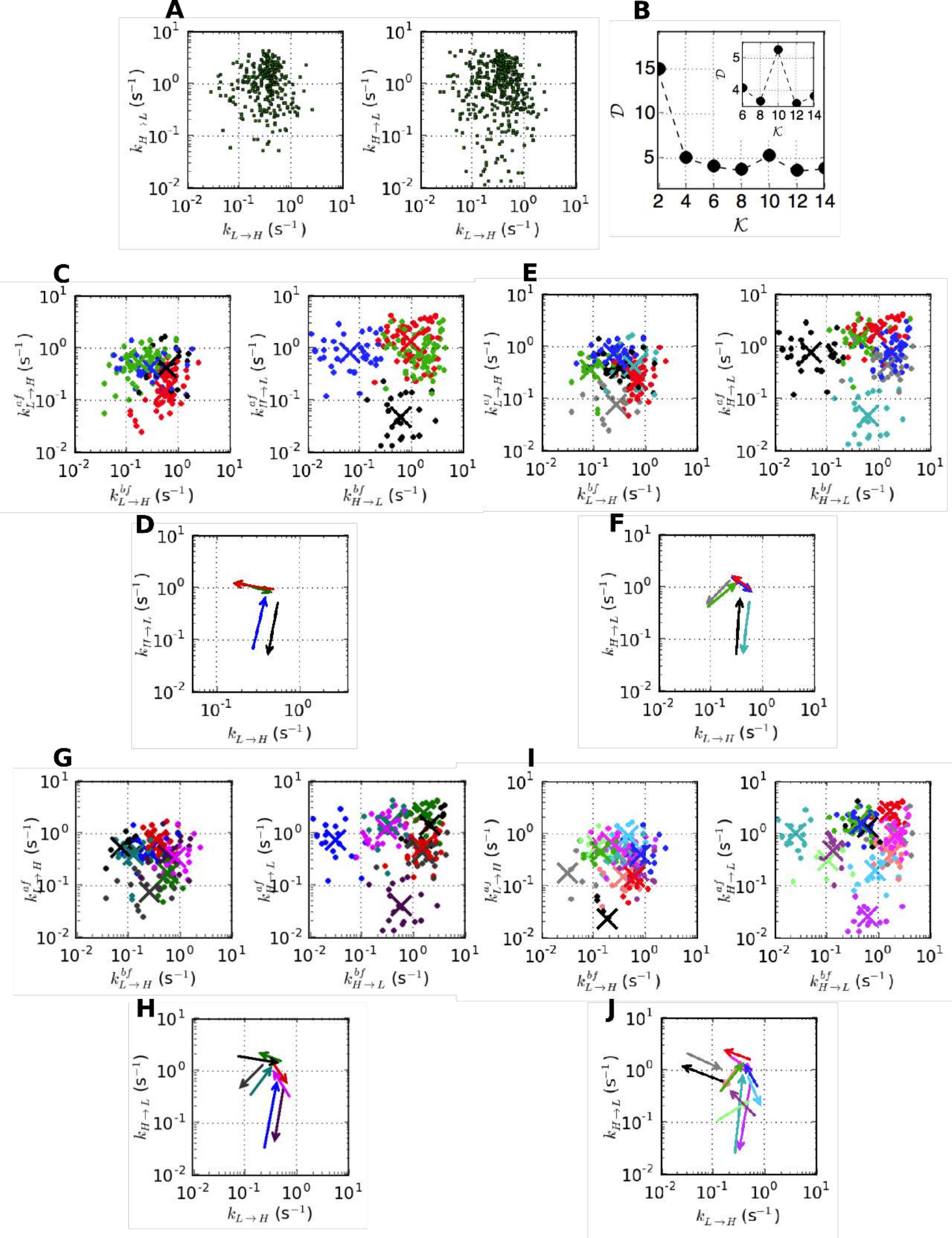}
        	\caption{Clustering H-DNA data ([Na$^+$] =  26 mM).
    			(A) The scatter plots of ($k_{L \rightarrow H}$, $k_{H \rightarrow L}$) before (left) and after (right) applying VB-DCMM from [Na$^+$] =  26 mM H-DNA data.
    			The inset shows the region around $\mathcal{K}=12$ of the same graph.
    			(B) The sum of pairing distances as a function of the number of centroids (See Methods).
    			The clustering analysis used in Fig. \ref{fig_r_clustering_100mM_6}C, D was performed by assuming $\mathcal{K}=4$ (C-D), $\mathcal{K}=6$ (E-F), $\mathcal{K}=8$ (G-H), or $\mathcal{K}=12$ (I-J).
    			Total 185 data points were used in each analysis.
        	}
        	\label{fig_r_clustering_26mM}
    \end{figure*}
    
    	\begin{figure*}[h]
		\centering \renewcommand{\thefigure}{S\arabic{figure}} 
		\includegraphics[scale=0.65]{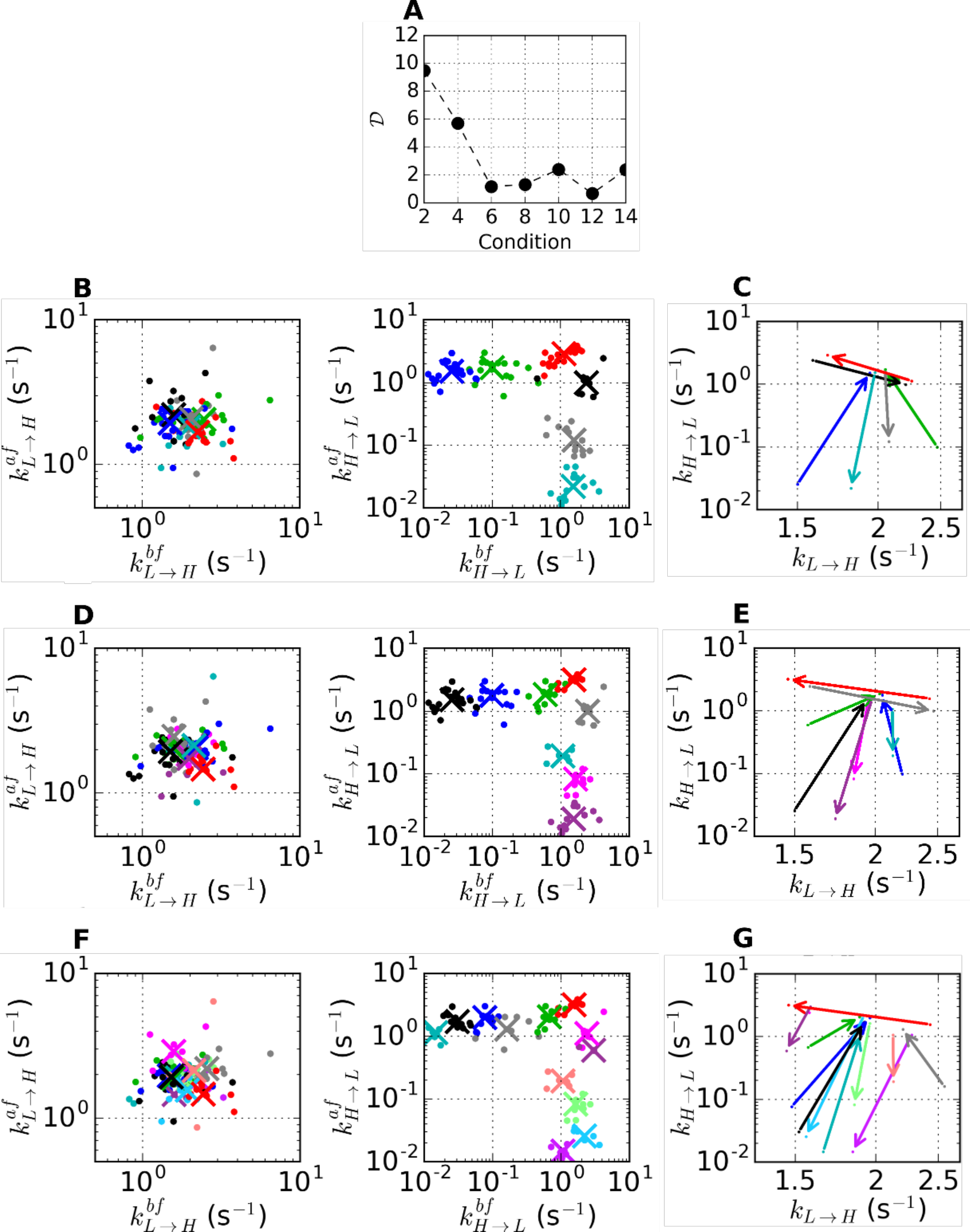}
		\caption{
			H-DNA data ([Na$^+$] =  100 mM) analyzed with k-means clustering algorithm using ``city block'' distance ($L1$-distance).
			(A) The sum of pairing distances as a function of the number of centroids (See Methods).
			The clustering analysis done in Fig. \ref{fig_r_clustering_100mM_6}C, D was performed again, but using ``city block'' distance. 
			The clustering results are presented for $\mathcal{K}=6$ (B-C), $\mathcal{K}=8$ (D-E), and  $\mathcal{K}=12$ (F-G).
			Although $\mathcal{D}$ is minimized at $\mathcal{K}=12$, 
			10 clusters out of 12 contain less than 10 data points, which is statistically not significant.
			Thus, $\mathcal{K}=6$, corresponding to the suboptimal point of $\mathcal{D}$, could still be considered as the best solution.
		}
		\label{fig_r_clustering_100mM_cityblock}
	\end{figure*}
%

	    \begin{figure*}[h]
		\centering \renewcommand{\thefigure}{S\arabic{figure}} 
		\includegraphics[scale=0.65]{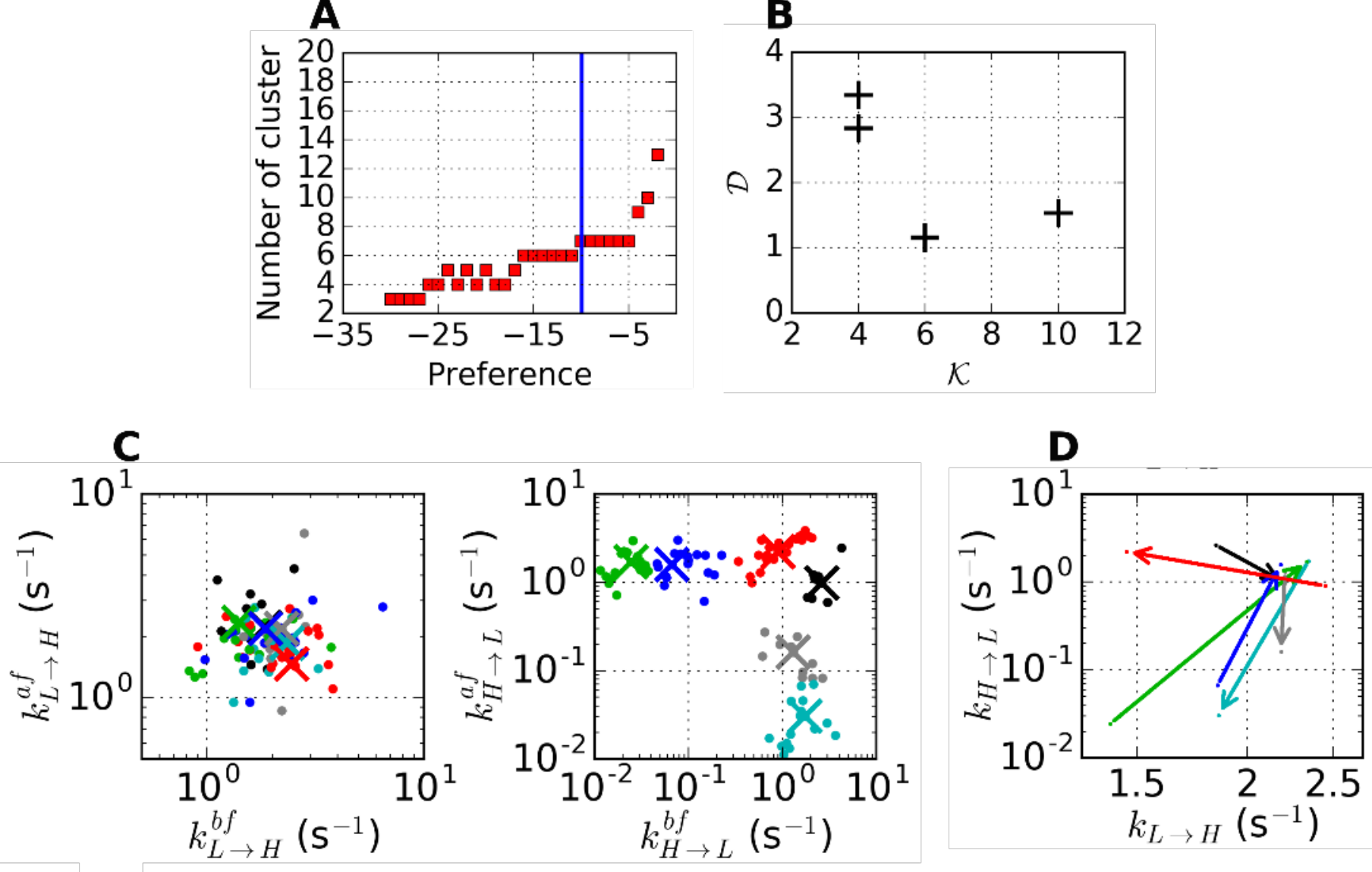}
		\caption{
			Clustering results of H-DNA data ([Na$^+$] =  100 mM) from ``affinity propagation'' \cite{Frey:Science:2007} that uses negative square-euclidean distance as the similarity metric.
			(A) The number of clusters calculated with varying ``preference" parameter, where the preference denotes an input parameter (self-similarity) in the ``affinity propagation'' method. 
			All the points were set to have the same preference value. 
			Blue vertical line denotes median value of similarities between data points.
			(B) The sum of pairing distances as a function of $\mathcal{K}$. 
			Only the results with even number of clusters in (A) are plotted. 
The clustering results at $\mathcal{K}=6$ are shown in (C-D).
		}
		\label{fig_r_af_100mM}
		
	\end{figure*}
	

	\begin{figure*}[h]
		\centering \renewcommand{\thefigure}{S\arabic{figure}} 
		\includegraphics[scale=0.65]{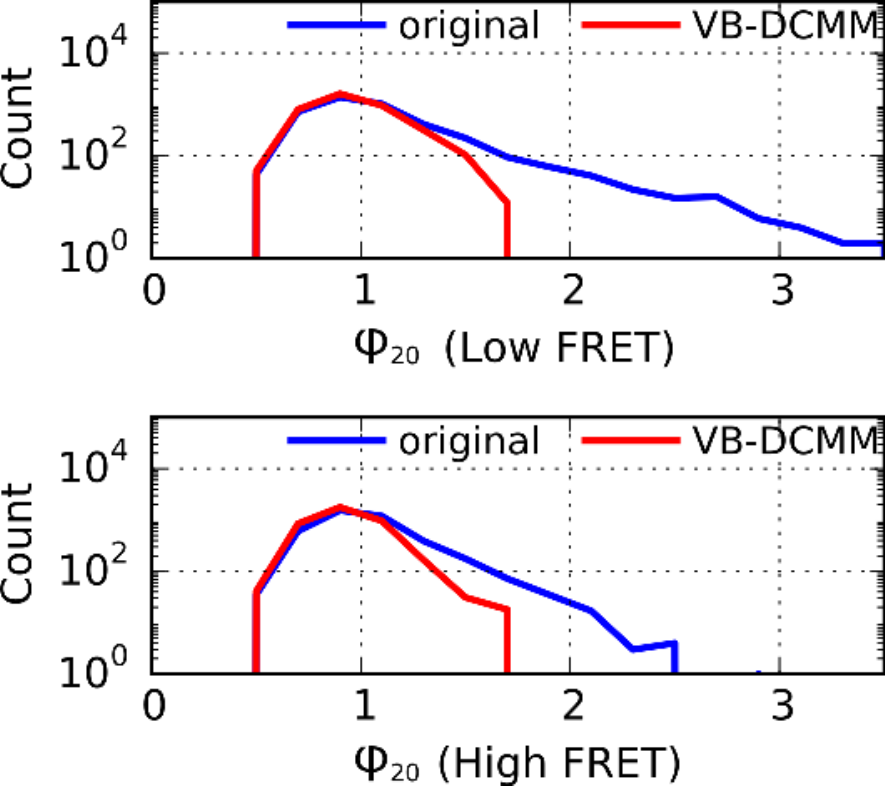}
		
		\caption{
			Comparison of $\varphi_{20} = \sigma_{20} / \mu_{20}$ histograms on synthetic data before and after removing dynamic heterogeneity (red) by decomposing the original traces into the pieces according to estimated internal state trace. The data used in Fig. \ref{fig_r_simul_K3} was analyzed.
			(Top): Comparison of $\varphi_{20} = \sigma_{20} / \mu_{20}$ histograms for low FRET dwell time data.
			(Bottom): Comparison of $\varphi_{20} = \sigma_{20} / \mu_{20}$ histograms for high FRET dwell time data.
		}
		\label{fig_r_simul_phi}
	\end{figure*} 

    \clearpage

\end{document}